                        \newcommand{\version}{
  Predrag       May 19 2012
                        }
\pdfoutput=1
    \newif\ifboyscout %\boyscouttrue   % default:  draft version
    \boyscoutfalse                   % uncomment for final version

\documentclass[
numberedheadings       % uncomment this option for numbered sections
  ]{aipproc}
\layoutstyle{6x9}
\usepackage{float}
\usepackage{amsmath,amsfonts,amssymb,amsbsy,amscd,amsgen}
\usepackage{dcolumn}% Align table columns on decimal point
\usepackage{bm}% bold math
\usepackage{color}

\graphicspath{{../figs/}{../Fig/}}   %% directories with graphics

\newcommand{\Fokker}{Fokker-Planck}
\newcommand{\diffCon}{\ensuremath{D}}         % diffusion constant
\newcommand{\diffTen}{\ensuremath{\Delta}}  % diffusion tensor
\newcommand{\covMat}{\ensuremath{Q}}             % covariance matrix
\newcommand{\Lnoise}[1]{{\cal L}^{#1}}    % noisy evolution operator, La Tesa
\newcommand{\Lmat}[1]{{{\bf L}_{#1}}}      % evolution matrix
\newcommand{\orbitDist}{\ensuremath{z}}     % Langevin distance from orbit point
\newcommand{\optPart}{optimal partition}

\newcommand{\timeStep}{{\delta t}}        %integration step
\newcommand{\Lop}{\ensuremath{{\cal L}_{det}}}  % evolution operator, La Tesa
\newcommand{\Uop}{\ensuremath{{\cal L}^{t \dagger}_{det}}}  % Koopman, adjoint notation

\newcommand{\Reynolds}{\textit{Re}}  % Reynolds number
\newcommand{\NS}{Navier-Stokes}

\newcommand{\Df}[1]{\map^{'}_{#1}}
\newcommand{\DDf}[1]{\map^{''}_{#1}}

\ifboyscout %%%%%%%% DISPLAY COMMENTS IN THE TEXT %%%%%%%%%%%%%%%%%%%%
\newcommand{\toCB}{\marginpar{\footnotesize 2CB}}  % compare with ChaosBook
    
  \newcommand{\PC}[1]{$\footnotemark\footnotetext{Predrag: #1}$}
  \newcommand{\DL}[1]{$\footnotemark\footnotetext{Domenico: #1}$}
  \newcommand{\PCedit}[1]{{\color{red}#1}}
  \newcommand{\DLedit}[1]{{\color{blue}#1}}
  \newcommand{\GMW}[1]{$\footnotemark\footnotetext{Gable: #1}$}
  \newcommand{\GMWedit}[1]{{\color{green}#1}}
  \newcommand{\DK}[1]{$\footnotemark\footnotetext{Daniel: #1}$}
  \newcommand{\DKedit}[1]{{\color{red}#1}}
\else % drop comments
  \newcommand{\toCB}{}
  \newcommand{\PC}[1]{}
  \newcommand{\PCedit}[1]{#1}
  \newcommand{\DL}[1]{}
  \newcommand{\DLedit}[1]{#1}
  \newcommand{\GMW}[1]{}
  \newcommand{\GMWedit}[1]{#1}
  \newcommand{\DK}[1]{}
  \newcommand{\DKedit}[1]{#1}
\fi  %%%%%%%%%%%% END OF ON/OFF COMMENTS SWITCH %%%%%%%%%%%%%%%%%%%%

       \newcommand{\href}[2]{{#2}}  % no hyperref
       \newcommand{\HREF}[2]{{#2}}
       \newcommand{\wwwcb}[1]{{\tt ChaosBook.org#1}}

\newtheorem{rmark}{{\small\textsf{\textbf{Remark}}}}[section]
\newcommand{\remark}[2]{
    \vspace*{1ex}

      \begin{rmark}
        {\small\em\noindent {\small\sf #1 ~} #2 } % 2010-12-12 experiment
      \end{rmark}
    \vspace*{1ex}
              }

\newcommand{\beq}{\begin{equation}}
\newcommand{\continue}{\nonumber \\ }
\newcommand{\nnu}{\nonumber}
\newcommand{\eeq}{\end{equation}}
\newcommand{\ee}[1] {\label{#1} \end{equation}}
\newcommand{\bea}{\begin{eqnarray}}
\newcommand{\ceq}{\nonumber \\ & & }
\newcommand{\eea}{\end{eqnarray}}
\newcommand{\barr}{\begin{array}}
\newcommand{\earr}{\end{array}}

\newcommand{\rf}     [1] {~\cite{#1}}
\newcommand{\refref} [1] {ref.~\cite{#1}}

\newcommand{\refrefs}[1] {refs.~\cite{#1}}

\newcommand{\refeq}  [1] {(\ref{#1})}

\newcommand{\reffig} [1] {figure~\ref{#1}}

\newcommand{\refFig} [1] {Figure~\ref{#1}}

\newcommand{\refsect}[1] {sect.~\ref{#1}}

\newcommand{\refappe}[1] {appendix~\ref{#1}}

\newcommand{\refrem} [1] {remark~\ref{#1}}

\newcommand{\etc}{{etc.}}       % APS
\newcommand{\etal}{{\em et al.}}    % etal in italics, APS too
\newcommand{\ie}{{i.e.}}        % APS

\newcommand{\evOper}{evolution oper\-ator}

\newcommand{\FPoper}{Perron-Frobenius oper\-ator} % Pesin's ordering
\newcommand{\statesp}{state space}

\newcommand{\jacobianM}{Jacobian matrix}  % back to Predrag's name 20oct2009
\newcommand{\JacobianM}{Jacobian matrix} %
\newcommand{\FloquetM}{Floquet matrix} % specialized to periodic orb
\newcommand{\stabmat}{stability matrix}     % stability matrix, velocity gradients
\newcommand{\monodromyM}{monodromy matrix} % monodromy matrix, Poincare cut
\newcommand{\Fd}{spec\-tral det\-er\-min\-ant}

\newcommand{\MatrixII}[4]{\left(
\begin{array}{cc}
{#1}  &  {#2} \\
{#3}  &  {#4} \end{array} \right)}

\newcommand{\obser}{\ensuremath{a}}     % an observable from phase space to R^n
\newcommand{\Obser}{\ensuremath{A}}     % time integral of an observable
\newcommand{\reals}{\mathbb{R}}
\newcommand{\complex}{\mathbb{C}}

\newcommand{\pde}{\partial}
\newcommand{\tr}{\mbox{\rm tr}\,}
\newcommand{\msr}{\ensuremath{\rho}}                % measure
\newcommand{\SRB}{{\rho_0}}             % natural measure
\newcommand{\prpgtr}[1]{\delta\negthinspace\left( {#1} \right)}
\newcommand{\expct}    [1]{\left\langle {#1} \right\rangle}
\newcommand{\pS}{\ensuremath{{\cal M}}}          % symbol for state space
\newcommand{\ssp}{\ensuremath{x}}                % state space point
\newcommand{\intM}[1]{{\int_\pS{\!d #1}\:}} %phase space integral
\newcommand{\PoincS}{\ensuremath{{\cal P}}}  % symbol for Poincare section
\newcommand{\matId}{\ensuremath{{\bf 1}}}      % matrix identity
\newcommand\map{f}                  % other people like \phi's etc
\newcommand\flow[2]{{f^{#1}(#2)}}
\newcommand{\vel}{\ensuremath{v}}   % state space velocity
\newcommand\velField[1]{{v(#1)}}    % ODE velocity field
\newcommand\xInit{{x_0}}        %initial x
\newcommand{\cl}[1]{{\ensuremath{n_{#1}}}}   % discrete length of a cycle, Predrag
\newcommand{\Mvar}{\ensuremath{A}}  % stability matrix
\newcommand{\monodromy}{\ensuremath{M}}   % monodromy matrix, full Poincare cut
\newcommand{\jEigvec}[1][]{\ensuremath{{\bf e}^{(#1)}}} % right jacobiam eigenvector
\newcommand{\oneMinJ}[1]
           {\left|\det\!\left(\matId-\monodromy_p^{#1}\right)\right|}
\newcommand{\ExpaEig}{\ensuremath{\Lambda}}
\newcommand{\Lyap}{\ensuremath{\lambda}}            %Lyapunov exponent

\newcommand{\eigExp}[1][]{
     \ifthenelse{\equal{#1}{}}{\ensuremath{\lambda}}{\ensuremath{\lambda^{(#1)}}}}
\newcommand{\eigRe}[1][]{
     \ifthenelse{\equal{#1}{}}{\ensuremath{\mu}}{\ensuremath{\mu^{(#1)}}}}
\newcommand{\eigIm}[1][]{
     \ifthenelse{\equal{#1}{}}{\ensuremath{\omega}}{\ensuremath{\omega^{(#1)}}}}

\newcommand{\po}{periodic orbit}

\newcommand{\eqv}{equi\-lib\-rium}

\newcommand{\costFct}{cost function}    % functional to minimize
\newcommand{\dmn}{-dimensional}  %  experimental 220ct2009
\newcommand{\cycle}[1]{\ensuremath{\overline{#1}}}
\newcommand\stagn{q}      %equilibrium/stagnation point suffix
\newcommand{\markGraph}{transition graph} % following Yorke
\begin{document}

\title{
Knowing when to stop: How noise frees us from determinism
}

%choose from http://www.aip..org/pacs/index.html


\author{Predrag Cvitanovi\'c}{
  address={Center for Nonlinear Science,
        School of Physics,
        \\
        Georgia Institute of Technology,
        Atlanta, GA 30332-0430}
}

\author{Domenico Lippolis}{
  address={Department of Physics, Pusan National University,
            Busan 609-735, South Korea}
}

\begin{abstract}
Deterministic chaotic dynamics presumes that the \statesp\ can be
partitioned arbitrarily finely. In a physical system, the inevitable
presence of some noise sets a finite limit to the finest possible
resolution that can be attained. Much previous research deals with what
this attainable resolution might be, all of it based on a global averages
over stochastic flow. We show how to compute the \emph{locally} optimal
partition, for a given dynamical system and given noise, in terms of
local eigenfunctions of the \Fokker\ operator and its adjoint. We first
analyze the interplay of the deterministic dynamics with the noise in the
neighborhood of a periodic orbit of a map, by using a discretized version
of Fokker-Planck formalism. Then we propose a method to determine the
`optimal resolution' of the state space, based on solving Fokker-Planck's
equation locally, on sets of unstable periodic orbits of the
deterministic system. We test our hypothesis on unimodal maps.
\end{abstract}

\date{May 20, 2012}
%\date{\today}
\maketitle

\section{Introduction}
\label{DL:intro}

The effect of noise on the behavior of a nonlinear dynamical system is a
fundamental problem in many areas of science%
\rf{vKampen92,LM94,Risken96}, and the interplay of noise and chaotic
dynamics is of particular interest\rf{gasp02,Fogedby05a,Fogedby06a}.
Our purpose here is two-fold. First, we address operationally the fact
that weak noise limits the attainable resolution of the \statesp\ of a
chaotic system by formulating the  {\em \optPart\ hypothesis.} In
\refref{LipCvi08} we have shown that the hypothesis enables us to define
the \optPart\ for a 1\dmn\ map; here we explain how it is implemented for
a high-dimensional \statesp\ flows with a few expanding directions, such
as the transitional \Reynolds\ number \NS\ flows. Second, we show that
the \optPart\ hypothesis replaces the \Fokker\ PDEs by finite,
low-dimensional matrix \Fokker\ operators, with finite cycle expansions,
optimal for a given level of precision, and whose eigenvalues give good
estimates of long-time observables (escape rates, Lyapunov exponents,
\etc).
  \GMW{footnote, \GMWedit{in-text edit.}}
  \DK{footnote, \DKedit{in-text edit.}}

%
%%%%%%%%%%%%%%%%%%%%%%%%%%%%%%%%%%%%%%%%%%%%%%%%%%%%%%%%%%%%%%%%%%
\begin{figure}%[tbp]
\begin{minipage}[t]{0.80\columnwidth}%
(a)~\includegraphics[width=0.37\textwidth]{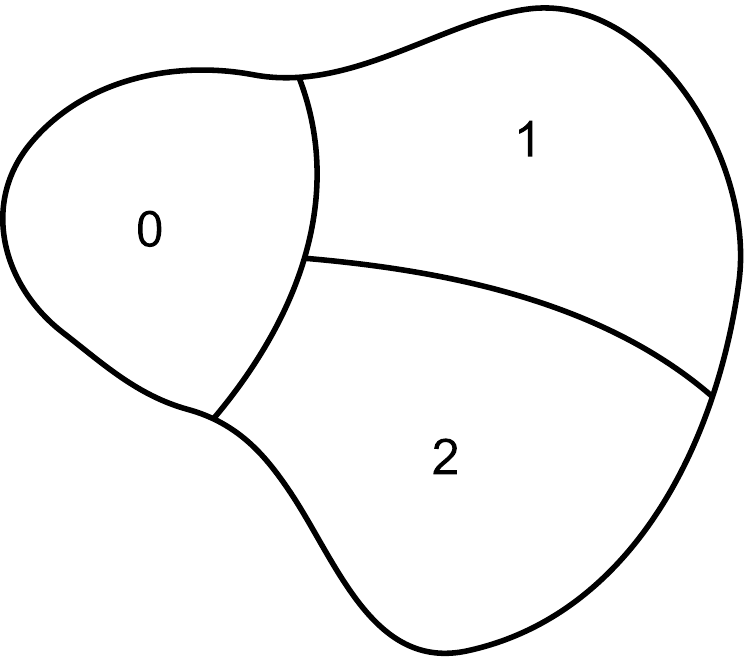}
(b)~\includegraphics[width=0.37\textwidth]{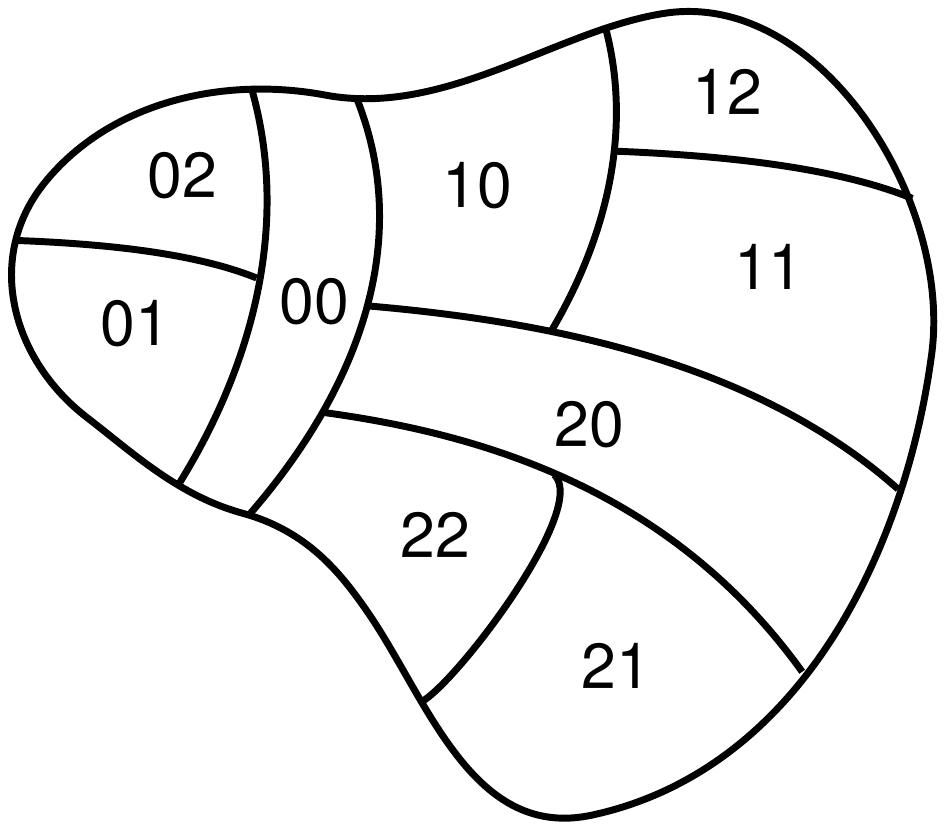}
\end{minipage}
\source{ChaosBook.org}
\caption{
(a) A coarse partition of \statesp\ $\pS$ into regions
$\pS_0$, $\pS_1$, and $\pS_2$, labeled by ternary alphabet
$ {\cal A} = \{1,2,3\}$.
(b) A 1-step memory refinement of the partition of
\reffig{f:kneadPartit}, with each region $\pS_i$ subdivided
into $\pS_{i0}$, $\pS_{i1}$, and $\pS_{i2}$, labeled by nine
`words' $\{00,01,02,\cdots,21,22\}$.
%Chaosbook{f:kneadRefinePartit}
}
\label{f:kneadPartit}
\end{figure}
%%%%%%%%%%%%%%%%%%%%%%%%%%%%%%%%%%%%%%%%%%%%%%%%%%%%%%%%%%%%%%%%%%
%

A chaotic trajectory explores a strange attractor, and
for chaotic flows
evaluation of long-time averages requires effective partitioning
of the \statesp\ into smaller regions.
In a hyperbolic, everywhere unstable deter\-mi\-ni\-stic
dynamical system, consecutive Poincar\'e section returns
subdivide the \statesp\ into exponentially growing number
of regions, each region labeled by a distinct finite symbol
sequence, as in  \reffig{f:kneadPartit}.
														\toCB
In the unstable directions these regions stretch, while in the stable
directions they shrink exponentially. The set of unstable \po s forms a
`skeleton' that can be used to implement such partition of the \statesp,
each region a neighborhood of a periodic point\rf{ruelle,inv}. Longer and
longer cycles yield finer and finer partitions as the neighborhood of
each unstable cycle $p$ shrinks exponentially with cycle period as
$1/|\ExpaEig_p|$, where  $\ExpaEig_p$ is the product of cycle's expanding
Floquet multipliers. As there is an exponentially  growing infinity of
longer and longer cycles, with each neighborhood shrinking asymptotically
to a point, a deter\-mi\-ni\-stic chaotic system can - in principle - be
resolved arbitrarily finely. But that is a fiction for any of the
following reasons:

\begin{itemize}
  \item
any physical system experiences (background, observational,
intrinsic, measurement, $\cdots$) noise
  \item
any numerical computation is a noisy
process
due to the finite precision of each
step of computation
  \item
any set of dynamical equations models nature up to a given finite
accuracy, since degrees of freedom are always neglected
  \item
any prediction only needs to be computed to a desired finite accuracy
\end{itemize}

\begin{figure}%[tbp]
\begin{minipage}[t]{0.80\columnwidth}%
(a)~\includegraphics[width=0.36\textwidth]{SymbolSeqRef}
~~
(b)~~~\includegraphics[width=0.36\textwidth]{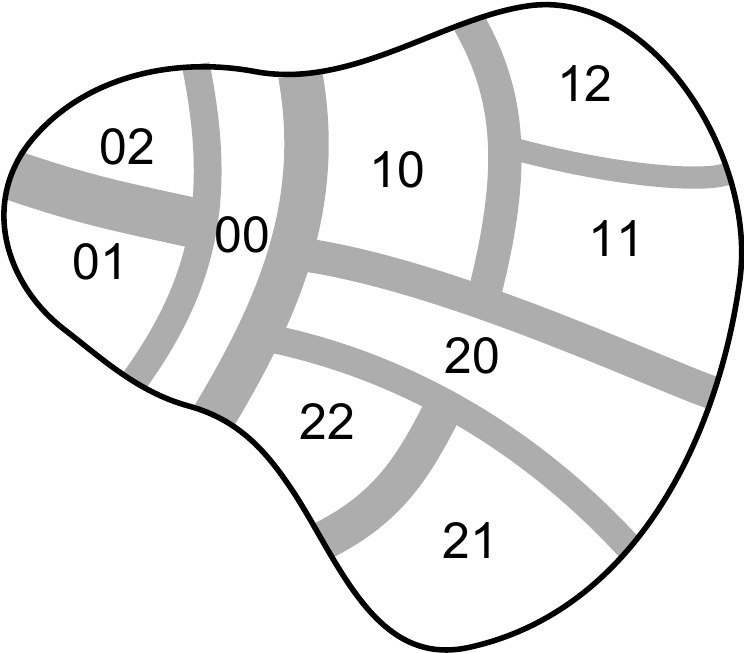}
    \\
(c)~\includegraphics[width=0.42\textwidth]{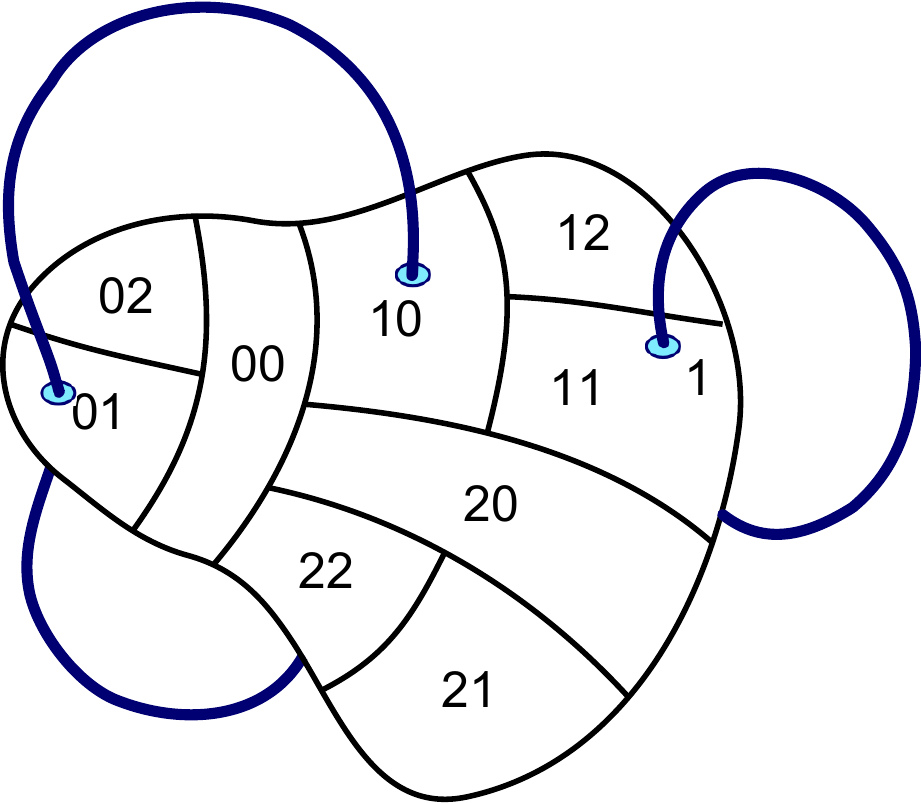}
(d)~\includegraphics[width=0.42\textwidth]{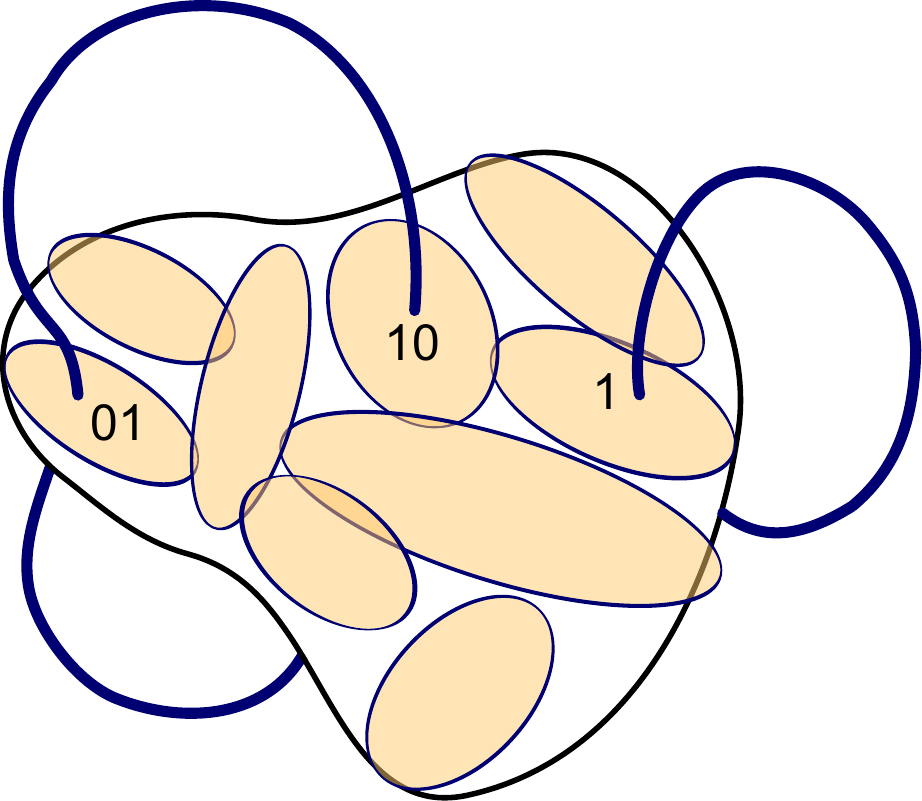}
\end{minipage}
\source{ChaosBook.org}
\caption{
(a) A deterministic partition of \statesp\ \pS.
(b) Noise blurs the partition boundaries. At some
level of deterministic partitioning some of the boundaries
start to overlap, preventing further local refinement of adjecent
neighborhoods.
(c) The fixed point $\cycle{1} = \{\ssp_{1}\}$
and the two-cycle $\cycle{01} = \{\ssp_{01},\ssp_{10}\}$ are
examples of the shortest period periodic points
within the partition regions of (a).
The {\em \optPart} hypothesis:
(d) Noise blurs periodic points into cigar-shaped
trajectory-centered densities explored by the Langevin noise.
The \optPart\ hypothesized in this paper
consists of the maximal set of resolvable periodic point
neighborhoods.
    }\label{f:SymbolSeqPart}
\end{figure}

The problem we address here is sketched in \reffig{f:SymbolSeqPart};
while a deterministic partition can, in principle, be made arbitrarily
fine, in practice any noise will blur the boundaries and render the best
possible partition finite. Thus our task is to determine the optimal
attainable resolution of the \statesp\ of a given hyperbolic dynamical
system, affected by a given weak noise. This we do by formulating the
{\em \optPart} hypothesis which we believe determines the best possible
\statesp\ partition for a desired level of predictive precision. We know
of no practical way of computing the `blurred' partition boundaries of
\reffig{f:SymbolSeqPart}\,(b). Instead, we propose to determine the
\optPart\ in terms of blurring of periodic point neighborhoods, as in
\reffig{f:SymbolSeqPart}\,(d). As we demonstrate in
\refsect{DL:OptPart}, our implementation requires determination of
only a small set of solutions of the deter\-mi\-ni\-stic equations of
motion.
%	\PC{Here we have addressed the criticisms of Mar 23 2008 	
%	NSF proposal review: `` The panel feels that there will be 	
%	difficulties in 	using noise to smooth singularities since that would 	
%	make it hard to 	define a sufficient number of useful \po s.
% 	'' }

Intuitively, the noise smears out the neighborhood of a periodic point,
whose size is now determined by the interplay between the diffusive
spreading parameterized\rf{einstein,VK79,gasp95} by a diffusion constant,
and its exponentially shrinking deter\-mi\-ni\-stic neighborhood. If the
noise is weak, the short-time dynamics is not altered significantly:
short \po s of the deter\-mi\-ni\-stic flow still coarsely partition the
\statesp. As the periods of \po s increase, the diffusion always wins,
and succes\-sive refinements of a deter\-mi\-ni\-stic partition of the
\statesp\ stop at the finest attainable partition,  beyond which the
diffusive smearing exceeds the size of any deter\-mi\-ni\-stic
subpartition.

There is a considerable literature (reviewed here in \refrem{rem:partHist})
on interplay of noise and chaotic deterministic dynamics, and the closely
related problem of limits on the validity of the semi-classical periodic
orbit quantization. All of this literature implicitly assumes uniform
hyperbolicity and seeks to define a single, globally averaged, diffusion
induced \emph{average resolution} (Heisenberg time, in the context of
semi-classical quantization). However, the local diffusion rate differs
from a trajectory to a trajectory, as different neighborhoods merge at
different times, so there is no one single time beyond which noise takes
over. Nonlinear dynamics interacts with noise in a nonlinear way, and
methods for implementing the \optPart\ for a given noise still need to be
developed. This paper is an attempt in this direction. Here we follow and
expand upon the \Fokker\ approach to the `{\optPart} hypothesis'
introduced in \refref{LipCvi08}.

What is novel here is that we show how to compute the \emph{locally}
\optPart, for a given dynamical system and given noise, in terms of local
eigenfunctions of the forward-backward actions of the \Fokker\ operator
and its adjoint. This is much simpler than it sounds: the Lyapunov equation
\[ %beq
\covMat \,=\, \monodromy \covMat \monodromy^T+\diffTen
% \,,
% \label{ddQfixed}
\] %eeq
(and its generalizations to periodic points of hyperbolic flows),
determines \covMat, the size of local neighborhood, as balance of
the noise variance \diffTen\ and the linearized dynamics $\monodromy$. The effort
of going local brings a handsome reward: as the \optPart\ is always
finite, the dynamics on this `best possible of all partitions' is encoded
by a finite transition graph of finite memory, and the \Fokker\
oper\-ator can be represented by a finite matrix.
In addition, while the \statesp\ of a generic deter\-mi\-ni\-stic flow is
an infinitely interwoven hierarchy of attracting, hyperbolic, elliptic
and parabolic regions, the noisy dynamics erases any structures finer
than the \optPart, thus curing both the affliction of long-period
attractors/elliptic islands with very small immediate basins of
attraction/ellipticity, and the power-law correlation decays caused by
marginally stable regions of \statesp.

The dynamical properties of high-dimensional flows are not just simple
extensions of lower-dimensional dynamics, and a persuasive application of
the Ruelle / Gutzwiller {\po} theory to high-dimensional dynamics would
be an important advance. If such flow has only a few expanding
directions, the above set of overlapping stochastic `cigars' should
provide an optimal, computable cover of the long-time chaotic attractor
embedded in a \statesp\ of arbitrarily high dimension.

The requisite Langevin / \Fokker\ description of noisy flows is reviewed
in \refsect{FP_evl}. This discussion leans heavily on the deterministic
dynamics and {\po} theory notation, summarized in \refappe{DL:POT}. In
\refsect{DL:oneFixP1} we derive the formulas for the size of
noise-induced neighborhoods of attractive fixed and periodic points, for
maps and flows in arbitrary dimension. These formulas are known as
Lyapunov equations, reviewed in \refappe{chap:LyapEq}.
In order to understand the effect on noise on the
hyperbolic, mixed expanding / contracting dynamics, we study the
eigenfunctions of the discrete time \Fokker\ oper\-ator in linear
neighborhood of a fixed point of a noisy one-dimensional map in
\refsect{DL:oneFixP}, and show that the neighborhood along unstable
directions is fixed by the evolution of a Gaussian density of
trajectories under the action of the adjoint \Fokker\ oper\-ator.
The continuous time formulation of the same problem, known as the
Ornstein-Uhlenbeck process, is reviewed in \refappe{DL:contFP_oper}.
Having defined the local neighborhood of every periodic point, we turn to
the global partition problem. Previous attempts at \statesp\ partitioning
are reviewed in \refrem{rem:partHist}.
We formulate our \optPart\ hypothesis in \refsect{DL:OptPart}: track the
diffusive widths of unstable \po s until they start to overlap.
We test the approach by applying it to a 1\dmn\ repeller, and in
\refsect{DL:corrections}, we assess the accuracy of our method by
computing the escape rate and the Lyapunov exponent, discuss
weak noise corrections, and compare the results with a discretization of
the \Fokker\ operator on a uniform mesh.
In \refsect{DL:flat_top} we address the problem of estimating the
\optPart\ of a non-hyperbolic map, where the linear approximation to the
\Fokker\ operator fails.
The results are summarized and the open problems discussed in
\refsect{DL:concl}.

\section{Noisy trajectories and their densities}
\label{FP_evl}

The literature on stochastic dynamical systems is vast, starting with the
Laplace 1810 memoir\rf{Laplace1810}. The material reviewed in this
section, \refsect{DL:oneFixP} and \refappe{DL:contFP_oper} is
standard\rf{vKampen92,Risken96,ArnoldL74}, but needed in order to set the
notation for what is new here, the role that \Fokker\ operators play in
defining stochastic neighborhoods of periodic orbits. The key result
derived here is the well known evolution law \refeq{AddVariances} for
the covariance matrix $Q_{a}$ of a linearly evolved Gaussian density,
\[
Q_{a+1}  =  \monodromy_{a} Q_{a} \monodromy_{a}^T+\diffTen_{a}
\,.
\]
To keep things simple we shall use only the discrete time dynamics in
what follows, but we do discuss the continuous time formulation in
\refappe{DL:contFP_oper}, as our results apply both to the continuous and
discrete time flows.

%\subsection{Discrete time \Fokker\ operator}
%\label{DL:discFP_oper}

Consider a noisy {\em discrete time} dynamical
system\rf{Kif74,FeHa82,Boy84}
\beq
{\ssp}_{n+1} = {f}({\ssp}_n) + \xi_n
\,,
\ee{DL:discrete}
where \ssp\ is a $d$\dmn\ state vector, and ${\ssp}_{n,j}$ is its $j$th
component at time $n$. In the \Fokker\ description  individual noisy
trajectories are replaced by the evolution of the density of noisy
trajectories, with the ${\ssp}_{n+1} - {f}({\ssp}_n)$  probability
distribution of zero mean and covariance  matrix (diffusion tensor)
$\diffTen$,
\beq
\expct{\xi_{n,j}} = 0
        \,,\qquad
\expct{\xi_{n,i} \, \xi^T_{m,j}}
        = \diffTen_{ij} \, \delta_{nm}
\,,
\ee{whiteDscr}
where $\expct{\cdots}$ stands for ensemble average over many realizations
of the noise.

The general case of a diffusion tensor $\diffTen(\ssp)$ which is a \statesp\
position dependent but time independent can be treated along the same
lines. In this case the stochastic flow \refeq{DL:discrete} is written as
\(
\ssp_{n+1} = {\ssp}_n + \sigma(\ssp)\,\xi_n
\,,
\)
where
\(
\expct{\xi_{n} \, \xi^T_{m}} = \mathbf{1} \, \delta_{nm}
\)
is white noise, $\diffTen = \sigma \, \sigma^T$,
$\sigma(\ssp)$ is called the `diffusion matrix', and the noise is
referred to as `multiplicative' (see Kuehn\rf{Kuehn11}).

The action of discrete one-time step {\em \Fokker\ oper\-ator} on the
density distribution $\msr$ at time $k$,
\bea
\msr_{k+1}({y}) &=& [\Lnoise{}  \msr_k](y)
	\,=\, \int d\ssp \, \Lnoise{}({y,\ssp})\,\msr_k(\ssp)
            \continue
    \Lnoise{}(y,\ssp)  &=& \frac{1}{N} \,
    e^{ %\exp\left\{
    -\frac{1}{2}
(y-f(\ssp))^T {} \frac{1}{\diffTen}{} (y-f(\ssp))
      } %\right\}
\label{DL:dscrt_FP}
\,,
\eea
is centered on the deter\-mi\-ni\-stic step $f(\ssp)$ and smeared out
diffusively by noise. Were diffusion uniform and isotropic,
$\diffTen(\ssp) = 2\, D \, \mathbf{1}$, the \Fokker\ oper\-ator would be
proportional to \DLedit{$\exp\left(-\{y - \flow{}{x}\}^2/2\diffTen\right)$},  \ie,
the penalty for  straying from the deterministic path is just a quadratic
error function.
                                                            \toCB
The $k$th iterate of $\Lnoise{}$ is a $d$-dimensional path integral over
the $k-1$ intermediate noisy trajectory points,
\beq
  \Lnoise{k}(\ssp_k,\ssp_0)
      \,=\, \int[d\ssp]\,
          e^{-\frac{1}{2}
            \sum_{n}(\ssp_{n+1}-f(\ssp_n))^T
            {} \frac{1}{\diffTen}{} (\ssp_{n+1}-f(\ssp_n))
          }
\,,
\label{IntDef}
\eeq
where the Gaussian normalization factor in \refeq{DL:dscrt_FP} is
absorbed into intermediate integrations by defining
\beq
[d\ssp]  =   \prod_{n=1}^{k-1} \frac{d \ssp^d_n}{N}
    \,,\qquad
      N = \sqrt{2\pi^{d} \det \diffTen}
      % N = (2\pi)^{d/2} (\det \diffTen)^{1/2}
\,.
\label{GaussMsr}
\eeq

We shall also need to determine the effect of noise accumulated along the
trajectory points \emph{preceding} $\ssp$. As the noise is additive
forward in time, one cannot simply invert the {\Fokker} operator;
instead, the past is described by the {\em adjoint \Fokker\ oper\-ator},
\bea
\tilde{\msr}_{k-1}(\ssp) &=&
[\Lnoise{\dagger} \tilde{\msr}_{k}](\ssp)
%	\continue
    = \int [dy]  \,\,
    e^{ -\frac{1}{2}
(y-\map(\ssp))^T {} \frac{1}{\diffTen}{} (y-\map(\ssp))
      } \, \tilde{\msr}_{k}(y)
\,,
\label{DL:dscrt_adj}
\eea
which transports a density concentrated around 
%the previous point $\ssp$
\DLedit{the point $\map(\ssp)$ 
to a density concentrated around the 
previous point $\ssp$ and adds noise to it}. In
the deterministic, vanishing noise limit this is the Koopman operator
\refeq{APPE3.14a}.

The \Fokker\ oper\-ator  \refeq{DL:dscrt_FP} is 
%non-selfadjoint. For
\DLedit{non-hermitian and non-unitary.} For
example, if the deterministic flow is contracting, the natural measure
(the leading right eigenvector of the \Fokker\ oper\-ator) will be
concentrated and peaked, but then the corresponding left eigenvector has
to be broad and flat, as backward in time the deterministic flow is
expanding. We shall denote by $ \msr_{\alpha}$ the right eigenvectors of
$\Lnoise{}$, and by $\tilde{\msr}_{\alpha}$ its left eigenvectors, \ie,
the right eigenvectors of the adjoint operator $\Lnoise{\dagger}$.

\section{All nonlinear noise is local}
% Linearized flow}
\label{DL:oneFixP1}
                        \renewcommand{\version}{
  Predrag                   May 5 2012
                        }
% Predrag                   Apr 24 2012
% Predrag  from stoch.tex   Sep 24 2011
% Predrag           Dec 31 2010
% Predrag           Sep  6 2010
% Predrag           Jul  4 2010
% Predrag           Mar 17 2010
% Domenico	        Nov 18 2009
% Predrag           Aug 26 2007
% Domenico          Jul 16 2007
% Domenico          Sep 29 2006
% Predrag           Sep  6 2006
% Domenico          Aug 21 2006
% Domenico          Aug 20 2005

Our first goal is to convince the reader that the diffusive dynamics of
nonlinear flows is \emph{fundamentally different from Brownian motion},
with the flow inducing a local, history dependent noise. In order to
accomplish this, we go beyond the standard stochastic literature and
generalize the notion of invariant deterministic recurrent solutions,
such as fixed points and \po s, to noisy flows. While a Langevin
trajectory \refeq{DL:discrete} cannot be periodic, in the \Fokker\
formulation \refeq{IntDef} a recurrent motion can be defined as one where
a peaked distribution returns to the initial neighborhood after time
$\cl{}$. Recurrence so defined not only coincides with the classical
notion of a recurrent orbit in the vanishing noise limit, but it also
enables us to derive exact formulas for how this local, history dependent
noise is to be computed.

As the function $\ssp_{n+1}-f(\ssp_n)$ is a nonlinear function, in
general the path integral \refeq{IntDef} can only be evaluated
numerically. In the vanishing noise limit the Gaussian kernel sharpens
into the  Dirac $\delta$-function, and the \Fokker\ oper\-ator reduces to
the deterministic {\FPoper} \refeq{TransOp1}. For weak noise the \Fokker\
oper\-ator  can be evaluated perturbatively%
\rf{noisy_Fred,conjug_Fred,diag_Fred,Ronc95} as an asymptotic series in
powers of the diffusion constant, centered on the
deterministic trajectory. Here we retain only the linear term in this
series, which has a particulary simple dynamics given by a covariance
matrix evolution formula (see \refeq{AddVariances} and \refeq{ddQpPoint}
below) that we now derive.

We shift local coordinates to the deterministic trajectory
$\{\ldots,\ssp_{-1}$, $\ssp_{0}$, $\ssp_1$, $\ssp_2,\ldots,\}$ centered
coordinate frames $\ssp=\ssp_a+\orbitDist_a$, Taylor expand $f(\ssp) =
f_a(\orbitDist_a) = \ssp_{a+1} + \monodromy_a \orbitDist_a + \cdots$, and
approximate the noisy map \refeq{DL:discrete} by its linearization,
\beq
\orbitDist_{a+1}= \monodromy_a \orbitDist_a + \xi_a
      \,, \quad
\monodromy_{ij} = \partial f_i / \partial \ssp_j
\ee{DL:linMap}
with the deterministic trajectory points at
$\orbitDist_a=\orbitDist_{a+1}=0$, and $\monodromy_a$ the one step
\jacobianM\ (see \refappe{DL:POT}). The corresponding linearized \Fokker\
oper\-ator \refeq{DL:dscrt_FP} is given in the local coordinates
\(
\msr_{a}(\orbitDist_{a}) = \msr(\ssp_{a}+\orbitDist_{a},a)
\)
by
\beq
\msr_{a+1}(\orbitDist_{a+1})  =  \int d\orbitDist_{a}\,
\Lnoise{}{}_a(\orbitDist_{a+1},\orbitDist_{a}) \,
\msr_{a}(\orbitDist_{a})
\label{DL:FP_za1}
\eeq
by the linearization \refeq{DL:linMap} centered on the deterministic
trajectory
    \PC{emphasize that in these coordinates the deterministic dynamics
    is linearized, not the global \Fokker\ operator
    }
\bea
\Lnoise{}{}_a (\orbitDist_{a+1},\orbitDist_{a}) &=&
    \frac{1}{N }\,
e^{-\frac{1}{2}(\orbitDist_{a+1}- \monodromy_a \orbitDist_{a})^T
                {} \frac{1}{\diffTen} {\,}
                (\orbitDist_{a+1}-\monodromy_a \orbitDist_{a})
                }
\,.
\label{DL:app_evol}
\eea
The subscript `$a$' in $\Lnoise{}{}_a$ distinguishes the local, linearized
\Fokker\ operator from the full operator \refeq{IntDef}.

The \DLedit{kernel of the} linearized \Fokker\ operator \refeq{DL:app_evol} is a Gaussian. As a
convolution of a Gaussian with a Gaussian is again a Gaussian, we
investigate the action of the linearized \Fokker\ operator on a
normalized, cigar-shaped Gaussian density distribution
\beq
\msr_a(\orbitDist)= \frac{1}{C_a}
e^{- \frac{1}{2} \orbitDist^T {}  \frac{1}{\covMat_a} {\,} \orbitDist}
\,,\qquad
C_a = (2\pi)^{d/2} (\det \covMat_a)^{1/2}
\,,
\label{DL:GaussDens}
\eeq
and the action of the linearized adjoint \Fokker\ operator on density
\beq
\tilde{\msr}_a(\orbitDist)= \frac{1}{C_a}
e^{- \frac{1}{2} \orbitDist^T {}  \frac{1}{\tilde{\covMat}_a} {\,} \orbitDist}
\,,\qquad
C_a = (2\pi)^{d/2} (\det \tilde{\covMat}_a)^{1/2}
\,,
\label{DL:AdjGaussDens}
\eeq
also centered on the deterministic trajectory, but with its own strictly
positive $[d\!\times\!d]$ covariance matrices $\covMat$,
$\tilde{\covMat}$. Label `$a$' plays a double role, and $\{a+1,a\}$
stands both for the $\{$next,~initial$\}$ space partition and for the
times the trajectory lands in these partitions (see
\refappe{DL:detPart}). The linearized \Fokker\ operator
\refeq{DL:app_evol} maps the Gaussian $\msr_{a}(\orbitDist_{a})$ into
the Gaussian
\bea
\msr_{a+1}(\orbitDist_{a+1})
    &=&
    \frac{1}{C_{a}}\int [d\orbitDist_a] \,
e^{-\frac{1}{2}\left[(\orbitDist_{a+1}- \monodromy_a \orbitDist_a)^T
         {} \frac{1}{\diffTen} {\,}
           (\orbitDist_{a+1}-\monodromy_a \orbitDist_a)
                +\orbitDist_a^T {} \frac{1}{\covMat_a}{\,} \orbitDist_a\right]}
\label{DL:stepLater}
\eea
one time step later.
Likewise, linearizing the adjoint \Fokker\ oper\-ator
\refeq{DL:dscrt_adj} around the $\ssp_{a}$ trajectory point yields:
\bea
\tilde{\msr}_{a}(\orbitDist_{a})
    &=&
    \frac{1}{C_{a+1}}\int [d\orbitDist_{a+1}] \,
e^{-\frac{1}{2}[(\orbitDist_{a+1}- \monodromy_a \orbitDist_a)^T
         {} \frac{1}{\diffTen_a} {\,}
           (\orbitDist_{a+1}-\monodromy_a \orbitDist_a)
                +\orbitDist_{a+1}^T \frac{1}{\tilde{\tilde{\covMat}}_{a+1}}{\,} \orbitDist_{a+1}]}
\,.
\label{adjointEvol}
\eea
Completing the squares, integrating and substituting
\refeq{DL:GaussDens}, respectively \refeq{DL:AdjGaussDens}
we obtain the formula for $\covMat$ covariance matrix evolution forward in time,
\beq
\covMat_{a+1}  =  \monodromy_{a} \covMat_{a} \monodromy_{a}^T+\diffTen_{a}
\,,
\label{AddVariances}
\eeq
and in the adjoint case, the evolution of the
$\tilde{\covMat}$ is given by
\beq
\monodromy_a \tilde{\covMat}_{a} \monodromy^T_a  = \tilde{\covMat}_{a+1} + \diffTen_a
\,.
\ee{DL:adjoint_cov}
The two covariance matrices differ, as the adjoint evolution
$\tilde{\covMat}_{a}$ is computed by going backwards along the
trajectory. These \emph{covariance evolution} rules are the basis of all
that follows, except for the `flat-top' of \refsect{DL:flat_top}.

Think of the initial covariance matrix \refeq{DL:GaussDens} as an error
matrix describing the precision of the initial state, a cigar-shaped
probability distribution $\msr_{a}(\orbitDist_{a})$. In one time step
this density is deterministically advected and deformed into density with
covariance $\monodromy \covMat \monodromy^T$, and then the noise
$\diffTen$ is added: the two kinds of independent uncertainties add up as
sums of squares, hence the covariance evolution law \refeq{AddVariances},
resulting in the Gaussian ellipsoid whose widths and orientation are
given by the singular values and singular vectors \refeq{SVD-j} of the
covariance matrix. After $n$ time steps, the variance $\covMat_{a}$ is
built up from the deterministically propagated $\monodromy_{a}^n
\covMat_{a-n} \monodromy_{a}^{n T}$ initial distribution, and the sum of
noise kicks at intervening times, $\monodromy_{a}^k \diffTen_{a-k}
\monodromy_{a}^{k T}$, also propagated deterministically.

The pleasant surprise is that the evaluation of this noise requires no
\Fokker\ PDE formalism. The width of a Gaussian packet centered on a
trajectory is fully specified by a deterministic computation that is
already a pre-computed byproduct of the periodic orbit computations; the
deterministic orbit and its linear stability. We have attached label
`$a$' to $\diffTen_a = \diffTen(\ssp_a)$ in \refeq{AddVariances} to
account for the noise distributions that are inhomogeneous, \statesp\
dependent, but time independent multiplicative noise. As we shall show,
in nonlinear dynamics the noise is \emph{never} isotropic and/or
homogeneous. For example, if the iterative system we are studying is
obtained by Poincar\'e sections of a continuous time flow, the
accumulated noise integrated over one Poincar\'e section return depends
on the return trajectory segment, even when the infinitesimal time step
noise \refeq{DL:anis_ev} is homogenous.

\remark{Covariance evolution.}{
\label{rem:CovEvo}
In quantum mechanics the linearized evolution operator corresponding to
the linearized \Fokker\ oper\-ator \refeq{DL:app_evol} is known as the
Van Vleck propagator, the basic block in the semi-classical periodic
orbit quantization\rf{gutbook,DasBuch}. $\covMat$ covariance matrix
composition rule \refeq{AddVariances} or its continuous time version
\refeq{ContTcigar} is called `covariance evolution' in \refref{TiCo01},
for example, but it goes all the way back to Lyapunov's 1892
thesis\rf{Lyapunov1892}, see \refappe{chap:LyapEq}. In the Kalman filter
literature\rf{Kalman60,ACFK09} it is called `prediction'.
} %end \remark{Covariance evolution}{

\ifboyscout
\subsection{Il buono, il brutto, il cattivo}
\else
\subsection{The attractive, the repulsive and the noisy}
\fi
% play on "The Good, the Bad and the Ugly"

For Browniam dynamics \({\ssp}_{n+1} = {\ssp}_n + \xi_n\), with
$\monodromy=\mathbf{1}$, we obtain $\covMat_{n} = \covMat_{0} +
n\,\diffTen$, \ie, the variance of a Gaussian packet of
$\msr_n(\orbitDist)$ noisy trajectories grows linearly in time, as
expected for the Brownian diffusion. What happens for nontrivial,
$\monodromy \neq\mathbf{1}$,  dynamics? The formulas \refeq{AddVariances}
and \refeq{DL:adjoint_cov} are exact for finite numbers of time steps,
but whether they have a long time limit depends on the stability of the
deterministic trajectory.

Here we shall derive the $n \to \infty$ limit for deterministic flows
that are either contracting or expanding in all eigen-directions, with
asymptotic stationary distributions concentrated either on fixed points
or periodic points. We shall consider the general hyperbolic flows, where
some of the eigen-directions are unstable, and other stable in another
publication\rf{LipCvi07}. In this context the description in terms of \po
s is very useful; the neighborhood of a periodic point will be defined as
the noise contracting neighborhood forward in time along contracting
eigen-directions,  backward in time along the unstable, expanding
eigen-directions. The short cycles will be the most important ones, and
only finite time, single cycle period calculations will be required.

If $\monodromy$ is contracting, with the multipliers
\(
1 > | \ExpaEig_1 | \geq | \ExpaEig_2 |
    \geq \ldots \geq | \ExpaEig_d |
\,,
\)
in $n$ time steps the memory of the covariance $\covMat_{a-n}$ of the
starting density is forgotten at exponential rate $\sim   | \ExpaEig_1
|^{-2n}$, with iteration of \refeq{AddVariances} leading to a limit
distribution:
\beq
\covMat_{a} =  \diffTen_{a}
    +\monodromy_{a-1} \diffTen_{a-1} \monodromy_{a-1}^T
    +\monodromy_{a-2}^2 \diffTen_{a-2}  (\monodromy_{a-2}^{2})^{T}
           + \cdots
\,.
\label{VariancHist}
\eeq
For fixed and periodic points we can give an explicit formula for the
$n \to \infty$ covariance.

Consider a noisy map \refeq{DL:discrete} with a deterministic fixed point
at $\ssp_\stagn$. In a neighborhood $\ssp=\ssp_\stagn+\orbitDist$ we
approximate the map $\map$ by its linearization \refeq{DL:linMap} with
the fixed point at $\orbitDist=0$, acting on a Gaussian density
distribution \refeq{DL:GaussDens}, also centered on $\orbitDist=0$. The
distribution is cigar-shaped ellipsoid, with eigen\-vectors of
$\covMat_n$ giving the orientation of various axes at time $n$, see
\reffig{f:repPart}\,(b). If the fixed point is \emph{attractive}, with
all multipliers of $\monodromy$ strictly contracting, any compact initial
measure (not only initial distributions of Gaussian form) converges under
applications of \refeq{AddVariances} to the unique invariant natural
measure $\SRB(\orbitDist)$ whose covariance matrix satisfies the
condition
\beq
\covMat \,=\, \monodromy \covMat \monodromy^T+\diffTen
\,.
\label{ddQfixed}
\eeq
For a \emph{repelling} fixed point the condition
\refeq{DL:adjoint_cov} on the adjoint eigenvector $\tilde{\msr}_{0}$ yields
\beq
\monodromy \tilde{\covMat} \monodromy^T  \,=\,\tilde{\covMat} + \diffTen
\,,
\ee{DL:fix_adj}
with a very different interpretation: as the \jacobianM\ $\monodromy$ has
only {\em expanding} Floquet multipliers, the deterministic dynamics
expands the fixed-point neighborhood exponentially, with no good notion
of a local neighborhood  in the large forward time limit. Instead,
its past defines the neighborhood, with $\tilde{\covMat}$  the  covariance of the
optimal distribution of points that can reach the fixed point in one time
step, given the diffusion tensor $\diffTen$.

These conditions are central to control theory, where the attracting
fixed point condition \refeq{ddQfixed} is called the \emph{Lyapunov
equation} (see \refappe{chap:LyapEq}), $\covMat$ and $\tilde{\covMat}$
are known respectively as controllability and observability Gramians, and
there is much wisdom and open source code available to solve these
(see \refrem{rem:LyapEq}), as well as the
more general hyperbolic equations.
In order to develop some intuition about the types of solutions we shall
encounter, we assume first, for illustrative purposes, that $[d\!\times\!d]$
{\jacobianM} $\monodromy$ has distinct real contracting Floquet
multipliers $\{\ExpaEig_{1},\ExpaEig_{2},\cdots,\ExpaEig_{d}\}$ and right
eigen\-vectors % $\jEigvec[j]$
    \PC{please confirm:  is $\tilde{\covMat}$ 'observability'?}
\( %\beq
\monodromy \, \jEigvec[j]
   = \ExpaEig_{j} \,\jEigvec[j]
\,.
\) % \ee{cplxExpaEig1}
Construct from the $d$ column eigenvectors a $[d\!\times\!d]$
similarity transformation
\[
S = \left(\jEigvec[1],\jEigvec[2],\cdots,\jEigvec[d]\right)
\]
that diagonalizes $\monodromy$,
\(
S^{-1} \monodromy S = \Lambda
\)
and its transpose
\(
S^T \monodromy^T (S^{-1})^T = \Lambda
\,.
\)
Define
\(
\hat{\covMat} = S^{-1} \covMat  (S^{-1})^T
\)
and
\(
\hat{\diffTen} = S^{-1} \diffTen  (S^{-1})^T
\,.
\)
The fixed point condition \refeq{ddQfixed} now
takes form
\( %beq
\hat{\covMat} - \Lambda \hat{\covMat} \Lambda = \hat{\diffTen}
\,.
\) % \ee{ddQfixed2}
The matrix elements are
\(
\hat{\covMat}_{ij}(1 - \ExpaEig_i \ExpaEig_j) = \hat{\diffTen}_{ij}
\,,
\)
so
\beq
\hat{\covMat}_{ij} = \frac{\hat{\diffTen}_{ij}}
                                    {1 - \ExpaEig_i \ExpaEig_j}
\,,
\ee{ddQfixed3}
and the attracting fixed point covariance matrix in the original
coordinates is given by
\beq
\covMat =  S \hat{\covMat} S^T
\,.
\ee{cplxExpaEig3}
For the adjoint case, the same algebra yields
\beq
\hat{\covMat}_{ij} = \frac{\hat{\diffTen}_{ij}}
                                    {\ExpaEig_i \ExpaEig_j-1}
\,,
\ee{adjQfixed}
for the matrix elements of $\hat{\covMat}$, with the covariance matrix
in the fixed coordinates again given by $\tilde{\covMat}=S\hat{\covMat}S^T$.

As \refeq{cplxExpaEig3} is not a similarity transformation, evaluation of
the covariance matrix $\covMat$ requires a numerical diagonalization,
which yields the singular values and singular vectors (principal axes) of
the equilibrium Gaussian `cigar' (see \refappe{DL:POT}). The singular
vectors of this symmetric matrix have their own orientations, distinct
from the left/right eigenvectors of the non-normal {\jacobianM}
$\monodromy$.

\remark{Hyperbolic flows.}{
\label{rem:HypFlows}
The methods to treat the cases where some of the eigen-directions are
unstable, and other stable are implicit in the Oseledec\rf{lyaos}
definition of Lyapunov exponents, the rigorous proof of existence of
classical spectral (Fredholm) determinants by Rugh\rf{hhrugh92}, and the
controllability and observability Gramians of control
theory\rf{ZhoSalWu99}: the flow at a hyperbolic fixed point or cycle
point can be locally factorized into stable and unstable directions, and
for unstable directions one needs to study noise evolution in the past,
by means of the adjoint operator \refeq{DL:dscrt_adj}.
} %end \remark{Hyperbolic

\subsection{In nonlinear world noise is never isotropic}
% Attractive periodic points:
%           For nonlinear flow noise is never isotropic
\label{sect:POvariance}

Now that we have established the exact formulas \refeq{AddVariances},
\refeq{DL:adjoint_cov} for the extent of the noise-smeared out
neighborhood of a fixed point, we turn to the problem of computing them
for periodic orbits. An attractive feature of the deterministic periodic
orbit theory is that certain properties of \po s, such as their periods
and Floquet multipliers, are {\em intrinsic}, independent of where they
are measured along the the periodic orbit, and invariant under all smooth
conjugacies, \ie, all smooth nonlinear coordinate transformations. Noise,
however, is specified in a given coordinate system and breaks such
invariances (for an exception, a canonically invariant noise, see
Kurchan\rf{CK97}). Each cycle point has a different memory and
differently distorted neighborhood, so we need to compute the \Fokker\
eigenfunction $\msr_a$ at each cycle point $\ssp_a$.

The basic idea is simple: A periodic point of an $\cl{}$-cycle is a fixed
point of the $\cl{}$th iterate of the map \refeq{DL:discrete}. Hence the
formula \refeq{VariancHist} for accumulated noise, together the fixed
point condition \refeq{ddQfixed} also yields the natural measure
covariance matrix at a periodic point $\ssp_a$ on a \po\ $p$,
\bea
\covMat_a &=&
          \monodromy_{p,a} \covMat_a \monodromy_{p,a}^T
        + \diffTen_{p,a}
\,,
\label{ddQpPoint}
\eea
where
\bea
\diffTen_{p,a} &=&  \diffTen_{a}
    +\monodromy_{a-1} \diffTen_{a-1} \monodromy_{a-1}^T
    +\monodromy_{a-2}^2 \diffTen_{a-2}  (\monodromy_{a-2}^{2})^{T}
    \ceq
    + \cdots
    +\monodromy_{a-\cl{p}+1}^{\cl{p}-1}
    \diffTen_{a-\cl{p}+1}  (\monodromy_{a-\cl{p}+1}^{\cl{p}-1})^{T}
\label{POnoiseHist}
\eea
is the noise accumulated per a single transversal of the \po,
$\monodromy_{p,a} = \monodromy_p(\ssp_a)$ is the cycle {\jacobianM}
\refeq{perPointM} evaluated on the periodic point $\ssp_a$, and we have
used the periodic orbit condition $\ssp_{a+\cl{p}}= \ssp_{a}$. Similarly,
for the adjoint evolution the fixed point condition
\refeq{DL:fix_adj} generalizes to
\beq
\monodromy_{p,a} \tilde{\covMat}_{a} \monodromy^T_{p,a}
    \,=\,\tilde{\covMat}_a + \tilde{\diffTen}_{p,a}
\,,
\ee{VariancHistPO}
where
\bea
\tilde{\diffTen}_{p,a} &=&  \diffTen_{a}
    +\monodromy_{a+1} \diffTen_{a+1} \monodromy_{a+1}^T
    +\monodromy_{a+2}^2 \diffTen_{a+2}  (\monodromy_{a+2}^{2})^{T}
    \ceq
    + \cdots
    +\monodromy_{a+\cl{p}-1}^{\cl{p}-1}
    \diffTen_{a+\cl{p}-1}  (\monodromy_{a+\cl{p}-1}^{\cl{p}-1})^{T}
\label{POadjHist}
\eea
is the noise accumulated per a single transversal of the \po\ backward in
time.

As there is no single coordinate frame in which different
$\monodromy_{a-k}^k \diffTen_{a-k} (\monodromy_{a-k}^{k})^{T}$ can be
simultaneously diagonalized, the accumulated noise is {\em never}
isotropic. So the lesson is that regardless of whether the external noise
$\diffTen$ is isotropic or anisotropic, the nonlinear flow always renders
the effective noise anisotropic and spatially inhomogeneous.

\section{One-dimensional intuition}
% Linearized flow in one dimension}
\label{DL:oneFixP}
                        \renewcommand{\version}{
  Predrag                   May 11 2012
                        }
% Predrag  from stoch.tex   Sep 24 2011
% Predrag                   Sep 6 2010
% Predrag                   Mar 14 2010
% Domenico          Jul 16 2007
% Predrag           Sep  6 2006
% Domenico          Aug 21 2006
% Domenico          Aug 20 2005

The very general, exact formulas that we have obtained so far (and so
easily), valid in any dimension, might be elegant, but it is a bit hard
to get one's head around a formula such as the expression for the
accumulated cycle noise \refeq{POnoiseHist}. These results are easier to
grasp by studying the effect of noise on 1\dmn\ systems, such as the
noisy linear map \refeq{DL:linMap},
\beq
\orbitDist_{n+1} = \map(\orbitDist_n)  + \xi_n
    \,, \qquad
\map(\orbitDist_n) = \ExpaEig \orbitDist_n
    % \,, \qquad  |\ExpaEig| \neq 1
    \,,
\ee{DL:one_d}
with the deterministic fixed point at $\map(\orbitDist) = \orbitDist =
0$, and additive white noise \refeq{whiteDscr} with variance
$\diffTen$. The density  $\msr(x)$ of trajectories evolves by the action
of the
\Fokker\ oper\-ator \refeq{DL:dscrt_FP}:
\beq
[\Lnoise{} \msr](x) =
\int [dy] \, e^{-\frac{1}{2}\,\frac{(x-\ExpaEig y)^2}{\diffTen}} \msr(y)
\,.
\ee{DL:evolution}

If a 1\dmn\ noisy linear map \refeq{DL:one_d} is \emph{contracting}, any
initial compact measure  converges under applications of
\refeq{DL:evolution} to the unique invariant natural measure
$\SRB(\orbitDist)$ concentrated at the deterministic fixed point
$\orbitDist = 0$ whose  variance \refeq{DL:GaussDens} is given by
\refeq{ddQfixed}:
\beq
\covMat = \frac{\diffTen}{1 - \ExpaEig^2}
	\,,\qquad
\msr_0(\orbitDist)= \frac{1}{\sqrt{2\pi\, \covMat}} \,
					e^{- {\orbitDist^2}/{2\,\covMat}}
%					\exp\left({- \frac{\orbitDist^2}{2\,\covMat}}\right)
\,.
\ee{DL:width}
The variance \refeq{ddQpPoint} of a periodic point $\ssp_a$ on an
attractive $\cl{}$-cycle $p$ is
\beq
\covMat_a =
        \frac{\diffTen_{p,a}}{1-\ExpaEig^2}
    \,,\qquad
        \ExpaEig = f^{\cl{}'}_a
\,,
\ee{DL:condition}
where the accumulated noise per a cycle traversal \refeq{POnoiseHist} is
given by
\beq
\diffTen_{p,a}  =  \diffTen \, (1+(f'_{a-1})^2
                +(f^{2'}_{a-2})^2+
                \cdots +(f^{n-1'}_{a-\cl{p}+1})^2)
\,.
\label{DL:nSteps1}
\eeq
Variance \refeq{DL:width} expresses a balance between contraction by
$\ExpaEig$ and diffusive smearing by $\diffTen$ at each time step. For
strongly contracting $\ExpaEig$, the width is due to the noise only. As
$|\ExpaEig| \to 1$ the width diverges: the trajectories are only weakly
confined and diffuse by Brownian motion into a broad Gaussian.

Consider next the adjoint operator acting on a \emph{repelling} noisy
fixed point, $ |\ExpaEig|> 1$. The stationary measure condition
\refeq{DL:fix_adj} yields
\beq
\tilde{\covMat} =
        \frac{\diffTen}{\ExpaEig^2-1}
    \,,\qquad
\tilde{\msr}_0(\orbitDist)= \frac{1}{\sqrt{2\pi\, \tilde{\covMat}}} \,
					e^{- {\orbitDist^2}/{2\,\tilde{\covMat}}}
\;\;.
\ee{DL:condAdj}
While the dominant feature of the attracting fixed point variance
\refeq{DL:width} was the diffusion strength $\diffTen$, weakly modified
by the contracting multiplier, for the unstable fixed point the behavior
is dominated by the expanding multiplier $\ExpaEig$; the more unstable
the fixed point, the smaller is the neighborhood one step in the past
that can reach it.

The variance \refeq{VariancHistPO} of a periodic point $\ssp_a$ on an
unstable $\cl{}$-cycle $p$ is
\DLedit{
\beq
\tilde{\covMat}_a =
\frac{\diffTen}{1-\ExpaEig_p^{-2}}
\left(\frac{1}{(f_{a}')^2}+ \frac{1}{(f_{a+1}^{2'})^2}
      \cdots + \frac{1}{\ExpaEig_p^{2}}\right)
\,.
\ee{DL:n_eig}
}
\DL{used to be $...\frac{1}{(f_{a-1}^{2'})^2}...$ in the sum, not consistent
with the notation in equation \refeq{POadjHist}. Please doublecheck as 
I may be wrong here.}   
For an unstable cycle typically all derivatives along the cycle are
expanding, $|f_{a+k}'| > 1$, so the dominant term in \refeq{DL:n_eig} is
the most recent one, $\tilde{\covMat}_a \approx \diffTen/(f_{a}')^2$. By
contrast, forward in time \refeq{DL:condition} the leading estimate of
variance of an attractive periodic point is ${\covMat}_a \approx
\diffTen$. These leading estimates are not sensitive to the length of the
\po, so all trajectories passing through a neighborhood of periodic point
$\ssp_a$ will have comparable variances.

\subsection{Ornstein-Uhlenbeck spectrum}
\label{DL:locEig1}

The variance \refeq{ddQfixed} is stationary under the action of
$\Lnoise{}$, and the corresponding Gaussian is thus an eigenfunction.
Indeed, as we shall now show, for the linear flow  \refeq{DL:evolution}
the entire eigenspectrum is available analytically, and as $Q_a$ can
always be brought to a diagonal, factorized form in its orthogonal frame,
it suffices to understand the simplest case, the Ornstein-Uhlenbeck
process (see \refappe{DL:locEig}) in one dimension.
The linearized \Fokker\ operator is a Gaussian, so it is natural to
consider the set of Hermite polynomials, $H_0(x)=1$, $H_1(x)=2\,x$,
$H_2(x)=4\,x^2-2$, $\cdots$, as candidates for its eigenfunctions.
$H_n(x)$ is an $n$th-degree polynomial, orthogonal with respect to the
Gaussian kernel
\beq
\frac{1}{2^n n! \,\sqrt{\pi}} \, \int \! dx  \;
H_m(x) \; e^{-x^2} H_n(x) \, =  \, \delta_{mn}
\,.
\ee{HermiteOrthon}
There are three cases to consider:

\medskip

\noindent
{\bf $|\ExpaEig|> 1$ expanding case:~~}
The form of the left  $\tilde{\msr}_0$ eigenfunction \refeq{DL:condAdj}
suggests that we rescale $x \to x/\sqrt{2\,\tilde{\covMat}}$ and absorb
the Gaussian kernel in \refeq{HermiteOrthon} into left eigenfunctions
$\tilde{\msr}_0$, $\tilde{\msr}_1$, $\cdots$,
\beq
    \tilde{\msr}_{k}(\orbitDist) =
    \frac{1}{\sqrt{2\pi}\, 2^{3k/2} k! \,\tilde{\covMat}^{(k+1)/2}} \,
         H_k((2 \tilde{\covMat})^{-1/2}\,\orbitDist) \,e^{-{\orbitDist^2}/{2\tilde{\covMat}}}
\,,
\label{DL:Herm1}
\eeq
The right eigenfunctions are then
\beq
     \msr_{k}(\orbitDist) = (2 \tilde{\covMat})^{k/2} H_k((2 \tilde{\covMat})^{-1/2}\,\orbitDist)
\,,
\label{DL:Herm}
\eeq
By construction the left, right eigenfunctions are orthonormal to each other:
%    \PC{recheck} \DL{checks out}
\beq
\int dx \, \tilde{\msr}_k(x)\,\msr_j(x) = \delta_{kj}
\,.
\ee{DL:orthonormality}
One can verify\rf{Risken96} that for the fixed point $\orbitDist = 0$,
these
    \PCedit{
are the right, left eigenfunctions
    }
of the adjoint \Fokker\ operator
\refeq{DL:dscrt_adj}, where the $k$th eigenvalue is
${1}/{|\ExpaEig|\ExpaEig^k}$.
    \PC{here we are not being nice to r=the reader: ``one can verify''
        is a bit of work}
Note that the
Floquet multipliers $\ExpaEig^k$ are {independent} of the noise strength,
so they are the {same} as for the $\diffTen \to 0$ deter\-mi\-ni\-stic
{\FPoper} \refeq{TransOp1}.

\medskip

\noindent
{\bf $|\ExpaEig|= 1 $ marginal case:~~}
This is the pure diffusion limit, and the behavior is not
exponential, but power-law. If the map is nonlinear, one needs to go to
the first non-vanishing nonlinear order in Taylor expansion
\refeq{TaylExp} to reestablish the control\rf{gasp95}. This we do in
\refsect{DL:flat_top}.

\medskip

\noindent
{\bf $|\ExpaEig|< 1 $ contracting case:~~}
In each iteration the map contracts the cloud of noisy trajectories by
Floquet multiplier $\ExpaEig$ toward the $x=0$ fixed point, while the
noise smears them out with variance $\diffTen$. Now what was the left eigenfunction
for the expanding case \refeq{DL:Herm1} is the peaked right eigenfunction of the \Fokker\
operator,
$\{\msr_0$, $\msr_1$, $\msr_2$,$\cdots\}$, with eigenvalues $\{1$, $\ExpaEig$,
$\ExpaEig^2$,$\cdots\}$\rf{VK79,gasp95}
%    \PC{please recheck: $\{1$, $\ExpaEig$ ,$\cdots\}$ correct?} \DL{correct}
\bea
    \msr_k(x) &=&
      N^{-1}_k  H_k((2Q)^{-1/2} x)\, e^{-x^2/2Q}
            %\continue
        \,,\quad
    Q = % \frac{\diffTen}{1-\ExpaEig^2}
             \diffTen/(1-\ExpaEig^2)
%        \,,\;\;    \mu^{-2} = 2Q
\,,
\label{DL:exp}
\eea
where $H_k(x)$ is the $k$th Hermite polynomial, and
$N^{-1}_k$ follows from the prefactor in \refeq{DL:Herm1}.

These discrete time results can be straightforwardly generalized to
continuous time flows of \refsect{DL:contFP_oper}, as well as to higher
dimensions. So far we have used only the leading eigenfunctions (the
natural measure), but in \refsect{DL:corrections} we shall see that
knowing the whole spectrum in terms of Hermite polynomial is a powerful
tool for the computation of weak-noise corrections.

\remark{Ornstein-Uhlenbeck process.}{
\label{rem:OrnUhlPro}
The simplest example of a continuous time stochastic flow
\refeq{DL:LangFlow} is the Langevin flow \refeq{DL:continuous} in one
dimension. In this case, nothing is lost by considering discrete-time
dynamics which is strictly equivalent to the continuous time
{Ornstein-Uhlenbeck process \refeq{DL:FPa}} discussed in
\refappe{DL:locEig}.
} %end \remark{Hyperbolic

% lippolis/Maribor/repeller.tex
% $Author: domenico $ $Date: 2012-05-20 12:43:06 -0400 (Sun, 20 May 2012) $

\section{Have no fear of globalization}
% \OptPart\ hypothesis}
\label{DL:OptPart}
                        \renewcommand{\version}{
  Predrag                   May  6 2012
                        }
%  Predrag                   Mar 14 2010
%  from lippolis/stoch/repeller.tex

We are now finally in position to address our challenge:
{\em Determine the finest possible partition for a given noise.}

\bigskip

%\subsection{Resolution of a one-dimensional chaotic repeller}
%\label{DL:repeller}

We shall explain our \emph{`the best possible of all partitions'}
hypothesis by formulating it as an algorithm. For every unstable periodic
point $x_a$ of a chaotic one-dimensional map, we calculate the
corresponding width $\tilde{\covMat}_a$ of the leading Gaussian
eigenfunction of the local adjoint \Fokker\ operator $\Lnoise{\dagger}$.
Every periodic point is assigned a one-standard deviation neighborhood
$[x_a-\sqrt{\tilde{\covMat}_a},x_a+\sqrt{\tilde{\covMat}}_a]$.
    %\DL{the confidence interval is $[x_a-\sigma_a,x_a+\sigma_a]$ (cf. our
    %PRL), now $\tilde{\covMat}_a=\sigma_a^2$, so there should be no extra
    %factor of $1/2$.}
We cover the
{\statesp} with neighborhoods of orbit points of higher and higher period
$\cl{p}$, and stop refining the local resolution whenever the adjacent
neighborhoods, say of $x_a$ and $x_b$, overlap in such a way that
$|x_a-x_b|<\sqrt{\tilde{\covMat}_a}+\sqrt{\tilde{\covMat}_b}$.
As an illustration of the method, consider the chaotic repeller on the
unit interval
\beq
x_{n+1} = \ExpaEig_0 \, x_n(1-x_n)(1-bx_n) + \xi_n
 \,, \quad
\ExpaEig_0 = 8,\; b=0.6
 \,,
\ee{DL:Ulam_rep}
with noise strength $\diffTen=0.002$.

\begin{figure}[tbp]
%\centering
(a) \includegraphics[width=0.4\textwidth]{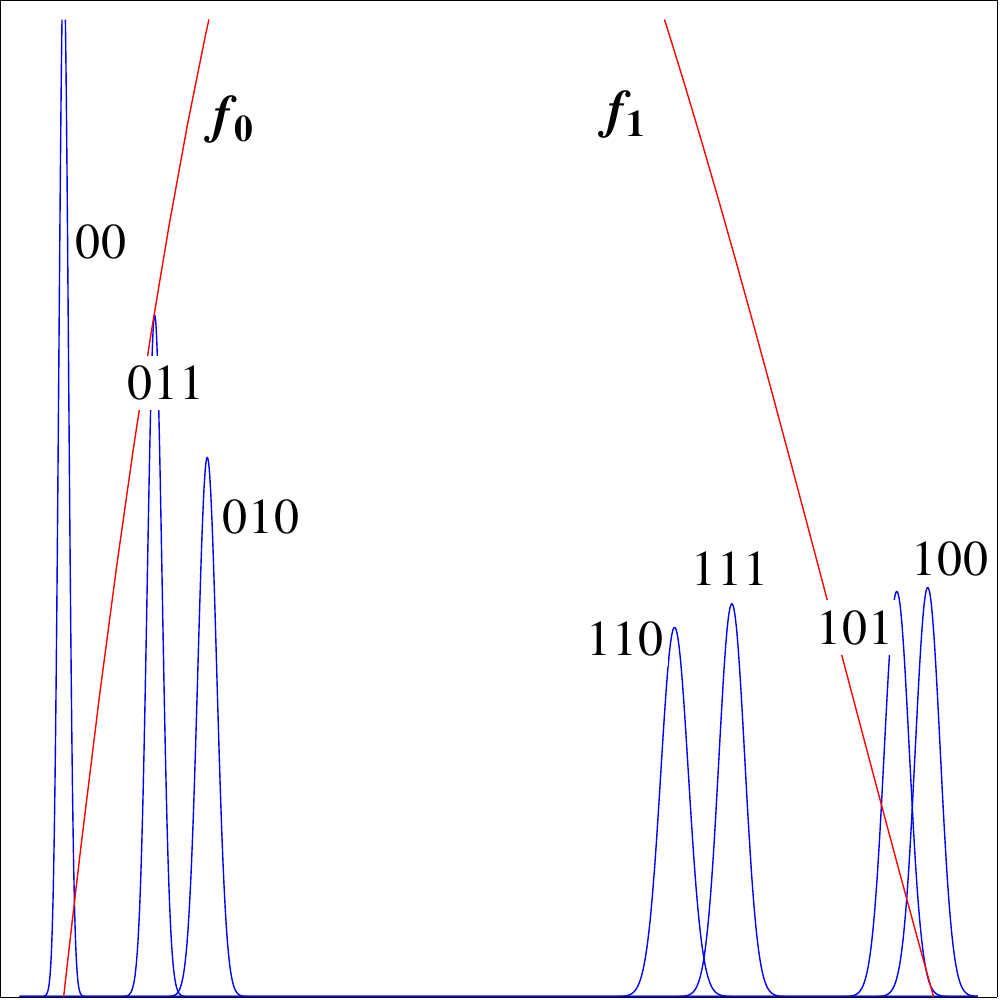}
~~~
(b) \includegraphics[width=0.4\textwidth]{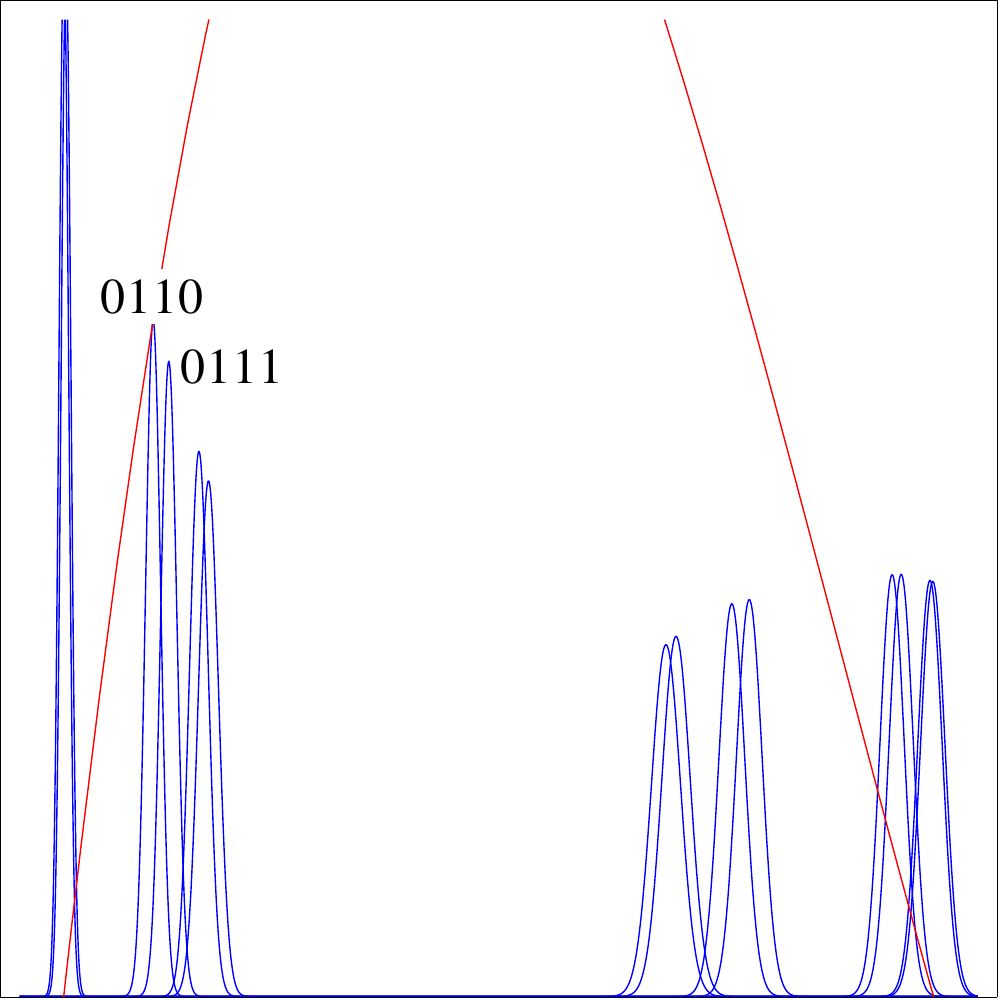}
\caption{
(a) $f_0,f_1$: branches of the deter\-mi\-ni\-stic map
\refeq{DL:Ulam_rep} for $\ExpaEig_0 = 8$ and $b=0.6$.
The local eigenfunctions $\tilde{\msr}_{a,0}$ with variances
given by \refeq{DL:n_eig} provide a \statesp\ partitioning by
neighborhoods of periodic points of period 3. These are
computed for noise variance $\diffTen=0.002$.
The neighborhoods $\pS_{000}$ and $\pS_{001}$
already overlap, so $\pS_{00}$ cannot be resolved further.
(b) The next generation of eigenfunctions
shows how the neighborhoods of the
\optPart\ cannot be resolved further.
For
periodic points of period 4, only $\pS_{011}$ can be resolved
further, into $\pS_{0110}$ and $\pS_{0111}$ (second and third peak
from the left), but that would not change the
transition graph of \reffig{f:rep_markov}.
}
\label{f:repOverlap}
\end{figure}
The map is plotted in \reffig{f:repOverlap}\,(a), together with the local
eigenfunctions $\tilde{\msr}_{a}$ with variances given by
\refeq{DL:n_eig}. Each Gaussian is labeled by the $\{f_0,f_1\}$ branches
visitation sequence of the corresponding deter\-mi\-ni\-stic periodic
point (a symbolic dynamics, however, is not a prerequisite for
implementing the method). \refFig{f:repOverlap}\,(b) illustrates the
overlapping of partition intervals: $\{\pS_{000},\pS_{001}\}$,
$\{\pS_{0101},\pS_{0100}\}$ overlap and so do all other neighborhoods of
the period $\cl{p}=4$ cycle points, except for $\pS_{0110}$ and
$\pS_{0111}$.
We find that in this case the \statesp\ (the unit interval) can be
resolved into 7 neighborhoods
\beq
 \{\pS_{00},\pS_{011},\pS_{010},
       \pS_{110},\pS_{111},\pS_{101},\pS_{100}\}
\,.
\label{TranMatD0.001}
\eeq
It turns out that resolving $\pS_{011}$ further into
$\pS_{0110}$ and $\pS_{0111}$ would not affect
our estimates, as it would produce the same
transition graph.

%%%%%%%%%%%%%%%%%%%%%%%%%%%%%%%%%%%%%%%%%%%%%%%%%%%%%%%%%%%%%%%%%%
\begin{figure}[tbp]
\centering
\includegraphics[width=0.8\textwidth,angle=0]{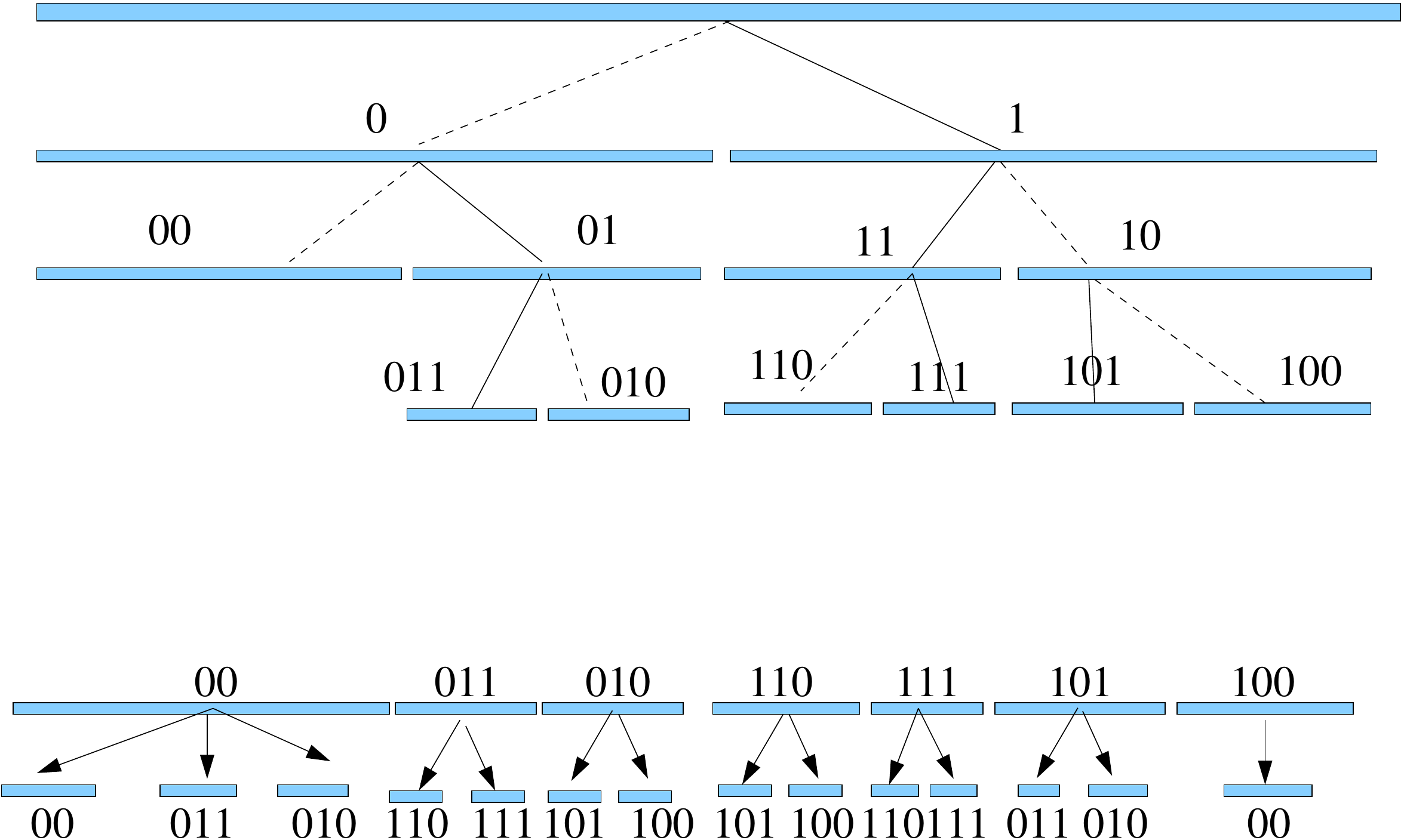}%
\caption{
(upper panel) The unit interval partitioned deter\-mi\-ni\-stically by a binary
tree. Due to the noise, the partitioning stops where the eigenfunctions
of \reffig{f:repOverlap} overlap significantly.
(lower panel) Once the \optPart\ is found, the symbolic dynamics is
recoded by relabeling the finite partition intervals, and refashioned
into the transition graphs of \reffig{f:rep_markov}.
        }
\label{f:rep_btree}
\end{figure}
%%%%%%%%%%%%%%%%%%%%%%%%%%%%%%%%%%%%%%%%%%%%%%%%%%%%%%%%%%%%%%%%%%
%

\begin{figure}[tbp]
(a) \includegraphics[width=0.42\textwidth]{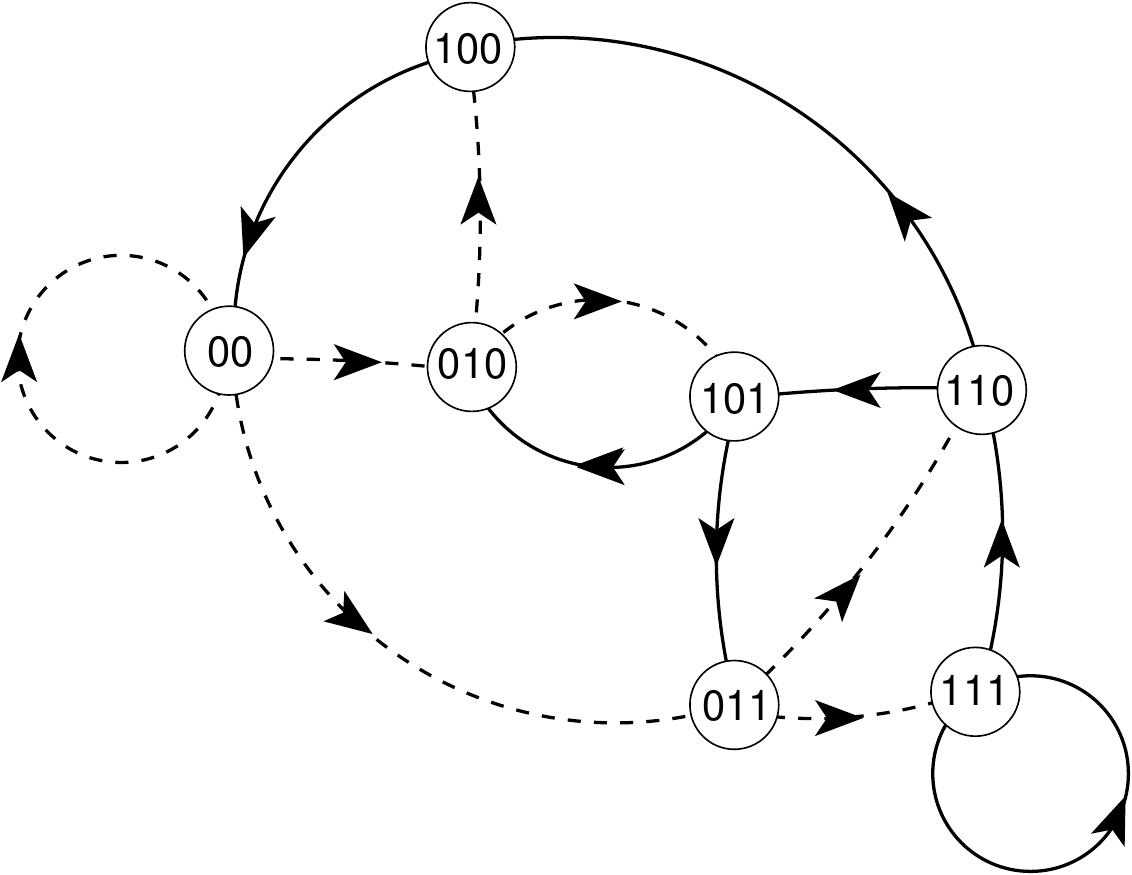}
~~~~
(b) \includegraphics[width=0.42\textwidth]{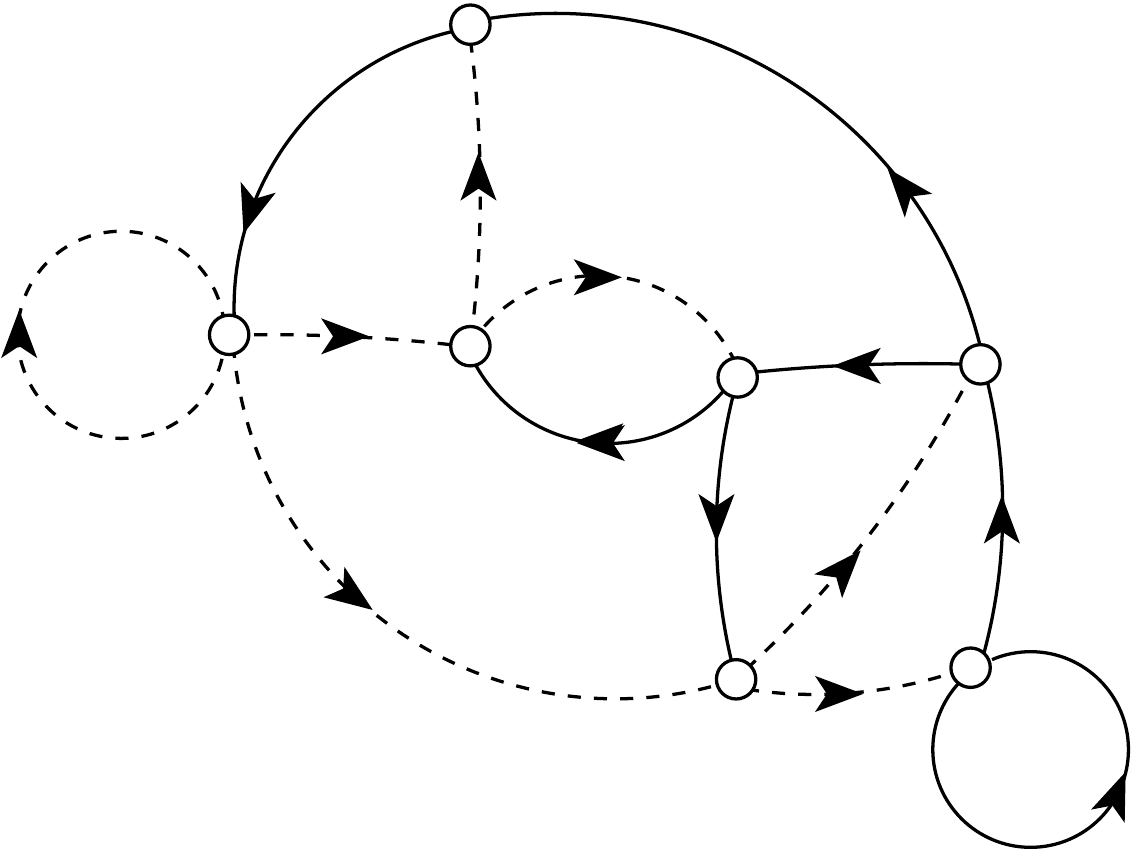}
\source{ChaosBook.org}
\caption{
(a) Transition graph (graph whose links correspond to the nonzero
elements of a transition matrix $T_{ba}$) describes which
regions $b$ can be reached from the region $a$ in one time
step. The 7 nodes correspond to the 7 regions of the {\optPart}
\refeq{TranMatD0.001}. Dotted links correspond to
symbol $0$, and the full ones to 1, indicating that the next
region is reached by the $f_0$, respectively $f_1$ branch of
the map plotted in \reffig{f:repOverlap}.
(b) The region labels in the nodes can be omitted, with links
keeping track of the symbolic dynamics.
}\label{f:rep_markov}
\end{figure}

Once the finest possible partition is determined, a finite binary tree
like the one in \reffig{f:rep_btree} is drawn: Evolution in time maps the
\optPart\ interval
$\pS_{011} \to \{\pS_{110},\pS_{111}\}$,
$\pS_{00} \to \{\pS_{00},\pS_{011},\pS_{010}\}$,
\etc. This is  summarized in the transition graph in \reffig{f:rep_markov},
which we will use to estimate the escape rate and the Lyapunov exponent
of the repeller.

%%%%%%%%%%%%%%%%%%%%%%%%%%%%%%%%%%%%%%%%%%%%%%%%%%%%%%%%%%%%%%%%%%
%%%%%%%%%%%%%%%%%%%%%%%%%%%%%%%%%%%%%%%%%%%%%%%%%%%%%%%%%%%%%%%%%%

  % lippolis/Maribor/partHist.tex
% $Author: domenico $ $Date: 2012-05-20 12:43:06 -0400 (Sun, 20 May 2012) $
%
%\section
\remark{A brief history of \statesp\ partitions.}{
\label{rem:partHist}
                        \renewcommand{\version}{
  Predrag                   May 5 2012
                        }
% Predrag  from stoch.tex     Jul 03 2010
%
There is considerable prior literature that addresses various
aspects of the `{\optPart}' problem.
Before reviewing it, let us state what is novel about the
{\optPart} hypothesis formulated here: Our estimates
of limiting resolution are {\em local}, differing from
region to region, while all of the earlier limiting resolution
estimates known to us are {\em global}, based on global averages such as
Shannon entropy or quantum-mechanical $\hbar$ `granularity' of phase
space.
We know of no published algorithm that sets a limit to the
resolution of a chaotic {\statesp} by studying the
interplay of the noise with the local stretching/contracting
directions of the deterministic dynamics, as we do here.

The engineering literature on optimal experimental
design\rf{Fedorov72,Kiefer85,AtFe75,FedKha86} employs
criteria such as `$D$-optimality,' the maximization of the
Shannon information content of parameter estimates. Purely
statistical in nature, these methods have little bearing on
the dynamical approach that we pursue here.
%

%
% Predrag transferred                           2010-03-29
%        from thesis/chapters/history.tex
In 1983 Crutchfield and Packard\rf{CruPack83} were the first
to study the problem of an {\optPart} for
a chaotic system in the presence of noise, and
formulate a \statesp\ resolution criterion in terms of a
globally averaged ``attainable information.''
The setting is the same
that we assume here: the laws governing deterministic
dynamics are given, and one studies the effects of noise (be it
intrinsic, observational or numerical) on the dynamics.
They
define the most efficient symbolic encoding of the dynamics
as the sequence of symbols that maximizes the metric entropy
of the entire system, thus their resolution criterion is based on
a global average. Once the maximum
for a given number of symbols is found, they refine the
partition until the entropy converges to some value.
They formulate their resolution criterion
in terms of \textit{attainable information}, a limiting value
for the probability to produce a certain sequence of symbols
from the ensemble of all possible initial conditions. Once
such limit is reached, no further refinements are possible.

Most of the dynamical systems literature deals with
estimating partitions from observed data\rf{DFT03}.
Tang and co-workers\rf{XZTang95} assume a noisy chaotic data
set, but with the laws of dynamics
assumed unknown. Their method is based on maximizing
Shannon entropy and at the same time minimizing an error
function with respect to the partition chosen. The same idea
is used by Lehrman \textit{et al.}\rf{Lehrm97} to encode
chaotic signals in higher dimensions, where they also detect
correlations between different signals by computing their
{conditional entropy}. For  a review of symbolic
analysis of experimental data up to 2001, see Daw, Finney and
Tracy\rf{DFT03}.

% Predrag merged with                           2010-04-07
%        from thesis/chapters/history.tex
Kennel and Buhl\rf{BuKe03,BuKe03a,BuKe05} estimate partitions
for (high-dimensional) flows from  noisy time-series  data by
minimizing a cost function which maximizes the correlation between
distances in the {\statesp} and in the symbolic space,
and indicates when to stop
adjusting their partitions and therefore what the {\optPart} is.
%
%\noindent {\bf PC} 13dec2009:
%is quite different from our idea. They say:
In \refref{BuKe03} their guiding principle for a good partition
is that
short sequences of consecutive symbols ought to localize
the corresponding continuous {\statesp} point as well as
possible. They embed symbol sequences into the unit square,
and minimize the errors in localizing the corresponding
{\statesp} points under candidate partitions.
    \PC{there is a long discussion of Kennel and Buhl
    papers on  refpage{?s:KennBuhl?}, summarize that here.
    In \refref{BuKe05} Kennel and Buhl minimize a cost function,
    and that is vaguely related to thing as we do.
    }
%
%{\bf PC} Aug 12, 2008:
Holstein and Kantz\rf{HolstKa08} present an
information-theoretic approach to deter\-min\-ation of
optimal Markov approximations from time series data based on
balancing the modeling and the statistical errors in
low-dimensional embedding spaces.
%
%{\bf PC} Apr 1, 2009:
Boland,  Galla  and McKane\rf{BoGaMcK09} study the effects of
intrinsic noise on a class of chemical reaction systems which
in the deter\-mi\-ni\-stic limit approach a limit cycle in an
oscillatory manner.

% Predrag transferred                           2010-03-29
%        from thesis/chapters/history.tex
A related approach to the problem of the optimal
resolution is that of the refinement of a transition matrix:
given a chaotic, discrete-time dynamical system, the
{\statesp} is partitioned, and the probabilities of points
mapping between regions are estimated, so as to obtain a
transition
matrix, whose eigenvalues and eigenfunctions are then used to
evaluate averages of observables defined on the chaotic set.
The approach was first proposed  in 1960 by Ulam\rf{Ulam60,LM94}, for
deterministic dynamical systems. He used a
uniform-mesh grid as partition, and conjectured that successive
refinements of such coarse-grainings would provide a
convergent sequence of finite-state Markov approximations to
the \FPoper.
Rechester and White\rf{RechWhi91,RechWhi91maps} have proposed
dynamics-based refinement strategies for
constructing partitions for chaotic maps in one and two
dimensions that would improve convergence
of Ulam's method.

%{\bf PC} Nov 10, 2007 -
Bollt \etal\rf{bollt06} subject a dynamical system to  a
small additive noise, define a finite Markov partition, and
show that the {\FPoper} associated
to the noisy system is represented by a finite-\-dimensional
stochastic transition matrix.
    \PC{I think they are inspired by $\hbar$ graining of
        quantum systems, different from us}
Their focus, however, is on
approximating the natural measure of a deter\-mi\-ni\-stic
dynamical system by the vanishing noise limit of a sequence
of invariant measures of the noisy system.
    \PC{read, summarize \refrefs{optp1,optp2} work on
        ``Optimal periodic orbits of chaotic systems''
    }

% Predrag transferred                           2010-03-29
%        from thesis/chapters/history.tex
In \refref{BuKe05} Kennel and Buhl approximate the distribution of the
points in each symbolic region by a Gaussian with mean $\mu$ and variance
$\tau$,
\beq
f(x|\mu,\tau) =
    \frac{1}{(2 \pi \tau)^{n/2}} \,
    \exp\left[\sum_i^n -\frac{1}{2 \tau} (x_i -\mu)^2 \right]
\,,
\ee{BuKe05:dscrt_FP}
and estimate ``code length'' by the {\em ad hoc} Rissanen prior on
$\tau$, defined with no reference to dynamics,  and thus morally
unrelated to our periodic orbits based {\optPart}.

Dellnitz and Junge\rf{DelJun98}, Guder and Kreuzer\rf{GuKr99},
Froyland\rf{FroyHD98}, and Keane \textit{et al.}\rf{Keane98} propose a
variety  of non-uniform refinement algorithms for such grids, reviewed in
a monograph by Froyland\rf{Froy-2}, who also treats their extension to
random dynamical systems. In all cases, the ultimate threshold for every
refinement is determined by the convergence of the spectrum of the
transition matrix.
    \PC{need to refer to some of the Graham articles:
\refrefs{Graham84,Graham84a,Graham85,Graham86,Graham90,Graham91,hamm_graham}.
I think at least one of these is in the \wwwcb{/library}.
    }

%
% {\bf PC} Oct 27, 2007:
Theoretical investigations mostly focus on deter\-mi\-ni\-stic limits of
stochastic models. The Sinai-Ruelle-Bowen\rf{sinai,bowen,Ruelle76} {or
natural measure} (also called {equilibrium measure}, SRB measure,
physical measure, invariant density, natural density, or natural
invariant) is singled out amongst all invariant measures by its
robustness to weak-noise perturbations, so there is considerable
literature that studies it as the deter\-mi\-ni\-stic limit of a
stochastic process.

} %end \remark{A brief history of \statesp\ partitions}{

% lippolis/Maribor/corrections.tex
% $Author: domenico $
% $Date: 2012-05-20 12:43:06 -0400 (Sun, 20 May 2012) $

\section{Finite \Fokker\ operator,
         and stochastic corrections}
\label{DL:corrections}

                        \renewcommand{\version}{
  Predrag                   Feb 11 2010
                        }

Next we show that the \optPart\ enables us to replace \Fokker\ PDEs by
finite-dimensional matrices. The variance \refeq{DL:n_eig} is stationary
under the action of $\Lnoise{\dagger\cl{p}}_{a}$, and the corresponding
Gaussian is thus an eigenfunction. Indeed, as we showed in
\refsect{DL:locEig1}, for the linearized flow the entire eigenspectrum is
available analytically. For a periodic point $x_a \in p$, the $\cl{p}$th
iterate $\Lnoise{\cl{p}}_a$ of the linearization \refeq{DL:app_evol} is
the discrete time version of the Ornstein-Uhlenbeck process,
with left $\tilde{\msr}_0$, $\tilde{\msr}_1$, $\cdots$, respectively
right ${\msr}_0$, ${\msr}_1$, $\cdots$ mutually orthogonal
eigenfunctions \refeq{DL:Herm1}.

The number of resolved periodic points determines the dimensionality
of the \Fokker\ matrix. Partition \refeq{TranMatD0.001} being the finest
possible partition, the \Fokker\ oper\-ator now acts as [$7\!\times\!7$]
matrix with non-zero $a \to b$ entries expanded in the Hermite basis,
\bea
[\Lmat{}_{ba}]_{kj} &=&
\left\langle \tilde{\msr}_{b,k} | \Lnoise{} |{\msr}_{a,j} \right\rangle
    \continue
    &=& \int \frac{d\orbitDist_b d\orbitDist_a \,\beta}
              {2^{j+1} j! \pi\sqrt{\diffTen/2}}
    e^{-(\beta \orbitDist_b)^2-\frac{(\orbitDist_b-f'_a(\orbitDist_a))^2}{2\diffTen}}
    \ceq \qquad \times\,
    H_k(\beta \orbitDist_b) H_j(\beta \orbitDist_a)
\,,
\label{DL:mtx_elem}
\eea
where $1/\beta=\sqrt{2\covMat_{a}}$, and $\orbitDist_a$ is the deviation
from the periodic point $x_a$.

Periodic orbit theory (summarized in \refappe{DL:POT}) expresses the
long-time dynamical averages, such as Lyapunov exponents, escape rates,
and correlations, in terms of the leading eigenvalues of the \Fokker\
operator. In our \optPart\ approach, $\Lnoise{}$ is
approximated by the finite-dimensional matrix $\Lmat{}$, and its
eigenvalues are determined from the zeros of $\det(1-z\Lmat{})$, expanded
as a polynomial in $z$, with coefficients given by traces of powers of
$\Lmat{}$. As the trace of the $n$th iterate of the \Fokker\ operator
$\Lnoise{n}$ is concentrated on periodic points $f^n(x_a)=x_a$, we
evaluate the contribution of {\po} $p$ to $\tr \Lmat{}^\cl{p}$ by
centering $\Lmat{}$ on the \po,
\beq
t_p = \tr_{p}\, \Lnoise{\cl{p}}
    =
\tr\Lmat{ad}\cdots\Lmat{cb}\Lmat{ba}
\,,
\ee{DL:tps}
where $x_a, x_b, \cdots x_d \in p$ are successive periodic points. To
leading order in the noise variance $\diffTen$, $t_p$ takes the
deter\-mi\-ni\-stic value $t_p =1/|\ExpaEig_p-1|$. The nonlinear
diffusive effects in \refeq{DL:mtx_elem} can be accounted
for\rf{noisy_Fred} by the weak-noise Taylor series expansion around the
periodic point $x_a$,
\beq
 e^{-\frac{(\orbitDist_b-f_a(\orbitDist_a))^2}{2\diffTen}}
    =
   e^{-\frac{(\orbitDist_{b}-\Df{a}\orbitDist_{a})^2}{2\diffTen}}
%    \times
%    \ceq
%    \qquad
\left(1-
\sqrt{2\diffTen}(f_a^{''}f_a^{'}\orbitDist_a^3 +
f_a^{''}\orbitDist_a^2\orbitDist_b) + O(\diffTen)\right).
\label{DL:appx_mtx}
\eeq
Such higher order corrections will be needed in what follows for a
sufficiently accurate comparison of different methods.

%%%%%%%%%%%%%%%%%%%%%%%%%%%%%%%%%%%%%%%%%%%%%%%%%%%%%%%%%%%%%%%%%%
\begin{figure}
\begin{minipage}[t]{0.90\columnwidth}%
(a)~%
\includegraphics[width=0.26\textwidth]{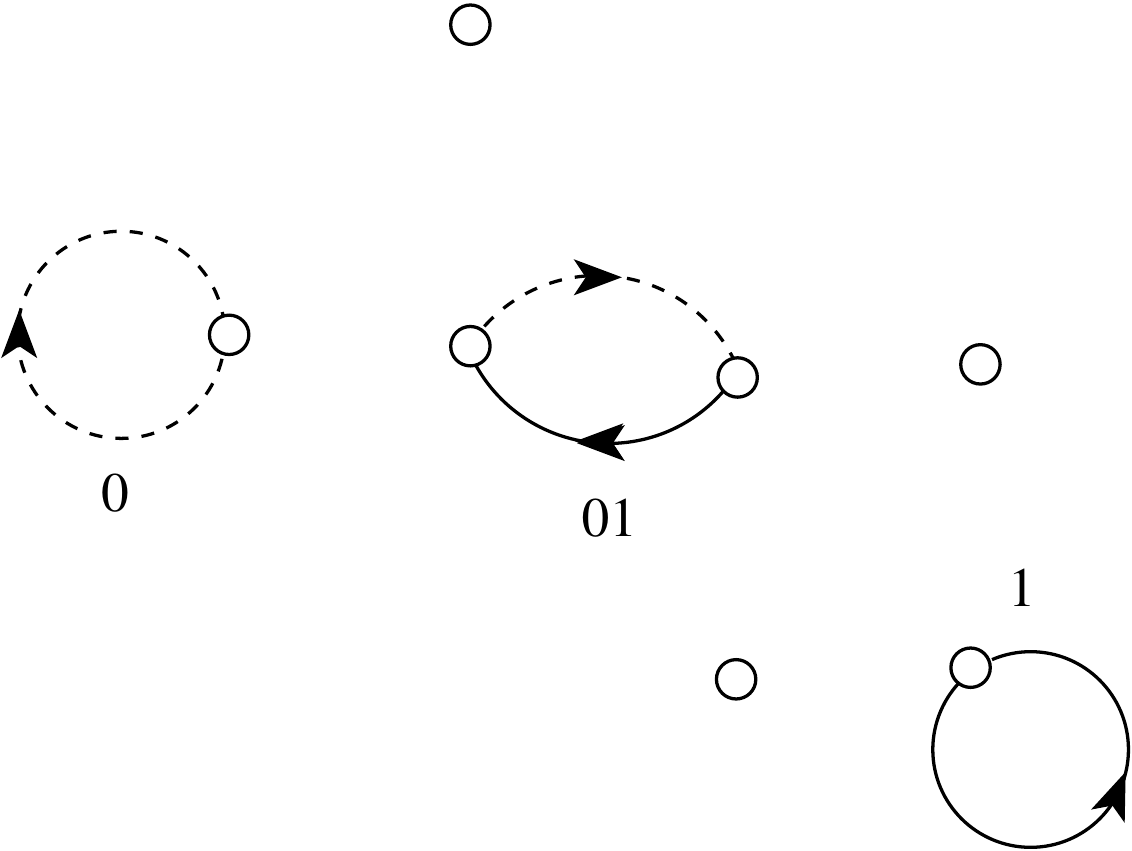}
\hskip 2ex
(b)~%
\includegraphics[width=0.22\textwidth]{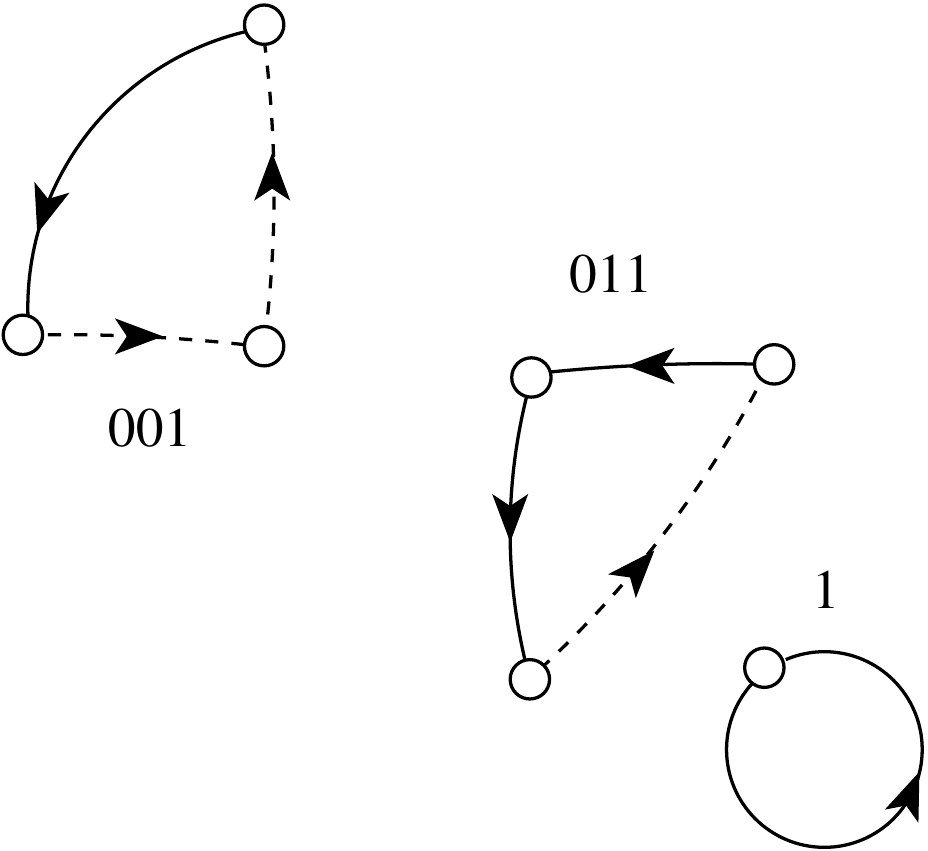}
\hskip 2ex
(c)~%
\includegraphics[width=0.22\textwidth]{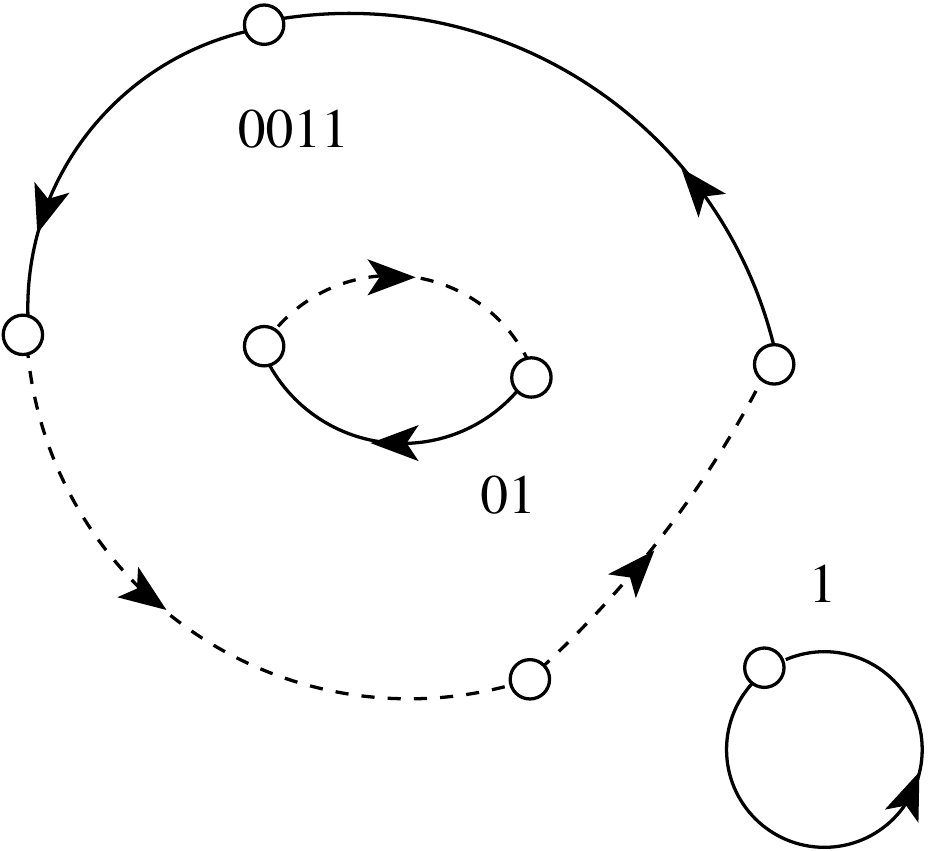}
\\
(d)~%
\includegraphics[width=0.26\textwidth]{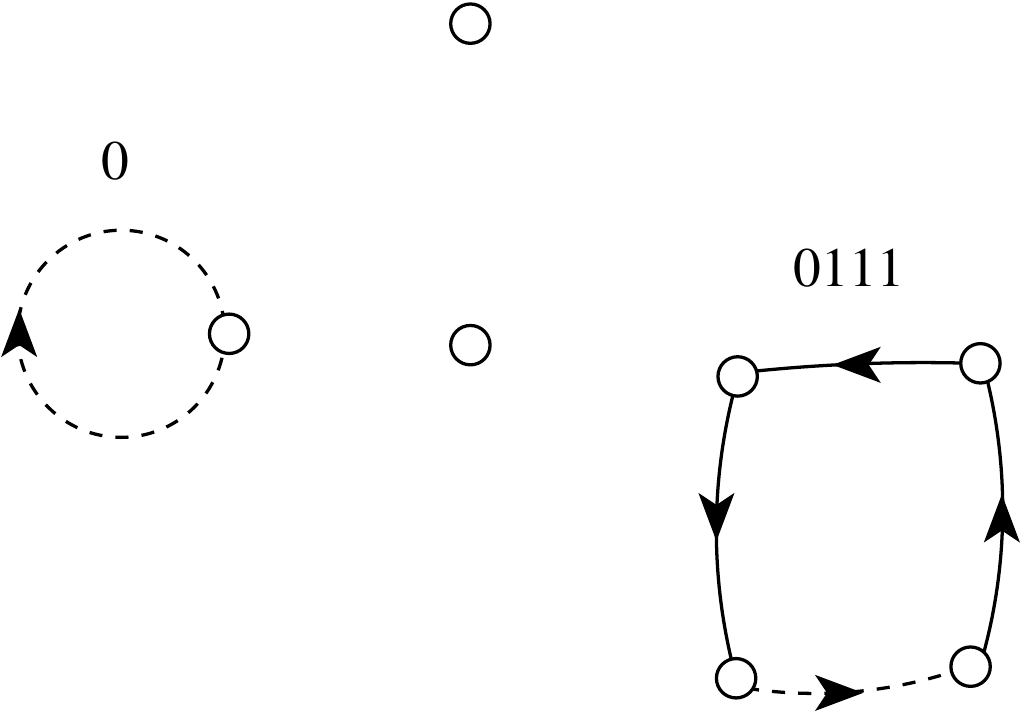}
\hskip 2ex
(e)~%
\includegraphics[width=0.22\textwidth]{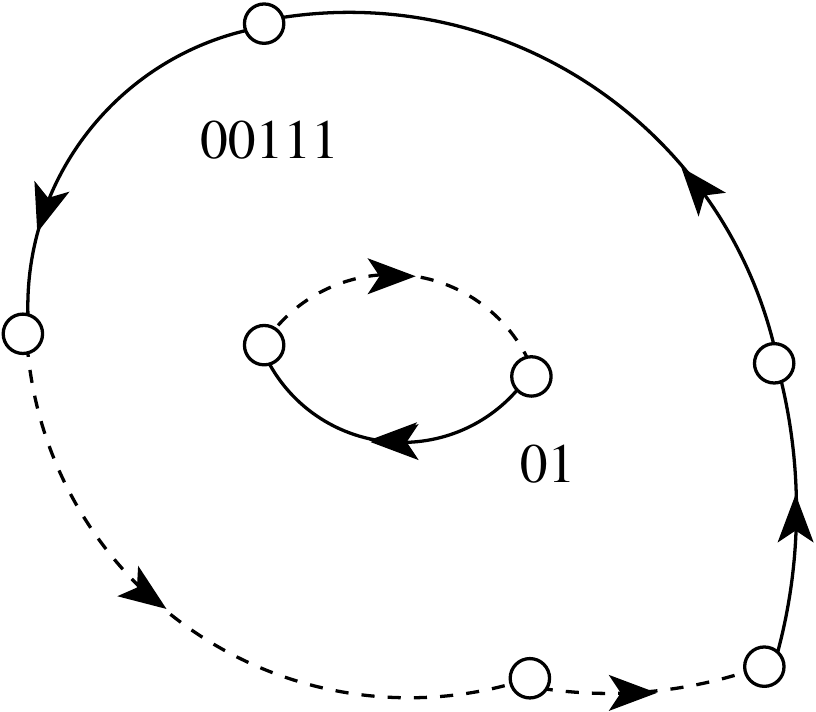}
\hskip 2ex
(f)~%
\includegraphics[width=0.22\textwidth]{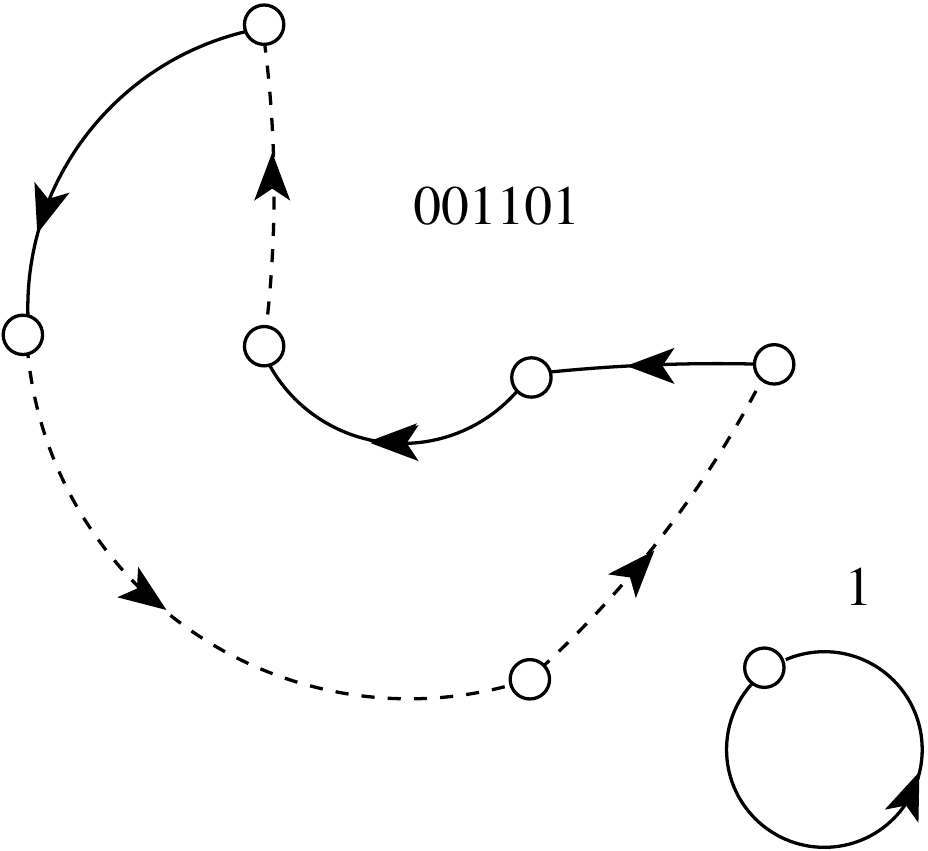}
\\
(g)~%
\includegraphics[width=0.24\textwidth]{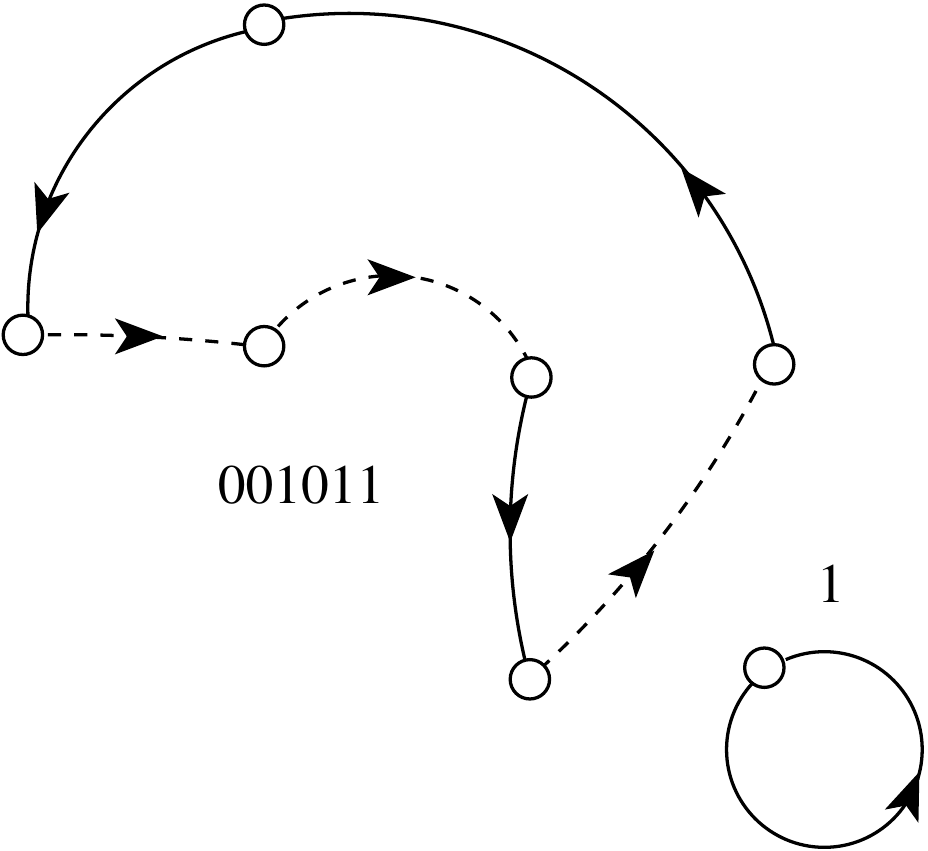}
\hskip 3ex
(h)~%
\includegraphics[width=0.22\textwidth]{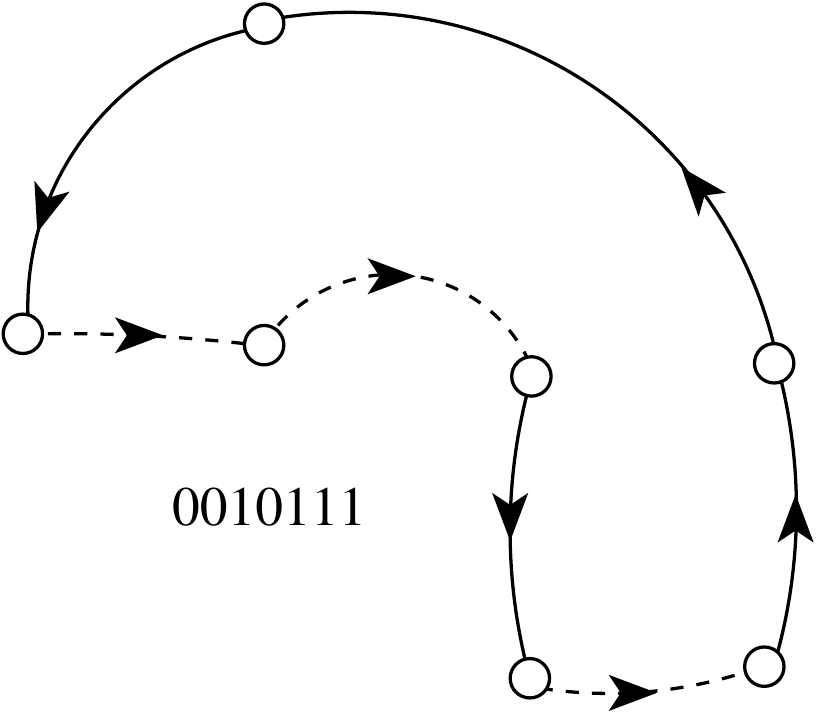}
\hskip 2ex
(i)~%
\includegraphics[width=0.22\textwidth]{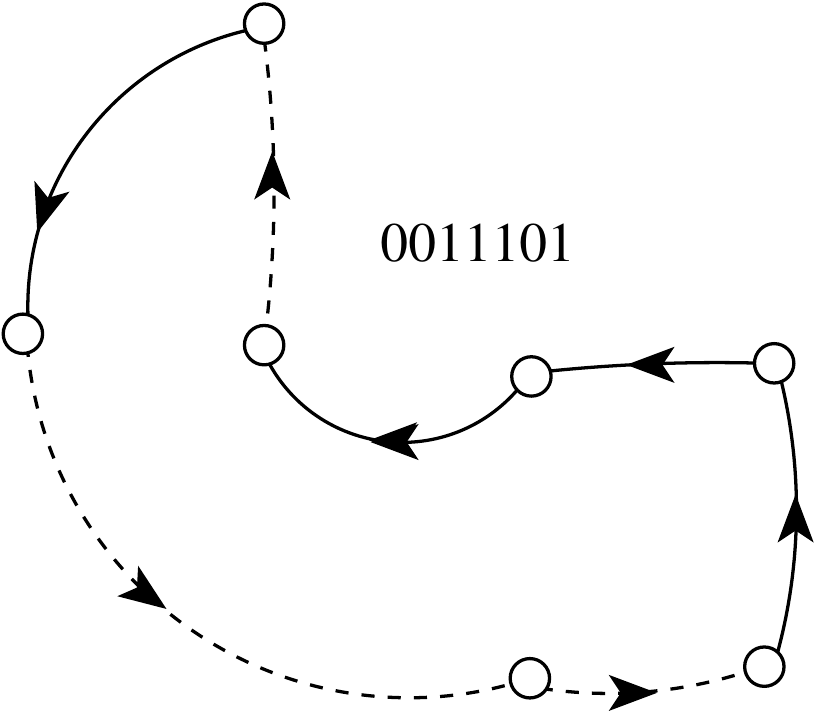}
\end{minipage}
\source{ChaosBook.org}
\caption{
(a)-(i) The {fundamental cycles} for the \markGraph\
\reffig{f:rep_markov}\,(b), \ie, the set of its non-self-intersecting
loops. Each loop represents a local trace $t_p$; together they form the
determinant \refeq{PC:rep_det}.
}
\label{f:repelLoops}
\end{figure}
%%%%%%%%%%%%%%%%%%%%%%%%%%%%%%%%%%%%%%%%%%%%%%%%%%%%%%%%%%%%%%%%%%

\begin{figure}[tbp]
(a) \includegraphics[width=0.46\textwidth]{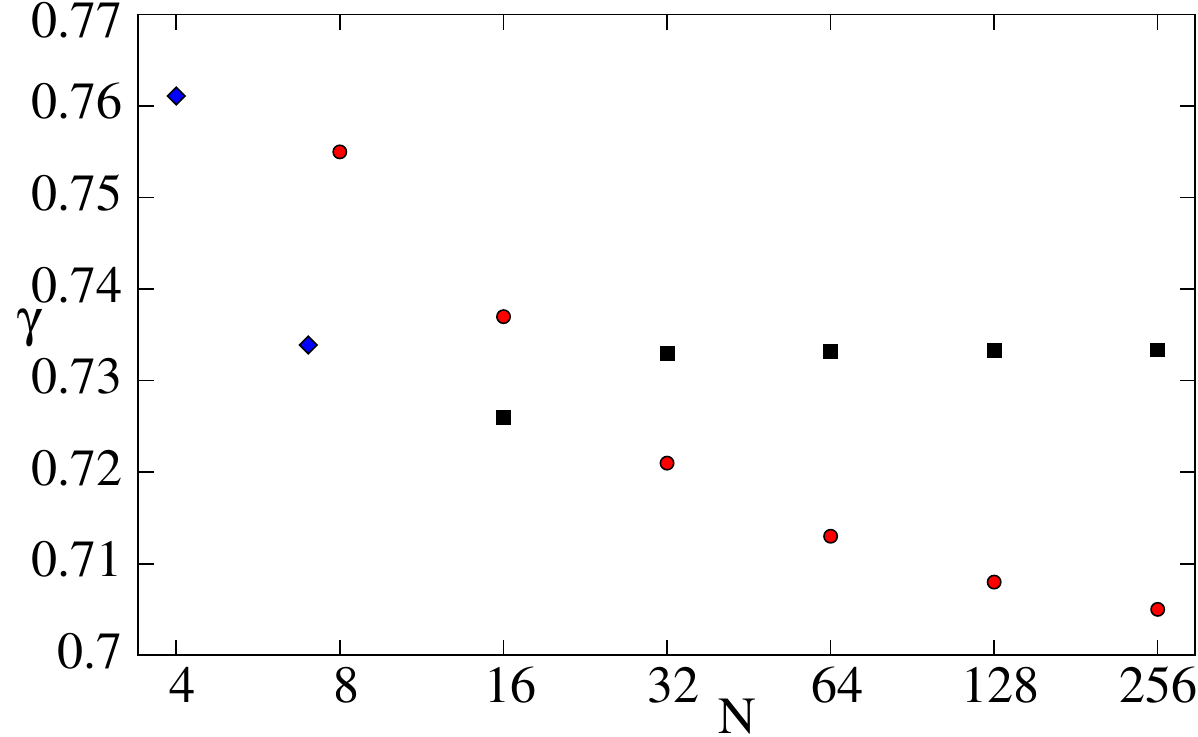}
(b) \includegraphics[width=0.46\textwidth]{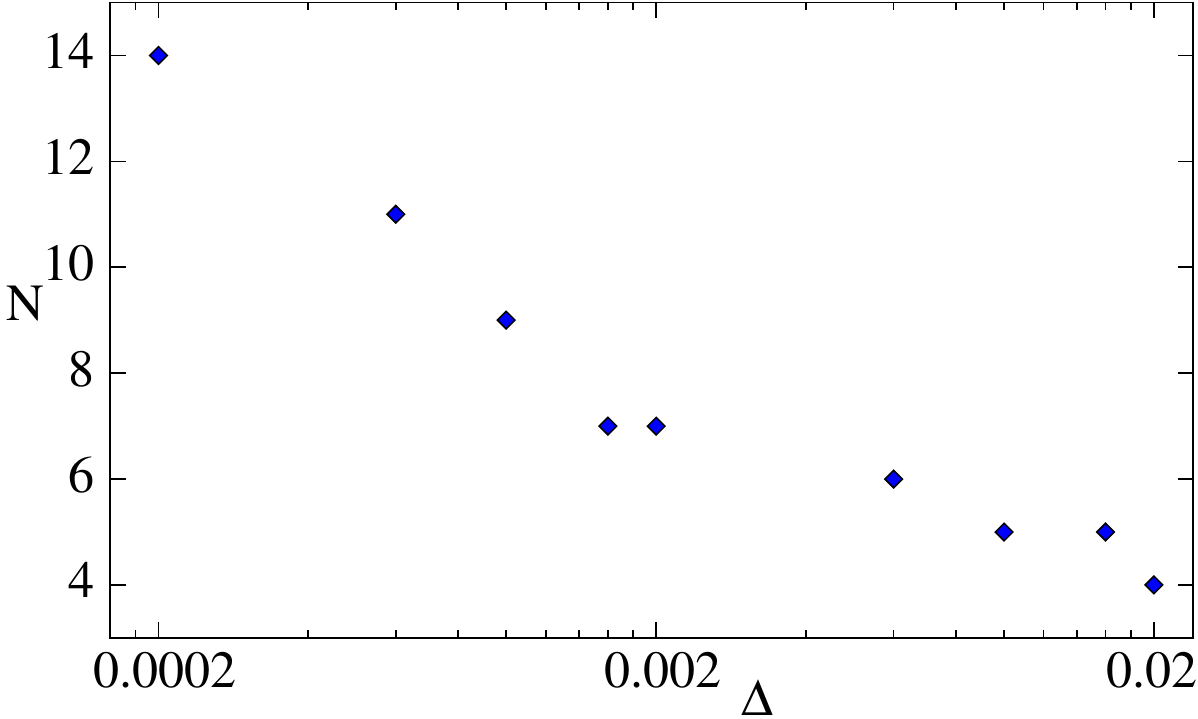}%
\caption{
(a) The escape rate $\gamma$ of the repeller in \reffig{f:repOverlap}
plotted as function of number of partition intervals $N$,
estimated using:
($\color{blue}\blacklozenge$) under-resolved 4-interval and
the  7-interval \optPart,
({%\Large
  $\color{red}\bullet$}) all \po s of periods
up to $n=8$ in the deter\-mi\-ni\-stic, binary symbolic dynamics,
with $N_i=2^n$ periodic-point intervals (the deter\-mi\-ni\-stic,
noiseless escape rate is $\gamma_{<>} = 0.7011$), and
({\small
  $\blacksquare$}) a uniform discretization
\refeq{DL:overl-intFP} in $N=16,\cdots, 256$ intervals. For $N=512$
discretization yields $\gamma_{\textrm{num}} = 0.73335(4)$.
(b) Number of neighborhoods required by the optimal partition
method vs. the noise strength $\diffTen$.
}
\label{f:escRates}
\end{figure}

We illustrate the method by calculating the escape rate $\gamma = - \ln
z_0$, where $z_0^{-1}$ is the leading eigenvalue of \Fokker\ operator
$\Lnoise{}$, for the repeller plotted in \reffig{f:repOverlap}. The \Fd\
can be read off the transition graph of \reffig{f:rep_markov} and
its loop expansion in \reffig{f:repelLoops},
\bea
&& \det(1-z\Lmat{}) =
    1 - (t_0 + t_1)z - (t_{01} - t_0 t_1)\,z^2
    \ceq\quad
      - (t_{001} + t_{011} -  t_{01} t_0 - t_{01} t_1)\,z^3
    % - \cdots
    \ceq
    \quad    -  (t_{0011} + t_{0111} -t_{001}t_1 -t_{011} t_0
         - t_{011} t_1 + t_{01} t_0 t_1)\,z^4
    \ceq
    \quad    - (t_{00111} -t_{0111} t_0 -t_{0011} t_1 + t_{011} t_0 t_1)\,z^5
    \ceq
    \quad    - (t_{001011} + t_{001101} - t_{0011} t_{01}
        - t_{001} t_{011})\,z^6
    \ceq
    \quad - (t_{0010111} + t_{0011101} - t_{001011} t_1 - t_{001101} t_1
    \ceq
    \quad \quad - t_{00111} t_{01} + t_{0011} t_{01} t_1 + t_{001} t_{011} t_1)\,z^7
    % \quad - (t_{0010111} + t_{0011101} - \cdots + t_{001} t_{011} t_1)\,z^7
.
\label{PC:rep_det}
\eea
The polynomial coefficients are given by products of non-intersecting
loops of the transition graph\rf{DasBuch}, with the escape rate given by
the leading root $z_0^{-1}$ of the polynomial.
Twelve \po s
\cycle{0}, \cycle{1}, \cycle{01}, \cycle{001}, \cycle{011},
\cycle{0011}, \cycle{0111}, \cycle{00111},
\cycle{001101}, \cycle{001011},
% $\dots$,
\cycle{0010111}, \cycle{0011101}
up to period 7 (out of the 41 contributing to the noiseless,
deter\-mi\-ni\-stic cycle expansion up to cycle period 7) suffice to
fully determine the \Fd\ of the \Fokker\ oper\-ator. In the evaluation of
traces ~\refeq{DL:tps} we include stochastic corrections up to order
$O(\diffTen)$ (an order beyond the term kept in \refeq{DL:appx_mtx}). The escape
rate of the repeller of \reffig{f:repOverlap} so computed is reported in
\reffig{f:escRates}.

Since our \optPart\ algorithm is based on a sharp overlap criterion,
small changes in noise strength $\diffTen$ can lead to transition graphs
of different topologies, and it is not clear how to assess the accuracy
of our finite \Fokker\ matrix approximations. We make three different
attempts, and compute the escape rate for:
(a) an under-resolved partition,
(b) several deter\-mi\-ni\-stic, over-resolved partitions, and
(c) a direct numerical discretization of the \Fokker\ operator.

(a) In the example at hand, the partition in terms of periodic points
\cycle{00}, \cycle{01}, \cycle{11} and \cycle{10}
is under-resolved; the corresponding escape rate is plotted in
\reffig{f:escRates}.
(b) We calculate the escape rate by over-resolved
{\po} expansions, in terms of \textit{all} deter\-mi\-ni\-stic
\po s of the map up to a given period, with $t_p$
evaluated in terms of \Fokker\ local traces \refeq{DL:tps},
including stochastic corrections up to order $O(\diffTen)$.
\refFig{f:escRates} shows how the escape rate varies as we
include all \po s up to periods 2 through 8.
Successive  estimates of the escape rate appear to converge to
a value different from the \optPart\ estimate.
\noindent
(c) Finally, we discretize the \Fokker\ operator $\Lnoise{}$
by a piecewise-constant approximation on a uniform mesh
on the unit interval\rf{Ulam60},
\beq
[\Lnoise{}]_{ij} \,=\, \frac{1}{|\pS_i|} \frac{1}{\sqrt{2\pi\diffTen}}
    \int_{\pS_i} \! dx  \int_{f^{-1}(\pS_j)} \!dy \,
    e^{-\frac{1}{2\diffTen}(y-f(x))^2}
\!,
\ee{DL:overl-intFP}
where $\pS_i$ is the $i$th interval in equipartition of the unit interval
into $N$ equal segments. Empirically, $N=128$ intervals suffice to
compute the leading eigenvalue of the discretized $[128\!\times\!128]$
matrix $[\Lnoise{}]_{ij}$ to four significant digits. This escape rate,
\reffig{f:escRates}, is consistent with the $N=7$ \optPart\ estimate to
three significant digits.

%%%%%%%%%%%%%%%%%%%%%%%%%%%%%%%%%%%%%%%%%%
%%%%%%%%%%%%%%%%%%%%%%%%%%%%%%%%%%%%%%%%%%

\begin{figure}[tbp]
\begin{minipage}[t]{1.00\columnwidth}%
(a) \includegraphics[width=0.46\textwidth]{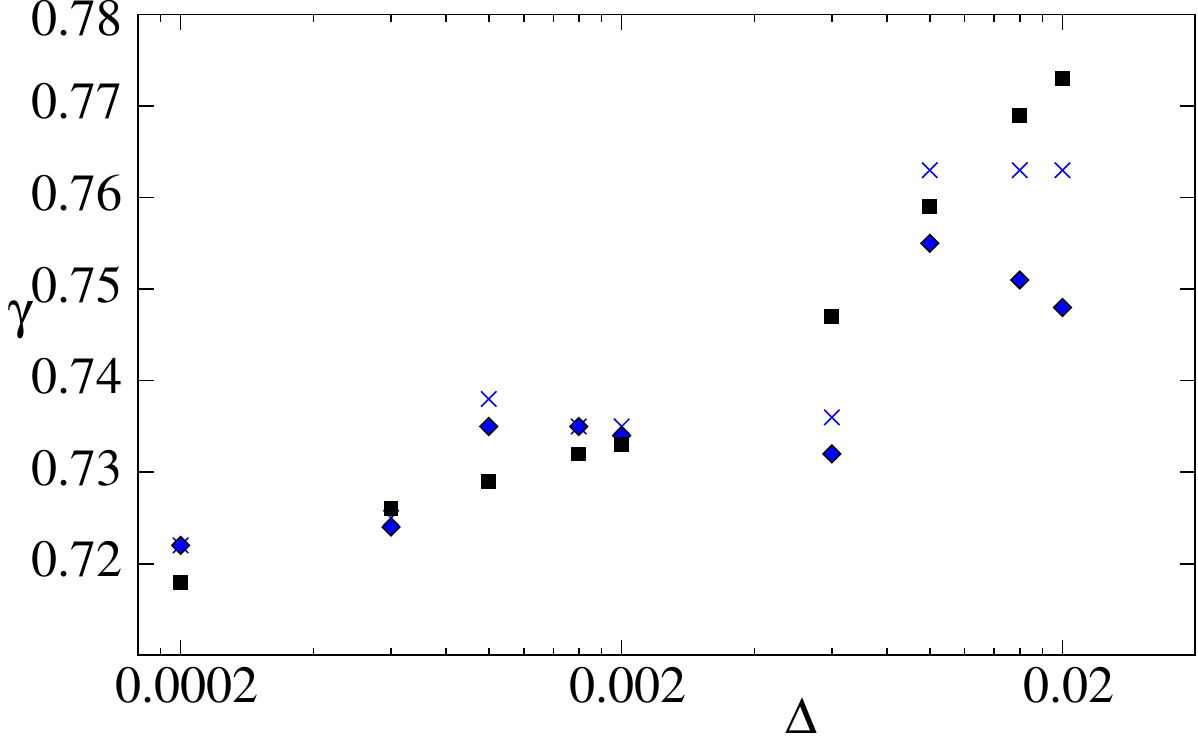}%
(b) \includegraphics[width=0.46\textwidth]{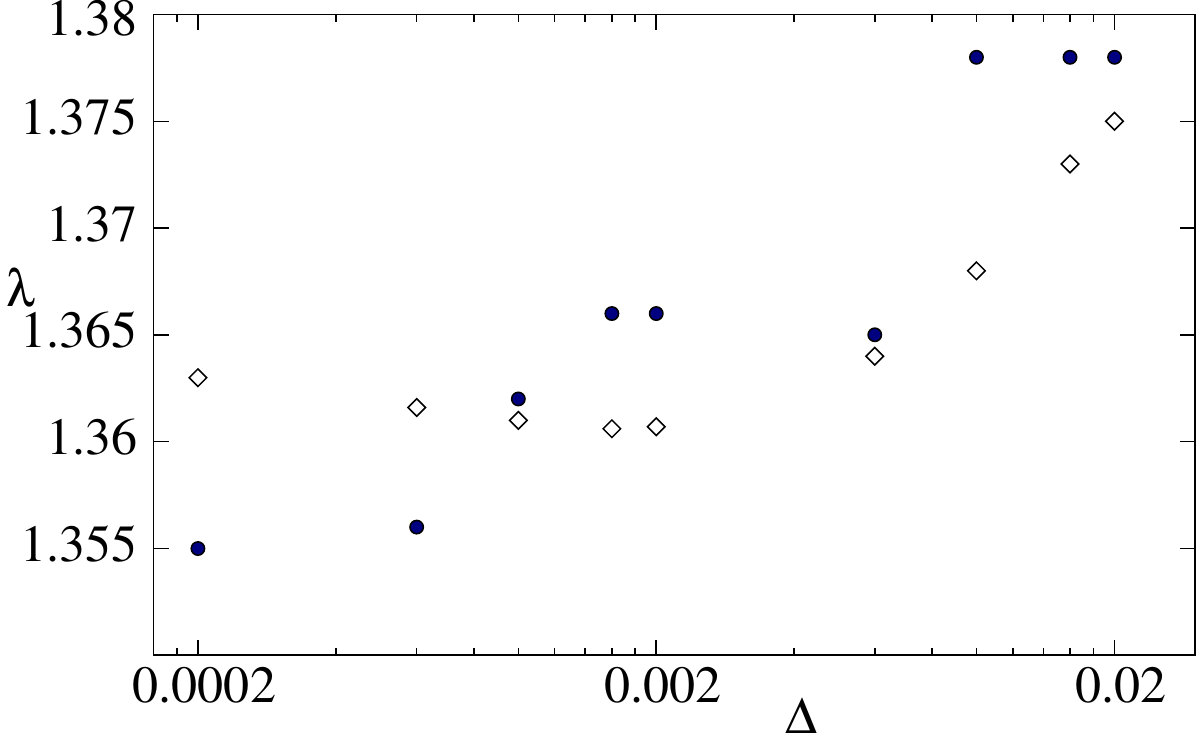}%
\end{minipage}
\caption{
(a)
Escape rates of the repeller \refeq{DL:Ulam_rep} vs. the noise strength
$\diffTen$, using: the \optPart\ method with ($\color{blue}\blacklozenge$) and
without ($\color{blue}\times$) stochastic corrections;
({%\scriptsize
$\blacksquare$}) a uniform discretization \refeq{DL:overl-intFP} in
$N=128$ intervals.
(b)
The Lyapunov exponent of the repeller \refeq{DL:Ulam_rep} vs. the noise
strength $\diffTen$, using: the \optPart\ method ($\bullet$) without stochastic
corrections, and ({$\diamond$}) a uniform discretization
\refeq{DL:overl-intFP} over $N=128$ intervals.
 }
\label{f:D_vs_gamma}
\end{figure}

We estimate the escape rate of the repeller \refeq{DL:Ulam_rep} for a
range of values of the noise strength $\diffTen$. The {\optPart} method
requires a different numbers of neighborhoods every time for different
noise strengths. The results are illustrated by \reffig{f:escRates}\,(b)
and \ref{f:D_vs_gamma}, with the estimates of the {\optPart} method
within $2\%$ of those given by the uniform discretization of \Fokker. One
can also see from the same table that the escape rates calculated with
and without higher order corrections to the matrix elements
\refeq{DL:mtx_elem} are consistent within less than $2\%$, meaning that
the stochastic corrections \refeq{DL:appx_mtx} do not make a significant
difference, compared to the effect of the optimal choice of the
partition, and need not be taken into account in this example.

%%%%%%%%%%%%%%%%%%%%%%%%%%%%%%%%%%%%%%%%%%
%%%%%%%%%%%%%%%%%%%%%%%%%%%%%%%%%%%%%%%%%%

The \optPart\ estimate of the Lyapunov exponent is given by $\Lyap =
\expct{\ln\,|\ExpaEig|}/\expct{n}$, where the cycle expansion average of
an integrated observable $A$~\rf{DasBuch}
\bea
\left<A\right> &=& A_0t_0+A_1t_1 +\,[A_{01}t_{01}-(A_0+A_1)\,t_0t_1]
\ceq\quad         + \,[A_{001}t_{001}-(A_{01}+A_0)t_{01}t_0] + \cdots
\ceq\quad                    +  \,[A_{011}t_{011}-(A_{01}+A_1)t_{01}t_1]
                                    + \cdots
\label{DL:average}
\eea
is the finite sum over cycles contributing to \refeq{PC:rep_det}, and
$\ln |\ExpaEig_p| = \sum\ln|f'(x_a)|$, the sum over the points of cycle
$p$, is the cycle Lyapunov exponent. On the other hand, we also use the
discretization \refeq{DL:overl-intFP} to cross check our estimate: this
way the Lyapunov exponent is evaluated as the average
\beq
\Lyap = \int dx \,e^\gamma \rho(x)\ln|f'(x)|
\,,
\ee{DL:num_liap}
where $\rho(x)$ is the leading eigenfunction of \refeq{DL:overl-intFP},
$\gamma$ is the escape rate, and $e^\gamma \rho$ is the normalized
repeller measure, $\int dx  \, e^{\gamma}\rho(x) = 1$.
\refFig{f:D_vs_gamma} shows close agreement ($<1\%$) between the Lyapunov
exponent estimated using the average \refeq{DL:average}, where  $t_p =
1/|\ExpaEig_p -1|$ (no higher-order stochastic corrections), and the same
quantity evaluated with \refeq{DL:num_liap}, by the discretization method
\refeq{DL:overl-intFP}.

\remark{Weak noise corrections.}{
\label{rem:WeakNoisCorr}
The weak-noise corrections to the spectrum of \evOper s were first
treated by Gaspard\rf{gasp02} for continuous-time systems, and in a
triptych of articles\rf{noisy_Fred,conjug_Fred,diag_Fred} for
discrete-time maps: they can be computed perturbatively to a remarkably
high order\rf{DettYear} in the noise strength $\diffTen$. However, as we
have shown here, the {\em eigenvalues} of such operators offer no
guidance to the `{\optPart}' problem; one needs to compute the {\em
eigenfunctions}.
} %end \remark{Weak noise corrections

% lippolis/Maribor/flatTop.tex
% $Author: predrag $ $Date: 2012-05-18 17:46:55 -0400 (Fri, 18 May 2012) $

\section{When the Gaussian approximation fails}
\label{DL:flat_top}
% former lippolis/stoch/flat_top.tex

    \PC{Preface by text from ChaosBook chapter on uniform approximations}
%    \PC{[addressed Reviewer 3 concerns]:
%A possible difficulty with this approach is that the density
%function can be highly singular, as typically observed in
%chaotic systems. It is not clear how this difficulty will be
%dealt with in the proposed research.
%    }
%	\PC{this paragraph is a verbatim repeat from intro.tex,
%    	    rewritten a bit...}
The \statesp\ of a generic deter\-mi\-ni\-stic flow is an infinitely
interwoven hierarchy of attracting, hyperbolic and marginal regions, with
highly singular invariant measures. Noise has two types of effects.
First, it feeds trajectories into \statesp\ regions that are
deterministically either disconnected or transient (``noise induced
escape,'' ``noise induced chaos'') and second, it smoothens out the
natural measure. Here, we are mostly concerned with the latter.
Intuitively, the noisy dynamics erases any structures finer than the
\optPart, thus -in principle- curing both the affliction of long-period
attractors/elliptic islands with very small immediate basins of
attraction/ellipticity, and the slow, power-law correlation decays
induced by marginally stable regions of \statesp. So how does noise
regularize nonhyperbolic dynamics?
    \PC{is Hypg\_Laguer.tex used? You added it to the repository, but do
        not call it anywhere. Seems to be included in flatTopAppe.tex.}

As a relatively simple example, consider the skew Ulam map\rf{AACII},
\ie, the cubic map \refeq{DL:Ulam_rep} with the parameter
$\ExpaEig_0=1/\map(\ssp_c)$. The critical point $\ssp_c$ is  the maximum
of $\map$ on the unit interval, with vanishing derivative
$\map'(\ssp_c)=0$. As this map sends the unit interval into itself, there
is no escape, but due to the quadratic maximum the (deterministic)
natural measure exhibits a spike $(\ssp_b - \ssp)^{-1/2}$ near the
critical value $\map(\ssp_c) = \ssp_b$ (see, for example,
\refref{Ruelle09} for a discussion). As explained in \refref{AACII}, a
close passage to the critical point effectively replaces the accumulated
Floquet multiplier by its square root. For example, for the skew Ulam map
\refeq{DL:Ulam_rep} $n$-cycles whose itineraries are of form
\cycle{0^{n-1}1} spend long time in the neighborhood of $\ssp_0=0$, and
then pass close to $\ssp_c$. In the neighborhood of  $\ssp_0=0$ the
Floquet multiplier gains a factor $\sim\ExpaEig_0 = \map'(\ssp_0)$ for
each of the first ${n\!-\!1}$ iterations, and then experiences a strong,
square root contraction during the close passage to the critical point
$\ssp_c$, resulting in the Floquet multiplier $\ExpaEig_{0\cdots01}
\propto \ExpaEig_{0}^{n/2} $, and a Lyapunov exponent that converges to
$\Lyap_0/2$, rather than $\Lyap_0$ that would be expected in a hyperbolic
flow for a close passage to a fixed point $\ssp_0$. The same strong
contraction is experienced by the noise accumulated along the trajectory
prior to the passage by the critical point, rendering, for example, the
period-doubling sequences more robust to noise than one would na\"{i}vely
expect\rf{Crutchfield81,Shraiman81,FeHa82}.

For the corresponding noisy map \refeq{DL:discrete} the critical point is
extended into the `flat top' region where $|\map'(x)| \ll 1$, and the
linearized, Gaussian approximation  \refeq{DL:app_evol} to the \Fokker\
operator does not hold. Thus, we should first modify our choice of
densities and neighborhoods, as the whole construction leading to the
\optPart\ algorithm was based on the Gaussian approximation.

The adjoint \Fokker\ operator acts on a Gaussian density centered at
$\ssp_a$, as in \refeq{DL:GaussDens}:
\bea
[\Lnoise{\dagger} \tilde{\msr}_a](\ssp) &=&
 \frac{1}{C_a} \int_{-\infty}^\infty\,
  e^{-\frac{(f(x)-y)^2}{2\diffTen}}e^{-(y-x_a)^2/2\covMat_a}[dy]
    \continue
    &=&
\frac{1}{C_{a-1}} \; e^{-\frac{(f(x)-x_a)^2}{\covMat_a+\diffTen}}
\,.
\label{DL:gss_evlnRpt}
\eea
Suppose the point $\ssp_{a-1} \ = \map^{-1}(\ssp_{a})$
%PC commented out  (cf. \refsect{DL:adjoint}),
around which we want to approximate the new density, is very
close to the critical point, so that we
can write
\beq
\msr_{a-1}(x) \,=\,
\frac{1}{C_{a-1}}e^{-\frac{(f(x)-x_a)^2}{2(\covMat_a+\diffTen)}}
\,=\, \frac{1}{C_{a-1}}e^{-\DDf{a-1}{}^2 \orbitDist_{a-1}^4/8(\covMat_a+\diffTen)}
\,.
\label{DL:flat_expnsn}
\eeq
%with $\DDf{a}$ defined in \refeq{DL:orbitDerv1d}.
During a close passage to the critical point, the variance does
not transform linearly, but as a square root:
    \PC{can ${\Gamma(3/4)}/{\Gamma(1/4)}$ be simplified?}
\beq
\covMat_{a-1} = \frac{\int
\orbitDist^2e^{-\DDf{a-1}{}^2 \orbitDist^4/8(\covMat_a+\diffTen)}d\orbitDist}
{\int e^{-\DDf{a-1}{}^2 \orbitDist^4/8(\covMat_a+\diffTen)}d\orbitDist}
= \frac{\Gamma(3/4)}{\Gamma(1/4)}\left(\frac{8(\covMat_a+\diffTen)}{\DDf{a-1}{}^2 }\right)^{1/2}
\,.
\ee{DL:flat_var}
We show in \refappe{DL:Hypg_Laguer} that in the next iteration the
variance of the density $\msr_{a-1}(\orbitDist_{a-1})$ transforms again
like the variance of a Gaussian, up to order $O(\diffTen)$ in the noise
strength. By the same procedure, one can again assume the next preimage
of the map $x_{a-3}$ is such that the linear approximation is valid, and
transform the density $\msr_{a-2}(\orbitDist_{a-2})$ (Eq.
\refeq{DL:hypergeom}, \refappe{DL:Hypg_Laguer}) up to $O(\diffTen)$ and
obtain the same result for the variance, that is
\beq
\covMat_{a-3} = \frac{\covMat_{a-1} + \diffTen(1+f_{a-2}^{'2})}
{f^{'2}_{a-2}f^{'2}_{a-3}}
\ee{DL:trd_var}
which is again the evolution
of the variances in the Gaussian approximation. In other words, the
evolution of the variances goes back to be linear, to $O(\diffTen)$,
although the densities transformed from the `quartic Gaussian'
\refeq{DL:flat_expnsn} are no longer Gaussians.

The question is now how to modify the definition of neighborhoods given
in \refsect{DL:OptPart}, in order to fit the new approximation. Looking
for eigenfunctions of $\Lnoise{\dagger}$  seems to be a rather difficult
task to fulfill, given the functional forms \refeq{DL:flat_expnsn} and
\refeq{DL:hypergeom} involved. Since we only care about the variances, we
define instead the following map
\beq
\covMat_{a-1} = \left\{
\begin{array}{ll}
C\left(\frac{\covMat_a+\diffTen}{\DDf{a-1}{}^2}\right)^{1/2}
& |f^{'2}_{a-1}<1| \\
\frac{\covMat_a+\diffTen}{f^{'2}_{a-1}} & \mbox{otherwise}
\end{array}
\right.
\,,
\ee{DL:var_map}
$C = 2\sqrt{2}\Gamma(3/4)/\Gamma(1/4)$, for the evolution of the
densities, and take its periodic points as our new neighborhoods. In
practice, one can compute these numerically, but we will not need orbits
longer than length $\cl{p}=4$ in our tests of the partition, therefore we
can safely assume only one periodic point of $f(x)$ to be close to the
flat top, and obtain analytic expressions for the periodic points of
\refeq{DL:var_map}:
\beq
\tilde{\covMat}_a \simeq
C\left(\frac{\diffTen\left(1+f^{'2}_{a-1}+...+(f_{a-n+1}^{n-1'})^2
\right)}{\tilde{\ExpaEig}_p}^2\right)^{1/2}
\ee{DL:prd_var1}
with $\tilde{\ExpaEig}_p=f_{a-n+1}^{n-1'}\DDf{a-1}{}^2 $, is valid when
the cycle starts and ends at a point $x_a$ close to the flat top.
Otherwise, take the periodic point $x_{a-k}$, that is the $k-$th preimage
of the point $x_a$. The corresponding periodic point variance has the
form
\beq
\tilde{\covMat}_{a-k} \simeq \frac{1}{(f_{a-1}^{k'})^2}\left(\diffTen(1+f_{a-1}^{'2}+...+
(f_{a-1}^{k-1'})^2) + \tilde{\covMat}_a \right)
\ee{DL:prd_var2}
both expressions \refeq{DL:prd_var1} and \refeq{DL:prd_var2} are
approximate, as we further assumed $\diffTen\tilde{\ExpaEig}_p^2 \gg 1$,
which is reasonable when $\diffTen\in [10^{-4},10^{-2}]$, our range of
investigation for the noise strength. As before, a neighborhood of width
$[x_a-\sqrt{\tilde{\covMat}}_a,x_a+\sqrt{\tilde{\covMat}}_a]$
is assigned to each periodic point $x_a$, and an \optPart\ follows.
However, due to the geometry of the map, such partitions as
{\small
\beq
 \{\pS_{000},\left[\pS_{001},\pS_{011}\right],\pS_{010},
       \pS_{110},\pS_{111},\pS_{10}\}
\ee{DL:conflict}
}
can occur. In this example the regions $\pS_{001}$ and $\pS_{011}$ overlap,
and the partition results in a transition graph with three loops (cycles)
of length one, while we know that our map only admits two fixed points.
In this case we decide to follow  the
deter\-mi\-ni\-stic symbolic dynamics and ignore the overlap.

Let us now test the method by estimating once again the escape rate of
the noisy map \refeq{DL:Ulam_rep}.
We note that the matrix elements
\bea
[\Lmat{}_{ba}]_{kj} &=&
\left\langle \tilde{\msr}_{b,k} | \Lnoise{} |{\msr}_{a,j} \right\rangle
    \continue
    &=& \int \frac{d\orbitDist_b d\orbitDist_a \,\beta}
              {2^{j} j! \pi\sqrt{2\diffTen}} \;
    e^{-(\beta \orbitDist_b)^2-\frac{(\orbitDist_b-f'_a\orbitDist_a)^2}{2\diffTen}}
    \ceq \qquad \times\,
    H_k(\beta \orbitDist_b) \, H_j(\beta \orbitDist_a)
\,,
\label{DL:mtx_elemRpt}
\eea
$1/\beta=\sqrt{2\covMat_{a}}$, should be redefined in the neighborhood of
the critical point of the map, where the Gaussian approximation to
$\Lnoise{}$ fails. We follow the approximation made in
\refeq{DL:flat_expnsn}:
\bea
[\Lmat{}_{ba}]_{kj} =
     \int \frac{d\orbitDist_b d\orbitDist_a \,\beta}
              {2^{j} j! \pi\sqrt{2\diffTen}}
    e^{-(\beta \orbitDist_b)^2-\frac{(\orbitDist_b-\DDf{a}
	\sqrt{2\diffTen}\orbitDist_a^2/2)^2}{2\diffTen}}
    H_k(\beta \orbitDist_b) H_j(\beta \orbitDist_a)
\,,
\label{DL:quad_elem}
\eea
However, as $\diffTen$ decreases, it also reduces the quadratic term in
the expansion of the exponential, so that the linear term
$f'_a\orbitDist_a$ must now be included in the matrix element:
\bea
[\Lmat{}_{ba}]_{kj} =
     \int \frac{d\orbitDist_b d\orbitDist_a \,\beta}
              {2^{j} j! \pi\sqrt{2\diffTen}}
    e^{-(\beta \orbitDist_b)^2-\frac{(\orbitDist_b- f'_a\orbitDist_a -\DDf{a}
        \sqrt{2\diffTen}\orbitDist_a^2/2)^2}{2\diffTen}}
    H_k(\beta \orbitDist_b) H_j(\beta \orbitDist_a)
\,,
\label{DL:cub_elem}
\eea

%%%%%%%%%%%%%%%%%%%%%%%%%%%%%%%%%%%%%%%%%%%%%%%%%%%%%%%%%%%%%%%%%%
\begin{figure}[tbp]
%\centering
(a) \includegraphics[width=0.48\textwidth,angle=0]{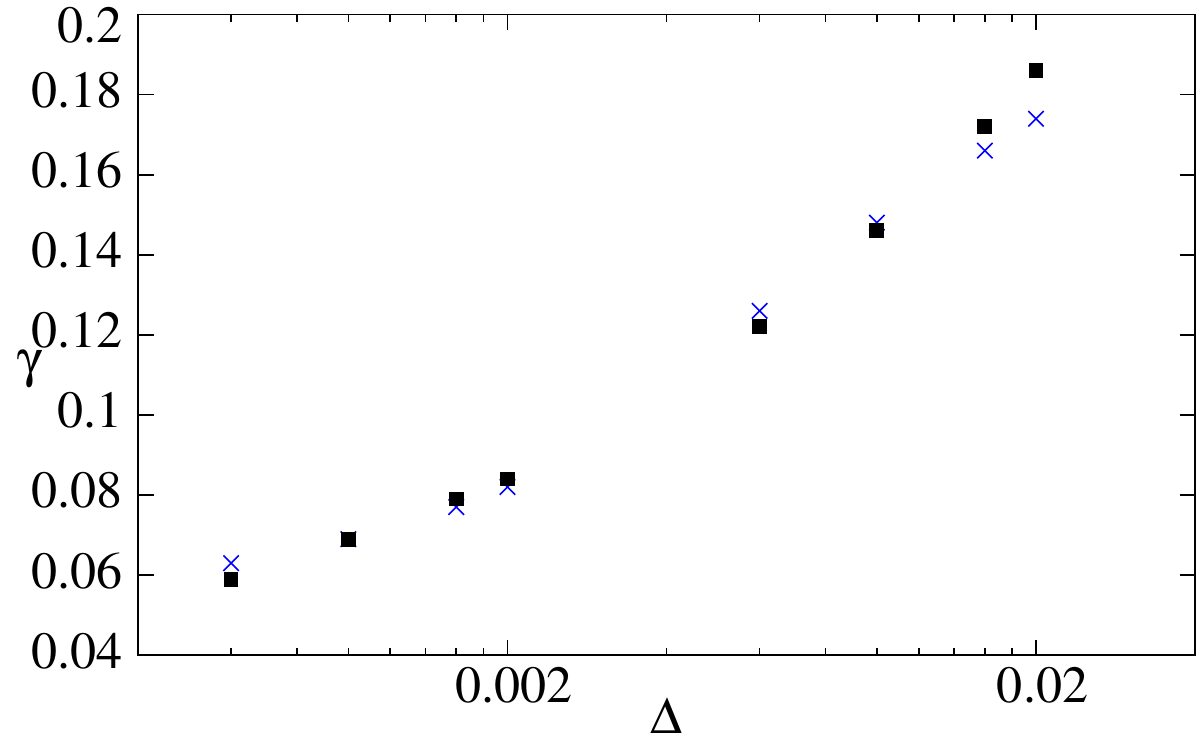}%
(b) \includegraphics[width=0.48\textwidth,angle=0]{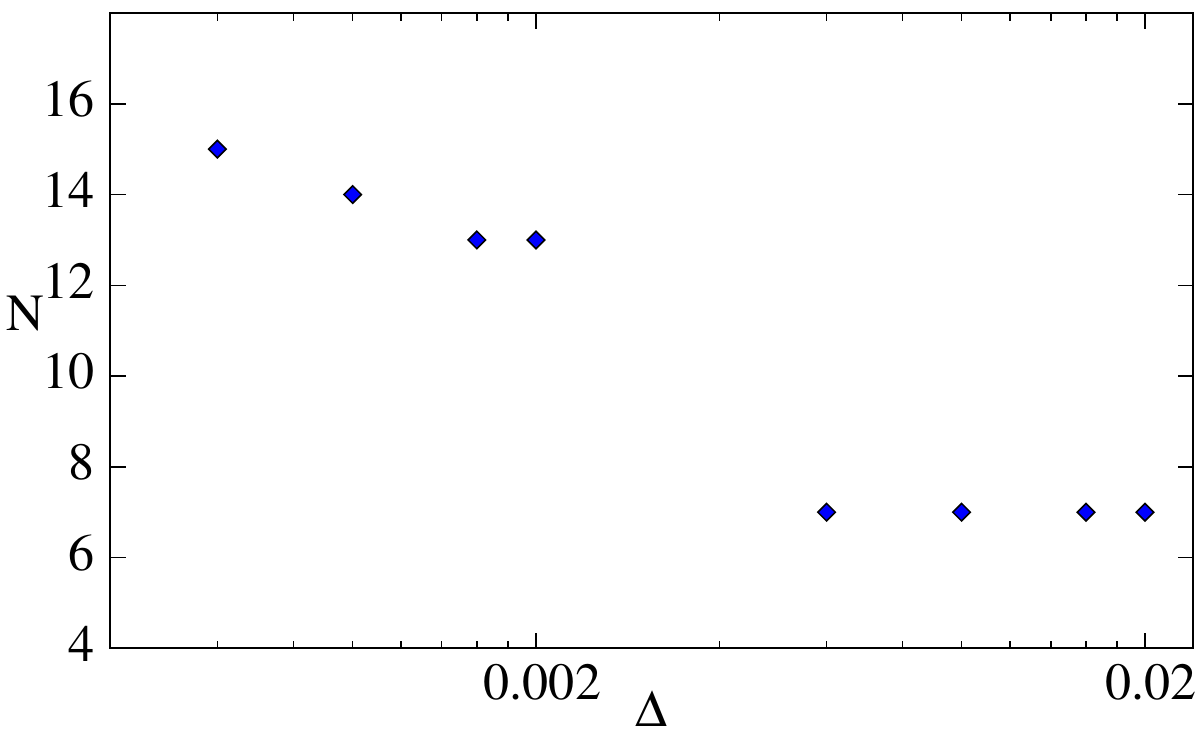}%
\caption{
(a) Escape rate $\gamma$ of the `skew Ulam' map vs. noise strength $\diffTen$, using:
($\color{blue}\times$)
the \optPart\ method;
({%\scriptsize
$\blacksquare$})
 a uniform discretization \refeq{DL:overl-intFP} in $N=128$ intervals.
(b) Number of neighborhoods required by the optimal partition method vs. the
noise strength $\diffTen$.}
\label{f:flat_gammas}
\end{figure}
%%%%%%%%%%%%%%%%%%%%%%%%%%%%%%%%%%%%%%%%%%%%%%%%%%%%%%%%%%%%%%%%%%

We find in our model that the \po s we use in our expansion have $x_a$'s
within the flat top, such that $f'_a\sim 10^{-1}$ and $\DDf{a} \sim 10$,
and therefore \refeq{DL:quad_elem} better be replaced with
\refeq{DL:cub_elem} when $\diffTen\sim 10^{-4}$. In order to know whether
a cycle point is close enough to the flat top for the Gaussian
approximation to fail, we recall that the matrix element
\refeq{DL:mtx_elemRpt} is the zeroth-order term of a series in $\diffTen$,
whose convergence can be probed by evaluating the higher order
corrections \refeq{DL:appx_mtx}: when the $O(\sqrt{\diffTen})$ and
$O(\diffTen)$ corrections are of an order of magnitude comparable or
bigger than the one of \refeq{DL:mtx_elemRpt}, we conclude that the Gaussian
approximation fails and we use \refeq{DL:quad_elem} or
\refeq{DL:cub_elem} instead. Everywhere else we use our usual matrix
elements \refeq{DL:mtx_elemRpt}, \textit{without} the higher-order
corrections, as they are significantly larger than in the case of the
repeller, and they are not accounted for by the \optPart\ method, which
is entirely based on a zeroth-order Gaussian approximation of the
{\evOper}. %  (see \refsect{DL:oneEigs}).
Like before, we tweak the noise
strength $\diffTen$ within the range $[10^{-4},10^{-2}]$ and compare the
escape rate evaluated with the \optPart\ method and with the uniform
discretization \refeq{DL:overl-intFP}. The results are illustrated in
\reffig{f:flat_gammas}: the uniform discretization  and the method
of the \optPart\ are consistent within a $5\%$ margin.

\section{Summary and conclusions}
\label{DL:concl}
% Predrag                   May 10 2012
% Predrag                   May  6 2012
% taken from stoch.tex \section{Summary and conclusions}
% Predrag                   Jul  4 2010
% Predrag                   Sep  6 2006

Physicists tend to believe that with time Brownian motion leads to
$x(t)^2 \approx \diffTen\,t$ broadening of a noisy trajectory
neighborhood. In nonlinear dynamics nothing of the sort happens; the
noise broadening is balanced by non-linear stretching and contraction,
and infinite length recurrent Langevin trajectories have finite noise
widths, not widths that spread $\propto$ time. Here  \po s play special
role: computable in finite time, they persist for for infinite time, and
are thus natural objects to organize \statesp\ partitions around. On the
other hand, computation of unstable \po s in high-dimensional \statesp s,
such as Navier-Stokes, is at the border of what is currently feasible
numerically\rf{KawKida01,CviGib10}, and criteria to identify finite sets
of the most important solutions are very much needed. Where are we to
stop calculating orbits of a given hyperbolic flow? Intuitively, as we
look at longer and longer \po s, their deterministic neighborhoods shrink
exponentially with time, while the variance of the noise-induced orbit
smearing remains bounded; there has to be a {\em turnover time}, a time
at which the noise-induced width overwhelms the exponentially shrinking
deter\-mi\-ni\-stic dynamics, so that no better resolution is possible.
Given a specified noise, we need to find, {\po} by \po, whether a further
sub-partitioning is possible.

We have described here the {\em {\optPart} hypothesis}, a method for
partitioning the \statesp\ of a chaotic repeller in presence of weak
Gaussian noise first introduced in \refref{LipCvi08}.
The key idea is that the width of the linearized adjoint \Fokker\
operator $\Lnoise{\dagger}{}$ eigenfunction computed on an unstable
periodic point $x_a$ provides the scale beyond which no further local
refinement of \statesp\ is feasible. This computation enables us to
systematically determine the {\optPart}, the finest \statesp\ resolution
attainable for a given chaotic dynamical system and a given noise. Once
the \optPart\ is determined, we use the associated transition graph to
describe the stochastic dynamics by a {\em finite dimensional} \Fokker\
matrix. An expansion of the \Fokker\ oper\-ator about periodic points was
already introduced in \refrefs{noisy_Fred,conjug_Fred,diag_Fred}, with
the stochastic trace formulas and deter\-min\-ants\rf{noisy_Fred,gasp02}
expressed as finite sums, truncated  at orbit periods corresponding to
the local turnover times. A novel aspect of the work presented here is its
representation in terms of the Hermite basis (\refsect{DL:locEig}),
eigenfunctions of the linearized \Fokker\ oper\-ator~\refeq{DL:app_evol},
and the finite dimensional matrix representation of the \Fokker\
oper\-ator.

It should be noted that our linearization of \Fokker\ operators does not
imply that the nonlinear dynamics is being modeled by a linear one. Our
description is fully nonlinear, with \po s providing the nonlinear
backbone of chaotic dynamics, dressed up stochastically by \Fokker\
operators local to each cycle. This is a stochastic cousin of
Gutzwiller's WKB approximation based semi-classical
quantization\rf{gutzwiller71} of classically chaotic systems, where in a
parallel effort to utilize quantum-mechanical $\hbar$ `graininess' of the
quantum phase space to terminate {\po} sums, Berry and Keating\rf{BK90}
have proposed inclusion of cycles of periods up to a single `Heisenberg
time'. In light of the stochastic dynamics insights gained here, this
proposal merits a reexamination - each neighborhood is likely to have its
own Heisenberg time.

% Predrag       sep 26 2005
% moved from \file{stoch.tex sep 26 2005}
% Domenico      sep 26 2005
%
% \subsubsection{Determinism {\em vs.} noise in chaotic dynamics}
%

%\PC{{\bf 2010-01-12 PC text for Lippolis thesis, ignored:}}
%
The work of Abarbanel \etal\rf{AbBrKe91,AbBrKe91a,ACFK09,QuAb09}
suggest one type of important application
beyond the low-dimensional {\Fokker} calculations
undertaken here.
    \PC{cite also (read siminos/blog 2008-07-02 entry) Ott
    talk about use of Kalman filters by the Maryland group
    weather prediction work.
    Explain Trevisan reference, just in case we need to do
    'turbulence prediction' in pipe experiments
    }
In data assimilation in weather prediction the convolution of noise
variance and trajectory variance \refeq{AddVariances} is a step in the
Kalman filter procedure. One could combine the \statesp\ charts of
turbulent flows of \refref{GHCW07} (computed in the full 3-dimensional
Navier-Stokes) with partial information obtained in experiments
(typically a full 3-dimensional velocity field, fully resolved in time,
but measured only on a 2-dimensional disk section across the pipe). The
challenge is to match this measurement of the turbulent flow with a
{\statesp} point in a $\approx 10^5$-dimensional ODE representation, and then
track the experimental observation to improve our theoretical prediction
for the trajectory in the time ahead. That would be the absolutely best
`weather prediction' attainable for a turbulent pipe flow, limited by a
combination of Lyapunov time and observational noise. In our parlance,
the `\optPart\ of {\statesp}.'

We have tested our \optPart\ hypothesis by applying it to evaluation of
the escape rates and the Lyapunov exponents of a $1d$ repeller in
presence of additive noise. In the $1d$ setting numerical tests indicate
that the `\optPart' method can be as accurate as the much finer grained
direct numerical \Fokker\ oper\-ator calculations. In higher
dimensions (and especially in the extreme high dimensions required by
fluid dynamics stimulations) such direct \Fokker\ PDE integrations are
not feasible, while the method proposed here is currently the only
implementable approach.

The success of the \optPart\ hypothesis in a one-dimensional setting is
encouraging, and use of noise as a smoothing device that eliminates
singularities and pathologies from clusterings of orbits is promising.
However, higher-dimensional hyperbolic maps and flows, for which an
effective \optPart\ algorithm would be very useful, present new, as yet
unexplored challenges of disentangling the subtle interactions between
expanding, marginal and contracting directions; the method has not yet
been tested in a high-dimensional hyperbolic setting.
%
%   \PC{incorporated  {\bf Reviewer 5}:
%
A limiting factor to applications of the periodic
orbit theory to high-dimensional problems ranging from fluid
flows to chemical reactions
might be the lack of a
good understanding of periodic orbits in more than three
dimensions, of their stability properties, their organization
and their impact on the dynamics.

In summary: Each periodic point owns a cigar, which for a
high-dimensional dissipative flow is shaped along a handful of expanding
and least-contracting directions. The remaining large (even infinite!)
number of the strongly contracting directions is limited by the noise;
the cigar always has the dimensionality of the full \statesp. Taken
together, the set of overlapping cigars, or the {\optPart}, weaves the
carpet (of the full dimensionality of \statesp) which envelops the entire
`inertial manifold' explored by turbulent dynamics.
    \PC{2012-05-11 Once the noisy \statesp\ partition reaches its limit,
          it looks like the current Baron Cohen's beard.}

%%%%%%%%%%%%%%%%%%%%%%%%%%%%%%%%%%%%%%%%%%%%%%%%%%%%%%%%%%%%%%%%%%%%%
\newpage %\bigskip

\noindent{\bf Acknowledgments.}

We are grateful to
D.~Barkley,
T.~Bartsch,
C.P.~Dettmann,
H.~Fogedby,
A.~Grigo,
A.~Jackson,
R.~Mainieri,
W.H.~Mather,
R.~Metzler,
E.~Ott,
S.A.~Solla,
N.~S\o ndergaard,
and
G.~Vattay
for many stimulating discussions.
Special thanks go to
M.~Robnik
who made it possible to present this work at the 8th Maribor Summer
School `Let's Face Chaos through Nonlinear Dynamics', which brought us
together with
T.~Prosen,
from whom we have learned that fixed points of the covariance evolution
equations are the Lyapunov equations.
P.C. thanks Glen P.~Robinson,~Jr. for support.
D.L. was supported by NSF grant DMS-0807574 and G.P.~Robinson,~Jr..

\appendix

  % lippolis/Maribor/POT.tex    pdflatex CviLip12
% $Author: predrag $
% $Date: 2012-05-13 11:45:07 -0400 (Sun, 13 May 2012) $

                        \renewcommand{\version}{
  Predrag   Apr 23 2012
                        }
%   Predrag  taken from stoch.tex  Jul  4 2010
%   based on dasbuch/book/chapters/maps.tex

  \section{Periodic orbit theory, deterministic dynamics}
  \label{DL:POT}

We offer here a brief review of deterministic dynamics  and {\po} theory.
All of this is standard, but needed to set the notation used above. The
reader might want to consult \refref{DasBuch} for further details.

\begin{figure}[tbp]
\begin{minipage}[t]{0.90\columnwidth}%
(a)~\includegraphics[width=0.48\textwidth]{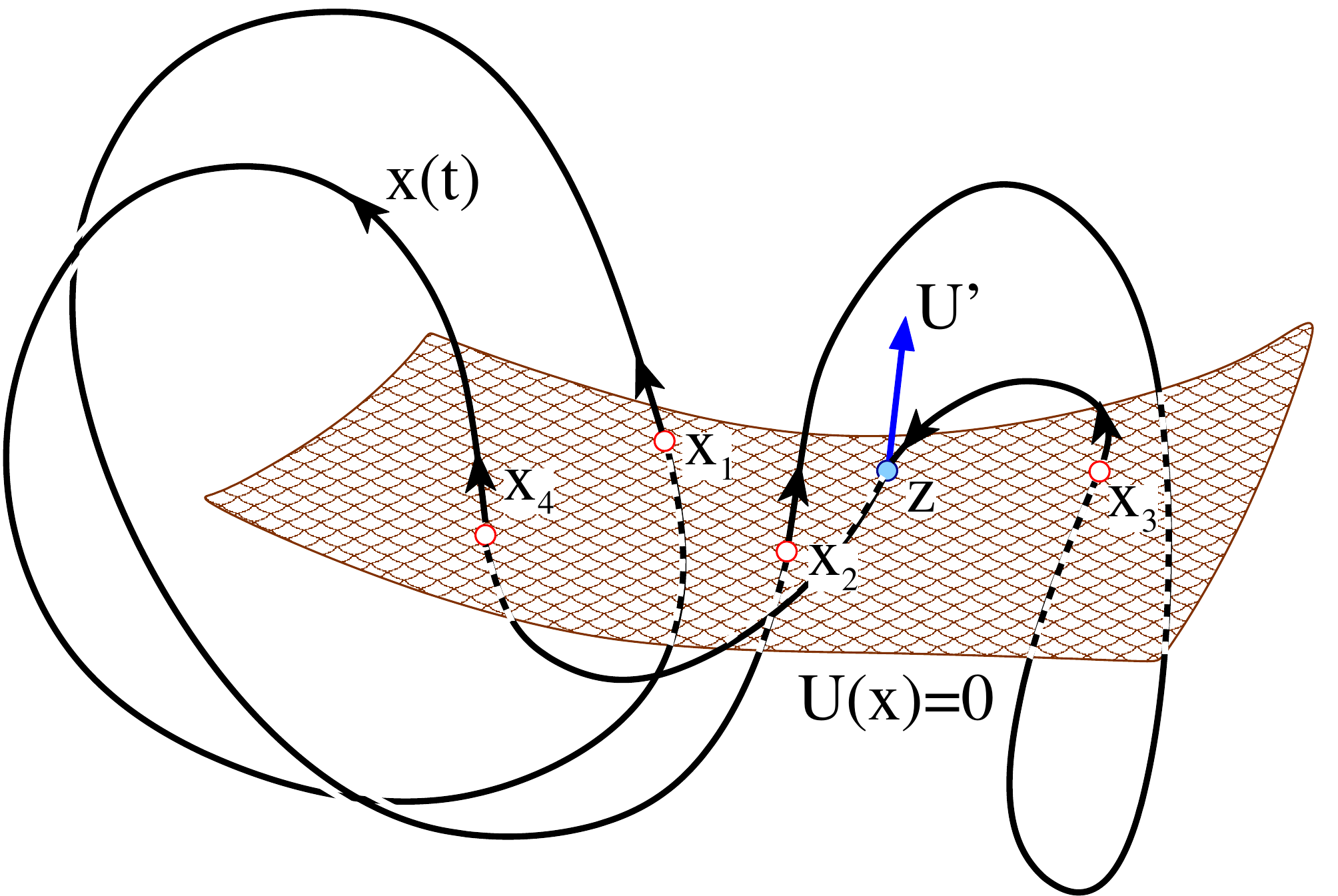}
~~~
(b)~\includegraphics[width=0.48\textwidth]{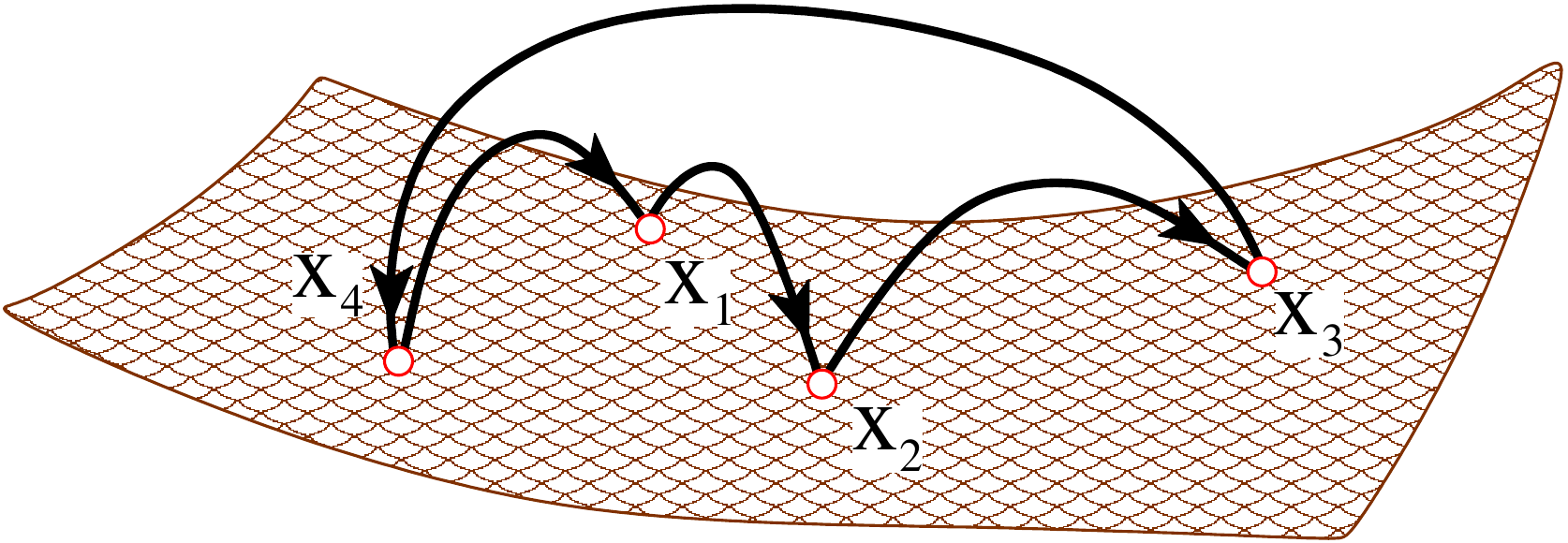}
\end{minipage}
\source{ChaosBook.org}
\caption{
(a) A Poincar\'e hypersurface $\PoincS$, defined by a condition
$U(\ssp)=0$, is intersected by the $\ssp(t)$ orbit
at times $t_1, t_2, t_3, t_4$, and
closes a cycle \( (\ssp_1,\ssp_2,\ssp_3,\ssp_4) \), \( \ssp_k =
\ssp(t_k) \in \PoincS \), of topological length 4 with respect
to this section. The crossing $z$ does not count, as it in
the wrong direction.
(b) The same orbit reduced to a
Poincar\'e return map that maps points in the Poincar\'e
section $\PoincS$ as \( \ssp_{n+1} = \map(\ssp_n) \,. \) In
this example the orbit of \( \ssp_1 \) is periodic
and consists of the four
periodic points \( (\ssp_1,\ssp_2,\ssp_3,\ssp_4) \).
% (from \wwwcb{}).
}
\label{f:PoincSct}
\end{figure}

Though the main applications we have in mind are to continuous flows, for
purposes at hand it will suffice to consider discrete time dynamics
obtained, for example, by reducing a continuous flow to mappings between
successive Poincar\'e sections, as in \reffig{f:PoincSct}. Consider
dynamics induced by iterations of a $d$-dimensional map
\(
\map: \pS \to \pS \,,
\)
where $\pS \subset \reals^d$ is the \statesp\ (or `phase space') of the
system under consideration. The discrete `time' is then an integer, the
number of applications of a map. We denote the $k$th iterate of map
$\map$ by composition
    \PC{remember to raise \reffig{f:PoincSct}\,(b)}
\beq
\flow{k}{\ssp} = \map \left(\flow{k-1}{\ssp} \right)
\,,\qquad
\flow{0}{\ssp} = \ssp
\,.
\ee{DL:det_map}
The {\em trajectory} of $\ssp = \ssp_{0}$
% (defined in \refsect{s_dyn_flows})
is the finite set of points  $\ssp_j = \flow{j}{\ssp}$,
\beq
\{\ssp_{0},\ssp_1,\ssp_2,\ldots,\ssp_k\} =
\left\{\ssp, \flow{}{\ssp}, \flow{2}{\ssp}, \ldots, \flow{k}{\ssp} \right\}
\,,
\ee{DL:determTraj}
traversed in time $k$, and the {\em orbit} of $\ssp$ is the subset
$\pS_x$ of all points of $\pS$ that can be reached by iterations of
$\map$. Here $\ssp_k$ is a point in the $d$-dimensional \statesp\  \pS,
and the subscript $k$ indicates time. While a trajectory depends on the
initial point $\ssp$, an orbit is a set invariant under dynamics. The
transformation of an infinitesimal neighborhood of an orbit point $\ssp$
under the iteration of a map follows from Taylor expanding the iterated
mapping at {finite} time $k$. The linearized neighborhood is transported
by the $[d\!\times\!d]$ {\jacobianM}
\beq
\monodromy^k_{ij}(\xInit)
  =  \left. { \pde \map^k_i(\ssp) \over \pde \ssp_j} \right|_{\ssp=\xInit}
\,. \label{hOdesMap}
\eeq
($J(\ssp)$ for Jacobian, or derivative notation $\monodromy(\ssp) \to D
\flow{}{\ssp}$ is frequently employed in the literature.) The formula for
the linearization of $k$th iterate
\beq
    \monodromy^k(\xInit) =
     % \prod_{m=n-1}^0  \monodromy(\ssp_m)
     \monodromy(\ssp_{k-1})
     \cdots
     \monodromy(\ssp_{1})
     \monodromy(\xInit)
     \,, \quad
    \monodromy_{ij} = \partial f_i / \partial \ssp_j
    \,,
\ee{jacoB1}
in terms of unit time steps
$\monodromy$
follows from the chain rule for functional composition,
\bea
\frac{\pde ~}{\pde \ssp_i} f_j(f(\xInit))
   &=& \sum_{k=1}^d \left. \frac{\pde ~}{\pde \ssp_k}  f_j(y)\right|_{y=f(\xInit)}
     \frac{\pde ~}{\pde \ssp_i}  f_k(\xInit)
     \continue
\monodromy^2(\xInit) &=&
     \monodromy(\ssp_{1})
     \monodromy(\xInit)
\,.
\nnu
\eea
We denote by $\ExpaEig_\ell$ the $\ell$th {\em eigen\-value} or {\em
multiplier} of the {\jacobianM} $\monodromy^k(\xInit)$, and by
$\eigExp[\ell]$ the $\ell$th {\em Floquet} or {\em characteristic}
exponent, with real part $\eigRe[\ell]$ and phase $\eigIm[\ell]$:
\beq
\ExpaEig_\ell
 =
e^{k \eigExp[\ell]}
 =
e^{k (\eigRe[\ell] +i \eigIm[\ell])}
\,.
\ee{stabExpon}
\JacobianM\ $\monodromy^{k}(\xInit)$ and its eigenvalues (Floquet
multipliers) depend on the initial point $\xInit$ and the elapsed time
$k$. For notational brevity we tend to omit this dependence, but in
general
\[
\ExpaEig = \ExpaEig_\ell = \ExpaEig_\ell(\xInit,k)
    \,,\;
\eigExp = \eigExp[\ell](\xInit,k)
    \,,\;
\eigIm = \eigIm[\ell](\xInit,k)
\,,\cdots\mbox{ \etc}
\,,
\]
depend on the traversed trajectory.

%
%%%%%%%%%%%%%%%%%%%%%%%%%%%%%%%%%%%%%%%%%%%%%%%%%%%%%%%%%%%%%%%
% dasbuch \SFIG{PoincSct} {f:PoincSct}
%         \SFIG{twoDmap}  {f:twoDmap}
%%%%%%%%%%%%%%%%%%%%%%%%%%%%%%%%%%%%%%%%%%%%%%%%%%%%%%%%%%%%%%%%

A {\em periodic point} ({\em cycle point}) $\ssp_k$ belonging
to a {\em {\po}} ({\em cycle}) of period $\cl{}$ is a real
solution of
\beq
   \flow{\cl{}}{\ssp_k} =  \map(\map( \ldots \map(\ssp_k) \ldots )) = \ssp_k
        \,,\quad k=0,1,2,\dots,\cl{}-1
\,.
      \label{e:perOrbDef}
\eeq
For example, the orbit of
\(
\ssp_1
\)
in  \reffig{f:PoincSct}
is the 4-cycle
\(
(\ssp_1,\ssp_2,\ssp_3,\ssp_4)
\,.
\)
The time-dependent $\cl{}$-periodic vector fields, such as the flow
linearized around a {\po}, are described by Floquet theory. Hence we
shall refer to the {\jacobianM}
\(
\monodromy_p(\ssp) = \monodromy^\cl{}(\ssp)
%\,,
\)
evaluated on a {\po} $p$ as the {\em monodromy} or {\em Floquet} matrix,
and to its eigenvalues $\{\ExpaEig_{p,1}$, $\ExpaEig_{p,2}$, $\dots$,
$\ExpaEig_{p,d}\}$ as Floquet multipliers. They are flow-invariant,
independent of the choice of coordinates and the initial point in the
cycle $p$, so we label them by their $p$ label. We number the
eigen\-values in order of decreasing magnitude
\(
| \ExpaEig_1 | \geq | \ExpaEig_2 |
    \geq \ldots \geq | \ExpaEig_d |
\,,
\) %ee{LambdaOrd}
sort them into sets $\{e,m,c\}$
\bea
\mbox{expanding:}
        &\quad \{ \ExpaEig \}_e &
        =\, \{ \ExpaEig_{p,j}: \left|\ExpaEig_{p,j}\right| > 1 \}
                \continue
\mbox{marginal:}
        &\quad \{ \ExpaEig \}_m &
        =\, \{ \ExpaEig_{p,j}: \left|\ExpaEig_{p,j}\right| = 1 \}
                \label{EigSorted}\\
\mbox{contracting:}
        &\quad \{ \ExpaEig \}_c &
        =\, \{ \ExpaEig_{p,j}: \left|\ExpaEig_{p,j}\right| < 1 \}
\,,
\nnu
\eea
and denote by $\ExpaEig_{p}$ (no $j$th eigenvalue index)
the product of {\em expanding} Floquet multipliers
\beq
\ExpaEig_p=\prod_e \ExpaEig_{p,e}
\,.
\ee{expVol}

The stretching/contraction rates per unit time
are given by the real parts of {Floquet exponents}
\beq
\eigRe[i]_p = {1 \over \cl{p}} \ln \left|\ExpaEig_{p,i}\right|
\,.
\ee{stabExps}
They can be loosely interpreted as Lyapunov exponents evaluated on
the prime cycle $p$.

A {\po} $p$ is {\em stable} if real parts of all of its
Floquet exponents are strictly negative, $\eigRe[i]_p<0$. If
{all} Floquet exponents are strictly positive, $ \eigRe[i]
\geq \eigRe_{min} > 0 $, the {\po} is {\em repelling}, and
unstable to any perturbation. If some are strictly positive,
and rest strictly negative, the {\po} is said to be {\em
hyperbolic} or a {\em saddle}, and unstable to perturbations
outside its stable manifold. Repelling and hyperbolic \po s
are unstable to generic perturbations, and thus said to be
{\em unstable}. If all $\eigRe[i]=0$, the orbit is said to be
{\em elliptic}, and if $\eigRe[i]=0$ for a subset of
exponents, the orbit is said to be {\em partially
hyperbolic}. If {\em all} Floquet exponents (other than the
vanishing longitudinal exponent) of {\em all} \po s of a flow
are strictly bounded away from zero, the flow is said to be
{\em hyperbolic}. Otherwise the flow is said to be {\em
nonhyperbolic}.

%\subsection{Singular value decomposition}
% [2012-05-09 Predrag] - return edits from here back to ChaosBook.org
%\cyclist
The \jacobianM\ $\monodromy$ is in general non-normal: it neither
symmetric, nor diagonalizable by a rotation, nor do its (left or right)
eigen\-vectors define an orthonormal coordinate frame (for brevity we
omit in what follows the time superscript, $\monodromy^k \to \monodromy$). As any
matrix with real elements, $\monodromy$ can be expressed in the singular
value decomposition form
\beq
\monodromy = {U} {D}  {V}^T
\,,
\ee{SVD-j}
where ${D}$ is diagonal and real, and ${U}$, ${V}$ are orthogonal
matrices. The diagonal elements $\sigma_{1}$, $\sigma_{2}$, $\dots$,
$\sigma_{d}$ of ${D}$ are called the \emph{singular values} of
$\monodromy$, namely the square root of the eigenvalues of
$\monodromy^{T}\monodromy = {V}{D}^{2}{V}^T$ (or
$\monodromy\monodromy^{T} = {U}{D}^{2}{U}^T$), which is a symmetric,
positive semi-definite matrix (and thus admits only real, non-negative
eigenvalues).

Singular values $\{\sigma_{j}\}$ are \emph{not related} to the
$\monodromy$ eigenvalues $\{\ExpaEig_{j}\}$ in any simple way. From a
geometric point of view, when all singular values are non-zero,
$\monodromy$ maps the unit sphere into an ellipsoid, \reffig{f:repPart}\,(b):
the singular values are then the lengths of the semiaxes of this
ellipsoid. Note however that the singular vectors of
$\monodromy^{T}\monodromy$ that determine the orientation of the semiaxes
are distinct from the $\monodromy$ eigenvectors $\{\jEigvec[j]\}$, and
that $\monodromy^{T}\monodromy$ satisfies no semigroup property along the
flow.  For this reason the $\monodromy$ eigenvectors $\{\jEigvec[j]\}$
are sometimes called `covariant' or `covariant Lyapunov vectors', in
order to emphasize the distinction between them and the singular value
decomposition semiaxes directions.
    \PC{add inertia references}

Eigenvectors / eigenvalues are suited to study of iterated forms of a
matrix, such as $\monodromy^k$ or exponentials $\exp(t \Mvar)$, and are
thus a natural tool for study of dynamics. Singular vectors are  not.
They are suited to study of $\monodromy$ itself, and the singular value
decomposition is convenient for numerical work (any matrix, square or
rectangular, can be brought to this form), as a way of estimating the
effective rank of matrix $\monodromy$ by neglecting the small singular
values.

\subsection{Deterministic \statesp\ partitions}
\label{DL:detPart}

We streamline the notation by introducing local coordinate
systems $\orbitDist_a$ centered on the trajectory points
$\ssp_a$, together with a trajectory-centered notation for
the map \refeq{DL:det_map}, its derivative, and, by the chain
rule, the derivative \refeq{jacoB1} of the $k$th iterate
$f^k$ evaluated at the point $\ssp_a$,
\bea
\ssp &=& \ssp_a+\orbitDist_a
    \,, \quad
\map_a(\orbitDist_a) = \map(\ssp_a+\orbitDist_a)
	\continue
\orbitDist_{a+1} &=& \monodromy_{a} \orbitDist_{a} + \cdots
    \label{DL:orbitDerv}\\
\monodromy_{a} &=& f'(\ssp_a)
    \,, \;\;
\monodromy_a^k{} =  \monodromy_{a+k-1} \cdots \monodromy_{a+1}\monodromy_{a}
    \,, \;\;
k \geq 2
\,.
    \nnu
\eea
The monodromy (or Floquet) matrix,
\beq
\monodromy_{p,a} = \monodromy^\cl{p}(\ssp_a)
\,,
\ee{perPointM}
evaluated on the periodic point $\ssp_a$ is position dependent, but its
eigenvalues, the Floquet multipliers $\{\ExpaEig_{p,1}$,
$\ExpaEig_{p,2}$, $\dots$, $\ExpaEig_{p,d}\}$ are invariant, intrinsic to
the \po.

For example, if $\map$ is a $1$-dimensional map, its Taylor expansion
about $\ssp_a = \ssp - \orbitDist_a$ is
\beq
\flow{}{\ssp} = \ssp_{a+1} + \Df{a} \orbitDist_a
                + \frac{1}{2} \DDf{a} \orbitDist_a^2 + \cdots
\,,
\ee{TaylExp}
where
\bea
\Df{a} &=& \map'(\ssp_a)
    \,,\qquad
\DDf{a} \,=\, \map''(\ssp_a)
    \continue
\map_a^k{}' &=& \Df{a+k-1} \cdots \Df{a+1}\Df{a}
    \,, \;\;
k \geq 2
\,.
\label{DL:orbitDerv1d}
\eea
A cycle point for which $\Df{a}=0$ is called a critical point.

%
%%%%%%%%%%%%%%%%%%%%%%%%%%%%%%%%%%%%%%%%%%%%%%%%%%%%%%%%%%%%%%%
% from \SFIG{cyclJac}   {f:cyclJac}
\begin{figure}[tbp]
\begin{minipage}[t]{0.90\columnwidth}%
(a)~~\includegraphics[width=0.40\textwidth]{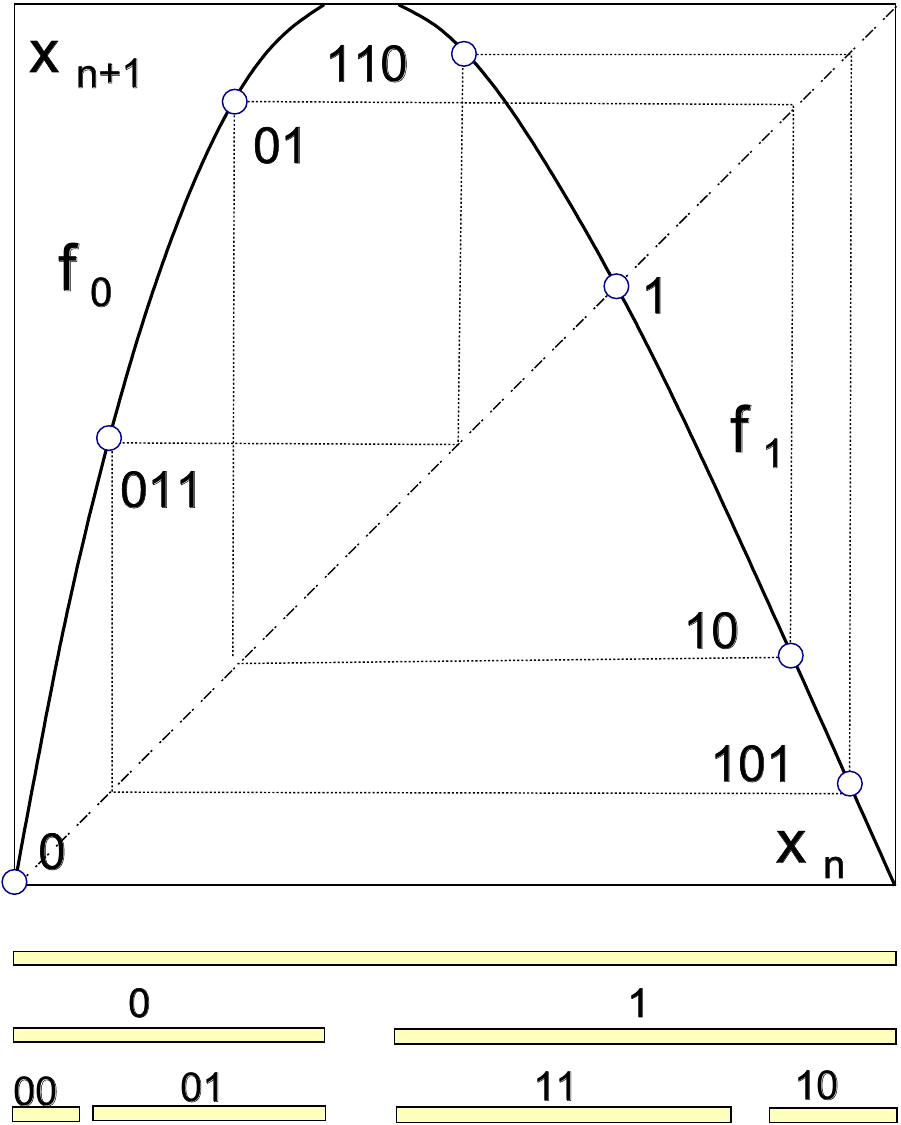}
~~~~~~~(b)~~\includegraphics[width=0.40\textwidth]{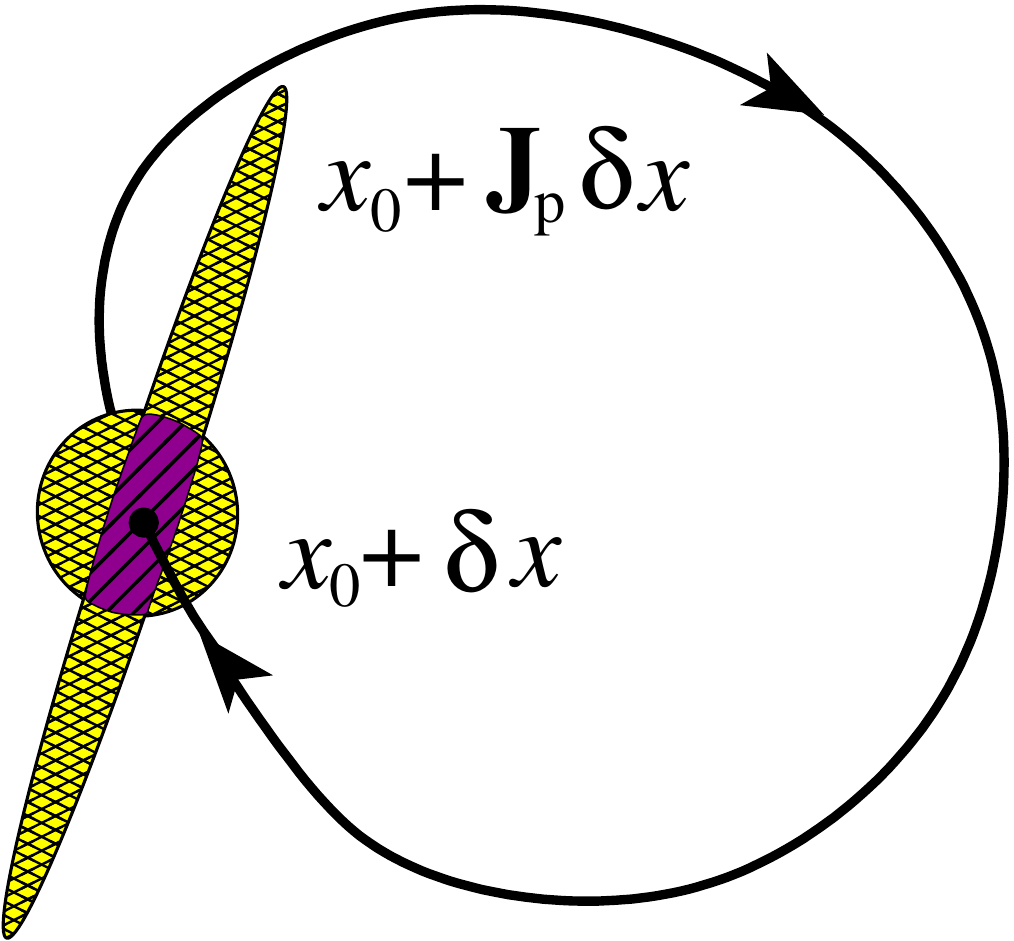}
\end{minipage}
\source{ChaosBook.org}
\caption{
(a) Periodic points of the map $f(x) = 6 \, x \, (1-x)(1-0.6 \, x)$
labeled according to the partition $\pS = \{\pS_0,\pS_1\}$.
(b)
For a prime cycle $p$, \FloquetM\ $\monodromy_{p}$ returns an
infinitesimal spherical neighborhood of $\xInit \in \pS_p$
stretched into a cigar-shaped ellipsoid, with principal axes given by the
eigen-directions $\jEigvec[i]$ of $\monodromy_p$, the  {\monodromyM}
\refeq{perPointM}.
}
\label{f:repPart}
\end{figure}

We label trajectory points by either $\ssp_n$, $n=1,2,\cdots$, in order
to emphasize as time evolution of $\xInit$, or by $\ssp_a$, to emphasize
that the trajectory point lies in the \statesp\ region labeled `$a$.'
Then the label $a\!+\!1 $ is a shorthand for the next \statesp\ region
$b$ on the orbit of $\ssp_a$, $\ssp_b=\ssp_{a+1}=\map(\ssp_a)$. For
example, in \reffig{f:repPart}\,(a) a periodic point is labeled $a=011$
by the itinerary with which it visits the regions of the partitioned
\statesp\
\(
\pS = \{\pS_0,\pS_1\}
\,,
\)
and as $\ssp_{110}=\map(\ssp_{011})$, the next point label is $b=110$.
The whole {\po}
\(
\cycle{011} =
(\ssp_{011},\ssp_{110},\ssp_{101})
\)
is traversed in 3 iterations.
    \PC{remember to (1) replace \reffig{f:repPart}\,(b) by a
    parallelepiped transported, (2) remove references to cigar, (c) raise
    the figure}

\subsection{Periodic orbit theory}
%
% from nsf07th/TEX/intro.tex   sep 26 2006
%

Since its initial formulations by Ruelle\rf{Ruelle76} and
Gutzwiller\rf{gutzwiller71}, the {{\po} theory} has developed into a
powerful theoretical and computational tool for prediction of quantities
measurable in chaotic dynamics. Schematically (a detailed exposition can
be found in \refrefs{DasBuch,PG97}), one of the tasks of a theory of
chaotic systems is to predict the long-time average of an experimentally
measurable quantity  $\obser(x)$ from the spatial and time averages
\beq
\left<\obser\right> = \lim_{n\to\infty} \frac{1}{n} \left<\Obser^n \right>
\,,\qquad
\Obser^n (\ssp_0) = \sum_{k=0}^{n-1}  \obser(\ssp_k)
\,.
\ee{timAver}

What makes evaluation of such averages difficult is chaotic dynamics'
sensitivity to initial conditions; exponentially unstable trajectories
can be tracked accurately only for finite times. The densities of
trajectories $\msr (\ssp, t)$, however, can be well behaved for $t \to
\infty$. Hence the theory is recast in the language of {\em linear}
{\evOper s}  (Liouville, Perron-Frobenius, Ruelle-Araki, $\cdots$)
\beq
\msr(y,t) = [\Lop^t  \msr] (y) =
        \intM{\ssp} \prpgtr{y - \flow{t}{\ssp}} \msr(\ssp,0)
%   = \msr(\flow{-t}{\ssp})
\,,
\label{3.14cc}
\eeq
This {\evOper} assembles the density $\msr (y, t)$ at time $t$ by going
back in time to the density $\msr (\ssp,0)$ at time $t=0$. Here we shall
refer to the integral operator with singular kernel \refeq{3.14cc}
\beq
\Lop^t (x,y) =  \prpgtr{x - \flow{t}{y}}
\ee{TransOp1}
as the {\em \FPoper}, with the subscript ${\it \small det}$ indicating
that this is deterministic, to distinguish it from the noisy \Fokker\
oper\-ator \refeq{DL:dscrt_FP}.

The {\em Koopman operator} action on a {\statesp} function
$\obser(\ssp)$ is to replace it by its downstream value time $t$ later,
$\obser(\ssp) \rightarrow \obser(\ssp(t))$ evaluated at the trajectory
point $\ssp(t)$:
    \index{observable}\index{Koopman operator}\index{operator!Koopman}
\bea
[{\Uop}{} \obser ](x) &=& \obser(\flow{t}{x})
 \,=\, \intM{y} \Uop (x,y) \, \obser(y)
  \continue
\Uop (x,y) &=& \prpgtr{y - \flow{t}{x}}
\,.
\label{APPE3.14a}
\eea
Given an initial density of representative
points $\msr(x)$, the average value of $\obser(x)$ evolves as
\bea
\langle \obser \rangle(t) &=&
 {1\over \left|\msr_\pS\right|} \intM{x} \obser(\flow{t}{x}) \; \msr(x)
    \,=
\, {1\over \left|\msr_\pS\right|}\intM{x} [{\Uop}{} \obser](x)
\,\msr(x)
  \continue
   &=&
 {1\over \left|\msr_\pS\right|}
   \intM{x} dy \,\obser(y)\, \prpgtr{y - \flow{t}{x}}\, \msr(x)
\,.
\nnu
\eea
The `propagator' $\prpgtr{y - \flow{t}{x}}$ can equally well be
interpreted as belonging to the \FPoper\ \refeq{TransOp1}, so the two
operators are adjoint to each other,
\beq
 \intM{x} [\Uop  \obser](x)\, \msr(x)
  = \intM{y} \obser(y)\, [\Lop^t  \msr](y)
\,.
\ee{APPEFPoperKoop}
This suggests an alternative point of view, which is to push dynamical
effects into the density. In contrast to the Koopman operator which
advances the trajectory by time $t$, the \FPoper\
depends on the trajectory point time $t$ in the past.
                                \toCB
Koopman operators are so cool, that it is no wonder that Igor Mezi\'c is
so enamored with Koopmania\rf{LevnMezi08,RoMeBaSchHe09}.

The \FPoper s are non-normal, not self-adjoint operators, so their left
and right eigenvectors differ. The right eigenvectors of a \FPoper\ are
the left eigenvectors of the Koopman, and vice versa. While one might
think of a Koopman operator as an `inverse' of the \FPoper, the
notion of `adjoint' is the right one, especially in settings where flow
is not time-reversible, as is the case for infinite dimensional flows
contracting forward in time and for stochastic flows.

If the map is linear, $\map(\ssp) = \ExpaEig \ssp$, the \FPoper\ is
\beq
\Lop  \circ \msr(x)
    = \int dy \, \delta(x-\ExpaEig y)\msr(y)
    = \frac{1}{|\ExpaEig|} \msr(\frac{x}{|\ExpaEig|})
\,.
\ee{DL:PF}
In the  $|\ExpaEig|>1$  expanding case the right, expanding
deter\-mi\-ni\-stic eigenfunctions\rf{gasp95,eigendensity} are monomials
% \PC{recheck - is it $ k=1,2,\cdots$ ?} \DL{rechecked}
\beq
\msr_k(x)\rightarrow {x^k}/{k!}
    \,,\qquad
    k=0,1,2,\cdots
\,,
\ee{detlim2}
with eigenvalues ${1}/{|\ExpaEig|\ExpaEig^n}$,
while the left, contracting eigenfunctions
are distributions
\beq \msr_k(x)\rightarrow (-1)^k\delta^{(k)}(x)
\,.
\ee{DL:detlim}

In discretizations $\Lop^t (y,\ssp)$ is represented by a matrix with
$y$, $\ssp$ replaced by discrete indices, and integrals over $\ssp$
replaced by index summation in matrix multiplication. Indeed, for
piece-wise linear mappings \FPoper\ can be a finite-dimensional matrix.
For example, consider the expanding 1\dmn\ 2--branch map $f(x)$ with
slopes $\ExpaEig_0>1$ and $\ExpaEig_1 = - \ExpaEig_0/(\ExpaEig_0-1) <-1$:
\beq
f(x)=\left\{ \begin{array}{ll}
  f_0(x) =  \ExpaEig_0 x \,, \quad & x \in \pS_0=[0,1/\ExpaEig_0)     \\
  f_1(x) =  \ExpaEig_1 (1-x) \,,
            % \frac{\ExpaEig_0}{\ExpaEig_0-1} \,,
                            \quad & x \in \pS_1 =(1/\ExpaEig_0,1]
\,.
        \end{array}\right.
\ee{GetU-12.1a}
As in \reffig{f:repPart}\,(a), the \statesp\ (\ie, the unit interval) is
partitioned into two regions $\pS = \{\pS_0,\pS_1\}$. If  density
$\msr(x)$ is a piecewise constant on each partition
 \beq
 \begin{array}{lccll}
 \msr(x) = \left\{
      \begin{array}{lll}
      \msr_0 & \mbox{if\ } x \in \pS_0
                                         \\
      \msr_1  & \mbox{if\ } x \in \pS_1
      \end{array}
              \right. ,
 \earr
 \label{skewUlamDensity}
 \eeq
the {\FPoper}
acts as a
$[2\!\times\!2]$
Markov matrix ${\bf L}$ with matrix elements
\beq
\left( \msr_0 \atop \msr_1 \right)
    \,\to\,
{\bf L} \, \msr
= \MatrixII{1\over |\ExpaEig_0|}{1\over |\ExpaEig_1|}
          {1\over |\ExpaEig_0|}{1\over |\ExpaEig_1|}
            \left( \msr_0 \atop \msr_1 \right)
\,,
\ee{TransfMatSkewUlam}
stretching both $\msr_0$ and $\msr_1$ over the whole unit interval
$\ExpaEig$. Ulam\rf{Ulam60,LM94} had conjectured that successive
refinements of such piece-wise linear coarse-grainings would provide a
convergent sequence of finite-state Markov approximations to the \FPoper.

The key idea of the {\po} theory is to abandon explicit construction of
natural measure (the density functions typically observed in chaotic
systems are highly singular) and instead compute chaotic spatial and time
average \refeq{timAver} from the leading eigenvalue
$
z_0 = z(\beta)
$
of an evolution operator by means of the {\em classical trace
formula}\rf{pexp,DasBuch}, which for map $\map$, takes form
    \PC{crosscheck, still errors}
\beq
    \expct{\obser}
         =
    \left.
        \partial z_0 \over {\partial \beta}
        \right |_{\beta=0}
\,,\qquad
        \sum_{\alpha=0}^\infty {1 \over z -z_\alpha }
        =
        \sum_p \cl{p} \sum_{r=1}^\infty
        { z^{r \cl{p}}e^{r \beta \cdot \Obser_p}
            \over  \oneMinJ{r} }
\,.
% copied from \label{tr-L-cont}
\label{tr-L1}
\eeq
or, even better, by deploying the associated {\Fd}
    \PC{now introduce cycle expansions}
\beq
    \det(1-z\Lop) =
    \exp \left(
        - {
    \sum_{p} \sum_{r=1}^\infty {1 \over r}
        {  z^{r \cl{p}}  e^{r \beta \cdot  \Obser_p}
    \over  \oneMinJ{r}}
    }
    \right)
    \,.
\label{det-tr}
\eeq
These formulas replace the chaotic, long-time uncontrollable flow by its
{\em {\po} skeleton}, decomposing the dynamical \statesp\ into regions,
with each region $\pS_p$ centered on an unstable \po\ $p$ of period
$\cl{p}$, and the size of the $p$ neighborhood determined by the
linearization of the flow around the \po. Here $\monodromy_p$ is the
{\monodromyM} \refeq{jacoB1}, evaluated in the {\po} $p$, the
deter\-mi\-ni\-stic exponential contraction/expansion is characterized by
its Floquet multipliers $\{\ExpaEig_{p,1},\cdots,\ExpaEig_{p,d}\}$, and
$p$ contribution to \refeq{tr-L1} is inversely proportional to its
exponentiated return time (cycle period $\cl{p}$), and to the product of
{expanding} eigenvalues of $\monodromy_p$. With emphasis on {\em
expanding:}~ in applications to dissipative systems such as fluid flows
there will be only several of these, with the contracting directions -
even when their number is large or infinite - playing only a secondary
role.

Periodic solutions (or `cycles') are important because they form the
{skeleton} of the invariant set of the long time dynamics\rf{inv,AACI},
with cycles ordered {hierarchically}; short cycles give dominant
contributions to \refeq{tr-L1}, longer cycles corrections. Errors due to
neglecting long cycles can be bounded, and for hyperbolic systems they
fall off exponentially or even super-exponentially with the cutoff cycle
length\rf{hhrugh92}. Short cycles can be accurately determined and global
averages (such as transport coefficients and Lyapunov exponents) can be
computed from short cycles by means of {\em cycle expansions}%
\rf{inv,AACI,PG97,DasBuch}.
    \PC{add uniform approximations references either here or in
        \refsect{DL:flat_top}}

%{\bf Successes of the periodic orbit theory}.
%
A handful of very special, completely hyperbolic flows are today
mathematically fully and rigorously under control. Unfortunately, very
few physically interesting systems are of that type, and the full picture
is more sophisticated than the cartoon \refeq{tr-L1}.

  % lippolis/Maribor/cont.tex
% $Author: domenico $ $Date: 2012-05-20 12:43:06 -0400 (Sun, 20 May 2012) $

\section{\Fokker\ operator, continuous time formulation}
  \label{DL:contFP_oper}

                        \renewcommand{\version}{
  Predrag                   Apr 24 2012
                        }
% Predrag  from stoch.tex   Sep 24 2011
% Predrag from FP.tex       Jul 21 2011
% Domenico                  Nov  3 2006

%stoch.tex \section{Noisy trajectories and their densities}
%stoch.tex \section{Langevin, \Fokker\ equations}
%stoch.tex \label{DL:FP_evl}

The material reviewed in this appendix is
standard\rf{vKampen92,Risken96,ArnoldL74}, but needed in order to set the
notation for what is new here, the role that \Fokker\ operators play in
defining stochastic neighborhoods of periodic orbits.

Consider a $d$-dimensional stochastic flow
% Langevin flow
\beq
\frac{d \ssp}{dt} = \velField{\ssp} + \hat{\xi}(t)
\,,
\ee{DL:LangFlow}
where the deterministic velocity field $\velField{\ssp}$ is called
`drift' in the stochastic literature, and  $\hat{\xi}(t)$ is additive
noise, uncorrelated in time. A way to make sense of $\hat{\xi}(t)$ is to
first construct the corresponding probability distribution for additive
noise $\xi$ at a short but finite time $\timeStep$. In time $\timeStep$
the deterministic trajectory advances by $\velField{\ssp_n} \,
\timeStep$. As $\timeStep$ is arbitrary, it is desirable that the
diffusing cloud of noisy trajectories is given by a distribution that
keeps its form as $\timeStep \to 0$. This holds if the noise is Brownian,
\ie, the probability that the trajectory reaches $\ssp_{n+1}$ is given by
a normalized Gaussian
\beq
\Lnoise{\timeStep}(\ssp_{n+1},\ssp_n) \,=\,
\frac{1}{N}
\exp\left[
 -\frac{1}{2\,\timeStep} (\xi_n^T {} \frac{1}{\diffTen}{} \xi_n)
    \right]
\,.
\ee{DL:anis_ev}
Here
\(
\xi_n  =  \delta \ssp_n - \velField{\ssp_n} \, \timeStep
\,,
\)
the deviation of the noisy trajectory from the deterministic
one, can be viewed either in terms of velocities
$\{\dot{\ssp},\velField{\ssp}\}$
(continuous time formulation), or finite time maps
$\{\ssp_n \rightarrow \ssp_{n+1},
\ssp_n  \rightarrow \flow{\timeStep}{\ssp_n}\}$
(discrete time formulation),
\beq
\delta \ssp_n = \ssp_{n+1} - \ssp_n \simeq \dot{\ssp}_n \, \timeStep
            \,,\qquad
\flow{\timeStep}{\ssp_n} - \ssp_n  \simeq \velField{\ssp_n} \, \timeStep
\,,
\label{DL:anis_ev1}
\eeq
where
\beq
\{\ssp_0,
\ssp_1,
\cdots,
\ssp_n,
\cdots,
\ssp_k
\} = \{
\ssp(0),
\ssp(\timeStep),
\cdots,
\ssp(n\timeStep),
\cdots,
\ssp(t)
\}
\ee{timeDiscTraj}
is a sequence of $k+1$ points  $\ssp_n = \ssp(t_n)$
along the noisy trajectory, separated by time increments
$\timeStep = t/k$, and the superfix ${}^T$ indicates a
transpose. The probability distribution
$\xi(t_n)$ is characterized by zero mean and covariance  matrix
(diffusion tensor)
% $\diffTen_{ij}=\diffTen_{ji}$
\beq
\expct{\xi_j(t_n)} = 0
        \,,\qquad
\expct{\xi_i(t_m) \, \xi_j^T(t_n)}
        = \diffTen_{ij} \, \delta_{n m}
\,,
\ee{DL:whiteDscr}
where $\expct{\cdots}$ stands for ensemble average over
many realizations of the noise.
For example, in one dimension the white noise $\xi_n=
\ssp_{n+1} -\ssp_n$ for a pure diffusion process (no advection,
$\velField{\ssp_n}=0$)
is
a {normally distributed random variable}, with
{standard normal} ({Gaussian}) probability distribution
function,
\beq
\Lnoise{t}(\ssp,\ssp_0) =
\frac{1}{\sqrt{2 \pi\diffTen t}}
\exp \left[ - \frac{(\ssp-\ssp_0)^2}{2\diffTen t} \right]
\,,
\ee{DL:WienerLevy}
of mean $0$, variance $\diffTen t$,
and standard deviation $\sqrt{\diffTen t}$, uncorrelated in time:
\beq
\expct{\ssp_{n+1} -\ssp_n} = 0
        \,,\qquad
\expct{(\ssp_{m+1} -\ssp_m)(\ssp_{n+1} -\ssp_n)} = \diffTen \, \delta_{mn}
\,.
\ee{DL:whiteDisc}
$\Lnoise{t}(\ssp,\ssp_0)$ describes the
diffusion at any time, including
the integer time increments
$\{t_n\} = \{\timeStep,2\timeStep,\cdots,n\timeStep,\cdots\}$,
and thus provides a bridge between the continuous and
discrete time formulations of noisy evolution.
We have set $\timeStep=1$ in \refeq{DL:whiteDisc}
anticipating
the discrete time formulation of \refsect{FP_evl}.

In physical problems  the diffusion tensor \diffTen\
is almost always anisotropic: for example, the original Langevin
flow\rf{Chandr43} is a continuous time flow in
\{configur\-at\-ion, velocity\} phase space, with white noise
probability distribution $\exp( - \mathbf{x}^2/2 k_B T )$
modeling random Brownian force kicks applied only to the
velocity variables $\mathbf{x}$. In this case one thinks of
diffusion coefficient $\diffCon = k_B T/2$ as temperature. For sake
of simplicity we shall sometimes assume that diffusion
in $d$ dimensions is uniform and isotropic,
$\diffTen(\ssp) = 2\, D \, \mathbf{1}$.
The more general case of a tensor \diffTen\ which is a \statesp\
position dependent but time independent can be treated along the same
lines, as we do in \refeq{AddVariances}. In this case the stochastic
flow \refeq{DL:LangFlow} is written as\rf{Doob42}
\(
d \ssp = \velField{\ssp}\,dt + \sigma(\ssp)\,d\hat{\xi}(t)
\,,
\)
$\sigma(\ssp)$ is called `diffusion matrix,' and the noise is referred to
as `multiplicative.'

% \subsection{Fokker-Planck operator}
% \label{DL:FP_oper}
%
The distribution \refeq{DL:anis_ev} describes how an initial density of
particles concentrated in a Dirac delta function at $\ssp_n$ spreads in
time $\timeStep$. In the \Fokker\ description  individual noisy
trajectories are replaced by the evolution of the density of noisy
trajectories. The finite time \Fokker\ evolution
$
\msr({\ssp,t}) =
\Lnoise{t}  \circ  \msr ({\ssp,0})
$
of an initial density $\msr(\ssp_0,0)$ is obtained by a sequence of
consecutive short-time steps \refeq{DL:anis_ev}
\beq
\Lnoise{t}(\ssp_k,\ssp_0)  =  \int [d\ssp] \,
    \exp\left\{- \frac{1}{2\diffTen\timeStep}
    \sum_{n=1}^{k-1} [\ssp_{n+1} - \flow{\timeStep}{\ssp_n}]^2
        \right\}
\,,
\label{DL:dscrtz_FP}
\eeq
where $t = k \, \timeStep$, and the Gaussian normalization factor in
\refeq{DL:anis_ev} is absorbed into intermediate integrations by defining
as
                                                            \toCB
\bea
[d\ssp] &=&  {N}^{-1}\prod_{n=1}^{k-1} {d \ssp^d_n}
    \continue
      N &=& (2\pi \timeStep)^{d/2} (\det \diffTen)^{1/2}
     \qquad \mbox{anisotropic diffusion tensor $\diffTen$}
    \continue
      &=& (2\pi \diffTen\timeStep)^{d/2}
     \qquad\qquad \mbox{isotropic diffusion}
\,,
\label{GaussMsr1}
\eea
The stochastic flow \refeq{DL:LangFlow} can now
be understood as the continuous time, $\timeStep \to 0$ limit, with the
velocity noise $\hat{\xi}(t)$ a Gaussian random variable of zero mean and
covariance matrix
\beq
\expct{\hat{\xi}_j(t)} = 0
        \,,\qquad
\expct{\hat{\xi}_i(t) \, \hat{\xi}_j(t')}
        = \diffTen_{ij} \, \delta(t-t')
\,.
\ee{DL:white}
It is worth noting that the continuous time flow noise $\hat{\xi}(t)$ in
\refeq{DL:LangFlow} and \refeq{DL:white} is dimensionally a velocity
$[x]/[t]$, while the discrete time noise $\xi_n$ in \refeq{DL:anis_ev},
\refeq{DL:whiteDscr} is dimensionally a length $[x]$. The continuous time
limit of \refeq{DL:dscrtz_FP}, $\timeStep = t/k \to 0$, defines formally
the \Fokker\ oper\-ator
\beq
\Lnoise{t}(\ssp,\ssp_0)  =
  \int [d\ssp] \,
    \exp\left\{- \frac{1}{2\diffTen}
\int_0^t[\dot{\ssp}(\tau)-\velField{\ssp(\tau)}]^2 d\tau\right\}
\ee{DL:path_int}
as a stochastic path (or Wiener)
integral\rf{Shraiman81,Martin73,Risken96} for a noisy flow,
and the associated continuous time \Fokker\ (or forward Kolmogorov)
equation\rf{vKampen92,Risken96,Oksendal03} describes the time evolution
of a density  of noisy trajectories \refeq{DL:LangFlow},
\beq
\partial_t \msr({\ssp},t) + \nabla \cdot \left( {v}({\ssp})
    \msr({\ssp},t)\right)
        \,=\, D\, \nabla^2 \msr({\ssp},t)
\,.
\ee{DL:FP}
The $\timeStep \to 0$ limit and the proper definition of
$\dot{\ssp}(\tau)$ are delicate
issues\rf{Ito,Stratonovich,ArnoldL74,Fox78} of no import for
the applications of stochasticity studied here.

In probabilist
literature\rf{BeGe06} the differential operator
    $-\nabla \cdot \left( {v}({x})\msr({x},t)\right)
    \,+ \, D\, \nabla^2 \msr({x},t)$
    is called `\Fokker\ operator;' here we
reserve the term exclusively for the finite time, `Green function'
integral operator \refeq{DL:path_int}.
The exponent
\beq
-\, \frac{1}{2\diffTen \timeStep}
\left[\ssp_{n+1} - \flow{\timeStep}{\ssp_n}\right]^2
    \,\simeq\,
-\, \frac{1}{2\diffTen}
\left[\dot{\ssp}(\tau)-\velField{\ssp(\tau)}\right]^2 \timeStep
\ee{DL:costFct}
can be interpreted as a {\costFct} which penalizes deviation of the noisy
trajectory $\delta \ssp$ from its deterministic prediction $\vel \,
\timeStep$, or, in the continuous time limit, the deviation of the noisy
trajectory tangent $\dot{\ssp}$ from the deterministic velocity $\vel$. Its
minimization is one of the most important tools of the optimal control
theory\rf{Pontr62,Bell57}, with velocity $\dot{\ssp}(\tau)$ along a trial
path varied with aim of minimizing its distance to the target
$\velField{\ssp(\tau)}$.

The finite time step formulation \refeq{DL:dscrtz_FP} of the \Fokker\
oper\-ator motivates the exposition of \refsect{FP_evl}, which starts by
setting $\timeStep = 1$. In the linearized setting, the two formulations
are fully equivalent.

\remark{A brief history of noise.}{
\label{rem:CostFct}
The {\costFct} \refeq{DL:costFct} appears to have been first
introduced by Wiener as the exact solution for a purely diffusive
Wiener-L\'evy process in one dimension, see \refeq{DL:WienerLevy}.
Onsager and Machlup\rf{Onsager53,Freidlin98} use it in their variational
principle to study thermodynamic fluctuations in a neighborhood of
single, linearly attractive equilibrium point (\ie, without any
dynamics). The dynamical `action' Lagrangian in the exponent of
\refeq{DL:path_int}, and the associated symplectic Hamiltonian were first
written down in 1970's by Freidlin and Wentzell\rf{Freidlin98}, whose
formulation of the `large deviation principle' was inspired by the
Feynman quantum path integral\rf{Feynman65}. Feynman, in turn, followed
Dirac\rf{Dirac33}  who was the first to discover that in the short-time
limit the quantum propagator (imaginary time, quantum sibling of the
Wiener stochastic distribution \refeq{DL:WienerLevy}) is exact.
Gaspard\rf{gasp02} thus refers to the `pseudo-energy of the
Onsager-Machlup-Freidlin-Wentzell scheme.' M.
Roncadelli\rf{Ronc95,DeRonc95} refers to the Fokker-Planck exponent in
\refeq{DL:path_int} as the `Wiener-Onsager-Machlup Lagrangian,'
constructs weak noise saddle-point expansion and writes transport
equations for the higher order coefficients.
} %end \remark{A brief history of noise

\subsection{Ornstein-Uhlenbeck process}
\label{DL:locEig}

%    \PC{eliminate repeated text between here and \refsect{DL:locEig1}}
The variance \refeq{ddQfixed} is stationary under the action
of $\Lnoise{}$, and the corresponding Gaussian is thus an eigenfunction.
Indeed, for the linearized flow the entire eigenspectrum is available
analytically, and as $\covMat_a$ can always be brought to a diagonal,
factorized form in its orthogonal frame, it suffices to understand
the simplest case, the Ornstein-Uhlenbeck process in one dimension.
This simple example will enable us to
show that the noisy measure along unstable directions is described
by the eigenfunctions of the adjoint \Fokker\ operator.

The simplest example of a stochastic flow \refeq{DL:LangFlow}
is the Langevin flow in one dimension,
\beq
\frac{d \ssp}{dt} = \Lyap \, \ssp + {\hat{\xi}(t)}
\,,
\ee{DL:continuous}
with `drift' $\velField{\ssp}$ linear in $\ssp$,
and the single deterministic equilibrium solution $\ssp=0$.
The associated {\Fokker} equation \refeq{DL:FP} is known as
the Ornstein-Uhlenbeck process\rf{OrnUhl30,VK79,gasp95,Wax,Risken96}:
\beq
\partial_t \msr(\ssp,t) + \partial_\ssp (\Lyap \, \ssp \, \msr(\ssp,t))
    \,=\, \diffCon \, \partial_{\ssp}^2 \msr(\ssp,t)
\,.
\ee{DL:FPa}
(Here $\diffTen = 2\diffCon$, $\diffCon =$ Einstein diffusion constant.)
One can think of this equation as the linearization of the {\Fokker}
equation \refeq{DL:FP} around an \eqv\ point. For negative constant
$\Lyap$ the spreading of noisy trajectories by random kicks is balanced
by the linear damping term (linear drift) $\velField{\ssp} = \Lyap \,
\ssp$ which contracts them toward zero. For this choice of
$\velField{\ssp}$, and this choice only, the {\Fokker} equation can be
rewritten as the Schr\"odinger equation for the quantum harmonic
oscillator, with its well-known Hermite polynomial
eigenfunctions\rf{AbrSte64,Takada01} discussed here in \refsect{DL:locEig1}.

The key ideas are easier to illustrate by the noisy, strictly equivalent
discrete-time dynamics of \refsect{DL:locEig1}, rather than by pondering
the meaning of the stochastic differential equation \refeq{DL:FPa}.

\remark{Quantum mechanical analogue.~~}{
\label{rem:QManal}
The relation between Ornstein-Uhlenbeck process and the Schr\"odinger
equation for the quantum harmonic oscillator is much older than quantum
mechanics: Laplace\rf{Laplace1810} wrote down in 1810 what is now known
as the Fokker-Planck equation and computed the Ornstein-Uhlenbeck process
eigenfunctions\rf{Jac97} in terms of Hermite polynomials
\refeq{HermiteOrthon}. According to L Arnold\rf{ArnoldL74} review of the
original literature, the derivations are  much more delicate: the noise
is {\em colored} rather than Dirac delta function in time. He refers only
to the linear case \refeq{DL:continuous} as the `Langevin equation'.
} %end \remark{Quantum mechanical an

  % lippolis/Maribor/% LyapEq.tex
% $Author: predrag $ $Date: 2012-05-12 18:14:16 -0400 (Sat, 12 May 2012) $

                        \renewcommand{\version}{
  Predrag                   Apr 24 2012
                        }

\section{Lyapunov equation}
\label{chap:LyapEq}
% Predrag from noisy blog   25 Feb 2012
% Predrag from noisy blog   Jul 21 2011

In his 1892 doctoral dissertation\rf{Lyapunov1892}
\HREF{http://en.wikipedia.org/wiki/Aleksandr_Lyapunov}
{A.~M.~Lyapunov} defined a dynamical flow to be ``stable in the sense of
Lyapunov'' if the scalar Lyapunov function
\[
V(\ssp) \geq 0
    \,,\qquad
V(0) = 0
\]
computed on the \statesp\ of dynamical flow
\(
\dot{\ssp} = \vel(\ssp)
\)
satisfies the `inflow' condition
\beq
\dot{V} = \frac{\partial V}{\partial \ssp_j}
\vel_j %\frac{d\ssp_j}{dt}
        \leq 0
\,.
\ee{LyapFuncEq}
While there is no general method for constructing Lyapunov functions,
for a linear $d$\dmn\ autonomous flow,
\(
\dot{\ssp} = \Mvar \ssp
\,,
\)
the Lyapunov function can be taken quadratic,
\beq
V(\ssp) = \ssp^T \frac{1}{Q} \ssp
    \,,\qquad Q = Q^T > 0
\,,
\ee{LyapFuncQuad}
and \refeq{LyapFuncEq} takes form
\[
\dot{V} =  \ssp^T \left(\Mvar^T \frac{1}{Q} + \frac{1}{Q} \Mvar\right) \ssp
\,.
\]
                                                            \toCB
Here $\Mvar^T$ is the transpose of the \stabmat\ $\Mvar$, and $Q > 0$ is
the shorthand for matrix being positive definite (or Hurwitz), \ie,
having the entire eigenvalue spectrum, $\sigma(Q) \in \complex_{+}$, in
the right-hand half of the complex plane. Strict positivity $Q > 0$
guarantees that $Q$ is invertible.

The Lyapunov differential equation for a time varying system is
\beq
\dot{Q} = \Mvar Q + Q \Mvar^T + \diffTen
    \,,\qquad
Q(t_0) = Q_0
\,.
\ee{contLyapODE}
For  steady state, time invariant solutions $\dot{Q} = 0$,
\refeq{contLyapODE} becomes the Lyapunov matrix equation
\beq
\Mvar Q + Q \Mvar^T + \diffTen = 0
\,.
\ee{contLyapEq}
The Lyapunov theorem\rf{Lyapunov1892,Bell60} states that a
$[d\!\times\!d]$ matrix $Q$ has all its characteristic roots with real
parts positive if, and only if, for any positive definite symmetric
matrix $\diffTen = \diffTen^T > 0$, there exists a unique positive definite
symmetric matrix $Q$ that satisfies the \emph{continuous Lyapunov
equation} \refeq{contLyapEq}.
The flow is then said to be  asymptotically  stable. In our application
we are given a noise correlation matrix $\diffTen$, and the theorem states
the obvious: the stationary state with a covariance matrix $Q > 0$ exists
provided that the \stabmat\ $\Mvar$ is stable, $\Mvar < 0$. We have to
require strict stability, as $\Mvar \leq 0$ would allow for a noncompact,
Brownian diffusion along the marginal stability eigen-directions.

For a linear discrete-time system
\(
\ssp_{n+1} = \monodromy \ssp_n
\)
the quadratic Lyapunov function \refeq{LyapFuncQuad} must satisfy
\[
V(\ssp_{n+1}) -V(\ssp_n) =
\ssp_n^T
\left(\monodromy^{T} \frac{1}{Q} \monodromy- \frac{1}{Q} \right)
\ssp_n \leq 0
\,,
\]
leading to the \emph{discrete Lyapunov equation}
\beq
Q = \monodromy Q \monodromy^{T} + \diffTen
\,,\qquad \mbox{for any } \diffTen = \diffTen^T > 0
\,.
\ee{discrLyapEq}
% where $\diffTen$ is a symmetric matrix and $\monodromy^{T}$ is the
% transpose of $\monodromy$.
                                                            \toCB
Lyapunov theorem now states that there is a unique solution $Q$, provided
the \jacobianM\ $\monodromy$ is a \emph{convergent matrix}, \ie, a matrix
whose eigenvalues (Floquet multipliers) are all less than unity in
magnitude,
\(|\ExpaEig_i| < 1 \,.\)

% PC note  for Domenico, Apr 28, 2006 -
We note in passing that the effective diffusive width is easily recast
from the discrete map formulation back into the infinitesimal time step
form of \refappe{DL:contFP_oper}:
\beq
\monodromy = e^{\Mvar\timeStep} % = 1+\Mvar\timeStep
    \,,\qquad
\monodromy^T \,=\, e^{\Mvar^T\timeStep} % = 1+\Mvar^T\timeStep
    \,,\qquad
\diffTen \,\to\,  \timeStep\diffTen
\,,
\label{ContTcigar}
\eeq
where $\Mvar = \partial \vel/ \partial \ssp$ is the \stabmat. Expanding
to linear order yields the differential version of the equilibrium
condition \refeq{ddQfixed}
\beq
    0=\Mvar \covMat + \covMat \Mvar^T + \diffTen
    %\,.
\ee{FixedPtCont}
(with the proviso that now $\diffTen$ is covariance matrix
\refeq{DL:white} for the velocity fluctuations). The condition
\refeq{FixedPtCont} is well known\rf{ArnoldL74,Keizer87,Risken96} and
widely used, for example, in the molecular and gene networks
literature\rf{Pauls04,Paul05,Hornung08,BBW09}. In one dimension, $ \Mvar
\to \Lyap$ (see \refeq{stabExpon}), $\diffTen$ diffusion tensor $\to
\diffTen$, and the diffusive width is given by\rf{ArnoldL74}
%	\DL{section 1.8 of Keizer's book,
%	section 3.2 of Risken's with a $-$ sign to be mindful of}
\beq
\covMat = - \diffTen/2\Lyap
\,,
\ee{DiffContT}
a balance between the diffusive spreading $\diffTen$ and the
deterministic contraction rate $\Lyap < 0$. If $\Lyap \to 0$, the measure
spreads out diffusively.

If $\Mvar$ has eigenvalues of both signs, the necessary and sufficient
condition for the existence of a unique solution to time invariant case
\refeq{contLyapEq} is that no two eigenvalues add up to zero\rf{LaTi85},
    \PC{find out what to do in the Hamiltonian case?}
\beq
\eigExp[i]  + \eigExp[j] \neq O \,,\qquad i,j = 1,2,\cdots,d
\,.
\ee{contLyapExist}
For the discrete Lyapunov equation \refeq{discrLyapEq} the condition for
the existence of a unique solution is that no two eigenvalues have
product equal to one,
\beq
\ExpaEig_i \ExpaEig_j  \neq 1 \,,\qquad i,j = 1,2,\cdots,d
\,.
\ee{discLyapExist}
This condition is obviously satisfied in the Lyapunov case, with
\jacobianM\ $\monodromy$ asymptotically stable in the discrete-time
domain (all eigenvalues of $\monodromy$ are strictly inside of a unit
circle).

If $\Mvar$ is stable, the continuous Lyapunov equation \refeq{contLyapEq}
has a unique solution
\beq
Q = \int_0^\infty dt \, e^{\Mvar t}\diffTen e^{\Mvar^{T} t}
\,,
\ee{contLyapEqInt}
and
\[
Q = \sum_{k=0}^\infty \monodromy{}^k  \diffTen (\monodromy^{T}){}^k
\]
is the unique solution of the discrete Lyapunov equation \refeq{discrLyapEq}.
    \PC{remember that in the hyperbolic case one will need the
resonance condition: ``
if and only if eigenvalues $\sigma(\monodromy)$ have no inverse pairs
\(
\ExpaEig_i \, \ExpaEig_j \neq 1
\mbox{ for } \{i, j = 1, \cdots, d\}
\,,
\)
'' and similarly in the continuous time case.
    }

                                                \toCB
In chaotic dynamics we are interested in saddles, \ie, hyperbolic points
with  both expanding and contracting eigen-directions. Given a
$[d\!\times\!d]$ square matrix $\Mvar$ with real elements, let the
numbers of eigenvalues in the left half of the complex plane, the
imaginary axis and the right half of the complex plane be denoted by the
integer triple
\[
\textrm{In}\,\Mvar
= (\textrm{In}\,\Mvar_{-},\textrm{In}\,\Mvar_{0},\textrm{In}\,\Mvar_{+})
%\,,
\]
which counts (stable, marginal, unstable) eigen-directions. Following
J.~J.~Sylvester (1852), this triple is called the \emph{inertia} of
matrix $\Mvar$. In the case of a nondegenerate symmetric bilinear form
(such as the symmetric matrix $Q$) the numbers of positive and negative
eigenvalues  are also known as the signature of the matrix. $\Mvar$ is
said to be (negative) stable if $\Mvar < 0$, \ie, $\textrm{In}\,\Mvar =
(d,0,0)$, and positive stable if $\Mvar > 0$, \ie, $\textrm{In}\,\Mvar =
(0,0,d)$.

The Lyapunov theorem generalized to hyperbolic fixed points is
known as the  Main Inertia Theorem\rf{Taussky61,OsSchn62,CarlSchn62}:
For a given \stabmat\ $\Mvar$ and a noise correlation matrix
$\diffTen = \diffTen^T > 0$, there exists a symmetric $Q$ such that
\(
\Mvar Q + Q\Mvar^T + \diffTen = 0
\)
and
\[
(\textrm{In}\,Q_{-},0,\textrm{In}\,Q_{+})
=
(\textrm{In}\,\Mvar_{+},0,\textrm{In}\,\Mvar_{-})
\,,
\]
if and only if $\Mvar$ has no marginal eigenvalues\rf{AnSo01},
$\textrm{In}\,\Mvar_{0} = 0$. The Lyapunov theorem is a special case: it
states that $Q>0$ if and only if $\textrm{In}\,\Mvar = (d,0,0)$.

We shall solve \refeq{contLyapEq} and diagonalize $Q$ numerically. The
Main Inertia Theorem guarantees that the covariance matrix $Q$ will have
the same number of expanding / contracting semi-axes as the number of the
expanding / contracting eigenvalues of the \stabmat\ $\Mvar$. The
contracting ones define the semi-axes for covariance evolution forward in
time, and the expanding ones the semi-axes for the adjoint evolution.

\remark{Lyapunov equation.~~}{
\label{rem:LyapEq}
The
% \HREF{http://en.wikipedia.org/wiki/Lyapunov_equation}
{continuous Lyapunov equation} \refeq{contLyapEq} is a special case of
the
\HREF{http://en.wikipedia.org/wiki/Sylvester_equation}{Sylvester equation},
\beq
A Q + Q B = C
\ee{SylvEq}
where $A,B,Q,C$ are $[d\!\times\!d]$ matrices, and the discrete Lyapunov
equation \refeq{discrLyapEq} is a special case of Stein's equation. The
Sylvester equation can be solved numerically with the Bartels-Stewart
algorithm\rf{BaSt72}, in full generality, with no assumptions on
degenerate eigenvalues or defective matrices (matrices for which there
are fewer eigenvectors than dimensions). It is implemented in
\texttt{LAPACK}, \texttt{Matlab} and \texttt{GNU Octave}.
Kuehn\rf{Kuehn11} reviews the available numerical methods for solving the
Lyapunov equation. Discrete Lyapunov equation is solved numerically with
the Kitagawa algorithm\rf{Kitagawa77}, using, for example, Mathematica
\HREF{http://reference.wolfram.com/legacy/applications/anm/FunctionIndex/DiscreteLyapunovSolve.html}
{\texttt{DiscreteLyapunovSolve}}.
%
% [2011-07-28 Domenico]
%  From Mathematica online documentation center, DiscreteLyapunovSolve:
% New numerical methods are implemented in Advanced Numerical Methods to
%
The Mathematica solvers for the continuous- and discrete-time Lyapunov equation
implement the Schur method by Bartels and Stewart\rf{BaSt72}, based on the
decomposition of to the real Schur form; and the Hessenberg-Schur method
by Golub, Nash, and Van Loan\rf{GoNaVanLo79} to solve the Sylvester equation,
based on the decomposition of the smaller of two matrices and to the real
Schur form and the other matrix to the Hessenberg form\rf{Datta03}.
% These methods are invoked on numerical matrices for the
% default setting SolveMethod -> Automatic.'.
For continuous Lyapunov equation see
Matlab function
\HREF{http://www.mathworks.com/help/toolbox/control/ref/lyap.html}
{\texttt{lyap}}, which cites
\refref{BrHo75,Barraud77,Hammar82,Higham88,Penzl98,GoNaVanLo79} as
sources.
Related is Matlab function
\HREF{http://www.mathworks.com/help/toolbox/control/ref/covar.html}
{covar} for output and state covariance of a system driven by white noise.
% ChaosBook.org \remark{Routh-Hurwitz criterion for stability
%       of a fixed point.}{ \label{rem:RouthHurwitz}
% Bart Ermentrout:
For a more practical criterion that matrix has roots with
negative real parts, see
%\HREF{http://en.wikipedia.org/wiki/Routh-Hurwitz_stability_criterion}
     {Routh-Hurwitz stability criterion}\rf{Mein95s} on
the coefficients of the characteristic polynomial. The criterion provides
a necessary condition that a fixed point is stable, and determines the
numbers of stable/unstable eigenvalues of a fixed point.
%    } %end \remark{Routh-Hurwitz criterion for stability
} %end \remark{Lyapunov equation

  % lippolis/stoch/flatTopAppe.tex
% $Author: predrag $ $Date: 2012-05-13 11:45:07 -0400 (Sun, 13 May 2012) $
                        \renewcommand{\version}{
  Predrag                   Apr 27 2010
                        }
% formerly known as lippolis/stoch/hypg_laguer.tex
% Domenico                   Apr 14 2010

  \section{Confluent hypergeometric functions and Laguerre polynomials}
  \label{DL:Hypg_Laguer}

Let now $\Lnoise{\dagger}$
transform this new density, around the next pre-image $x_{a-2}=f^{-1}(x_{a-1})$, as
\beq
\Lnoise{\dagger}e^{-\alpha^2 \orbitDist_{a-1}^4} =
\int e^{-\frac{(y-f'_{a-2} \orbitDist_{a-2})^2}{2\diffTen}-\alpha^2y^4}[dy]
\ee{DL:trsf_quartc}
where $\alpha^2=\DDf{a-1}{}^2 /8(Q_a+\diffTen)$. Now change the variable
$\xi=y\sqrt{\alpha/2\diffTen}$, and write the density $\msr_{a-1}(y)$ as a power series,
so that the previous integral reads
    \PC{is Hypg\_Laguer.tex used? You added it to the repository, but do not call it anywhere.
        Seems to be included in flatTopAppe.tex.}
\bea
\Lnoise{\dagger}e^{-\alpha^2 \orbitDist_{a-1}^4} =
\sqrt{\frac{2\diffTen}{\alpha}}\int[d\xi]
e^{-\left(\frac{\xi}{\sqrt{\alpha}}-\frac{f'_{a-2} \orbitDist_{a-2}}{\sqrt{2\diffTen}}\right)^2}
\sum_{n=0}^{\infty}(-1)^n\frac{\left[(2\diffTen)^2\xi^4\right]^n}{n!}
\continue
= \sum_{n=0}^\infty\frac{(-1)^n(4n)!}{n!}(\sqrt{\alpha}f'_{a-2} \orbitDist_{a-2})^{4n}
\sum_{k=0}^{2n}\frac{1}{(4n-2k)!k!}\left(\frac{2\diffTen}{4(f'_{a-2} \orbitDist_{a-2})^2}\right)^k
\label{DL:trsf_quartc2}
\eea
We then group all the terms up to order $O(\diffTen)$ and neglect
$O(\diffTen^2)$ and higher, and call $\eta=\sqrt{\alpha}f'_{a-2} \orbitDist_{a-2}$
\bea
\sum_{n=0}^{\infty}\frac{(-1)^n}{n!}\left(\eta\right)^{4n}
+ 2\diffTen\sum_{n=0}^\infty\frac{(-1)^n(4n)!}{4\left[n!(4n-2)!\right]}\alpha^{1-2n}
\eta^{4n-2} = \continue
e^{-\eta^4} - 2\diffTen\left[3\alpha\eta^2\right]
\Phi\left(\frac{7}{4},\frac{3}{4},-\eta^4\right)
\label{DL:hypergeom}
\eea
where $\Phi$ (also
sometimes called $M$ or ${}_1 F {}_1$ in the literature) is a confluent hypergeometric function of the first kind\rf{Gradshteyn65}, which can be expressed as
% in terms of the Laguerre polynomial $L_1^{-1/4}(\eta^4)$
% and an exponential
%(see \refappe{DL:Hypg_Laguer}):
\beq
\Phi\left(\frac{7}{4},\frac{3}{4},-\eta^4\right)
= \frac{4}{3}e^{-\eta^4}\left(\frac{3}{4}-\eta^4\right)
\ee{DL:hyper_def}
We now want to evaluate the variance of the density
\refeq{DL:hypergeom}
\beq
Q_{a-2} = \frac{\int d\orbitDist_{a-2} \orbitDist_{a-2}^2\msr_{a-2}( \orbitDist_{a-2})}
{\int d \orbitDist_{a-2}\msr_{a-2}(\orbitDist_{a-2})}
\,.
\ee{DL:hyper_var}
It is useful to know, when computing the denominator of \refeq{DL:hyper_var}, that
\beq
\int \eta^2\Phi\left(\frac{7}{4},\frac{3}{4},\eta^4\right)d\eta = 0
\ee{DL:int_zero}
so that
\bea
Q_{a-2} &=& \frac{(\alpha^2f_{a-2}^{'4})^{-1/2}\int\eta^2 e^{-\eta^4}d\eta
- 4D[3f_{a-2}^{'-2}]\int\eta^4\Phi\left(\frac{7}{4},\frac{3}{4},-\eta^4\right)d\eta}
{\int e^{-\eta^4}d\eta}
\continue &=&
\frac{1}{f_{a-2}^{'2}}\left(\frac{\Gamma(3/4)}{\Gamma(1/4)}\frac{1}{\alpha}
 + \diffTen\right) = \frac{Q_{a-1}+\diffTen}{f^{'2}_{a-2}}
\label{DL:eval_var}
\eea
in the last identity we used the definition of $\alpha$ and \refeq{DL:flat_var}.

Next we derive \refeq{DL:hyper_def}, which expresses the a confluent hypergeometric function in terms of an exponential and a Laguerre polynomial.
Start with the identity\rf{Gradshteyn65}:
\beq
\Phi\left(a,b,z\right) = e^z\Phi\left(b-a,b,-z\right)
\ee{DL:hyp_exp}
in particular, the hypergeometric function in \refeq{DL:hypergeom} becomes
\beq
\Phi\left(\frac{7}{4},\frac{3}{4},-\eta^4\right) = e^{-\eta^4}
\Phi\left(-1,\frac{3}{4},\eta^4\right)
\ee{DL:hyp_exp2}
A confluent hypergeometric function can be written in terms of a Laguerre polynomial
\rf{Gradshteyn65}:
\beq
L_n^\alpha(x) = {n+\alpha \choose n}\Phi\left(-n,\alpha +1,x\right),
\,,\qquad
% \ee{DL:hyp_lag}
% where
% \beq
L_n^\alpha(x) = \sum_{m=0}^n (-1)^m {n+\alpha \choose n-m} \frac{x^m}{m!}
\ee{DL:laguerre}
Thus, in our case
\beq
L_1^{-1/4}(\eta^4) =
{1-1/4 \choose 1}
\Phi\left(-1,\frac{3}{4},\eta^4\right)
\ee{DL:hyp_lag2}
and
\beq
\Phi\left(-1,\frac{3}{4},\eta^4\right)e^{-\eta^4}
 = \frac{4}{3}\left(\frac{3}{4}-\eta^4\right)e^{-\eta^4}
 \,.
\ee{DL:hyp_lag_exp}

\bibliographystyle{aipproc}   % if natbib is available
\bibliography{../bibtex/lippolis}

\ifboyscout
\newpage
\section{flotsam}
\label{DL:flotsam}
\input ../noisy/flotsam.tex
\fi

\end{document}